\documentclass[ 
    aps,
    pra,
    twocolumn,
    letterpaper, 
    10pt,
    superscriptaddress, 
    showpacs,
    showkeys,
    notitlepage,
    amsmath, 
    amssymb, 
    floatfix 
]{revtex4-1} 

\usepackage{graphicx}
\usepackage{dcolumn}
\usepackage{bm}
\usepackage{color}
\usepackage{amssymb}
\usepackage{amsmath}
\usepackage{hyperref}
\usepackage{textcomp}

\usepackage{amsfonts}
\usepackage{amsmath}
\usepackage{amssymb}
\usepackage{dsfont}

\newcommand{\bel}{\begin{equation}}
\newcommand{\eel}{\end{equation}}

\renewcommand{\d}[1]{\!d#1\,}

\newcommand{\skyp}[1]{}

\newcommand{\te}{\text}

\newcommand{\fr}{\frac}

\newcommand{\vare}{\varepsilon}

\newcommand{\ee}{\end{equation}}
\newcommand{\be}{\begin{equation}}
\newcommand{\mbf}{\mathbf}

\newcommand{\bal}{\begin{eqnarray} }
\newcommand{\eal}{\end{eqnarray}}
\newcommand{\ba}{\begin{eqnarray*}}
\newcommand{\ea}{\end{eqnarray*}}

\newcommand{\reffig}[1]{Fig.~\ref{#1}}

\newcommand{\ev}[1]{\langle #1 \rangle}
\newcommand{\ket}[1]{| #1 \rangle}
\newcommand{\bra}[1]{\langle #1 |}

\newcommand{\Ket}[1]{\left| #1 \right\rangle}
\newcommand{\Bra}[1]{\left\langle #1 \right|}
\newcommand{\+}{^\dagger}
\newcommand{\s}{^\ast}

\newcommand{\br}{{\mathbf r}}

\newcommand{\pa}{\partial}

\newcommand{\eps}{\varepsilon}

\renewcommand{\d}[1]{\! d#1\,}

\newcommand{\refeq}[1]{Eq.~(\ref{#1})}

\newcommand{\threevector}[3]{
	\left[
		\begin{array}{c}
		#1 \\
		#2 \\
		#3
		\end{array}
	\right]
}

\begin{document} 

\title{ Quantum Optics in Maxwell's Fish Eye Lens with Single Atoms and Photons}

\author{J. Perczel}
\affiliation{Physics Department, Massachusetts Institute of Technology, Cambridge, MA 02139, USA}
\affiliation{Physics Department, Harvard University, Cambridge,
MA 02138, USA}

\author{P. K\'{o}m\'{a}r}
\affiliation{Physics Department, Harvard University, Cambridge,
MA 02138, USA}

\author{M. D. Lukin}
\affiliation{Physics Department, Harvard University, Cambridge,
MA 02138, USA}


\date{\today}

\bigskip
\bigskip
\bigskip
\begin{abstract} 

We investigate the quantum optical properties of Maxwell's two-dimensional fish eye lens at the single-photon and single-atom level. We show that such a system mediates effectively infinite-range dipole-dipole interactions between atomic qubits, which can be used to entangle multiple pairs of distant qubits. We find that the rate of the photon exchange between two atoms, which are detuned from the cavity resonances, is well described by a model, where the photon is focused to a diffraction-limited area during absorption. We consider the effect of losses on the system and study the fidelity of the entangling operation via dipole-dipole interaction. We derive our results analytically using perturbation theory and the Born-Markov approximation and then confirm their validity by numerical simulations. We also discuss how the two-dimensional Maxwell's fish eye lens could be realized experimentally using transformational plasmon optics.

\end{abstract}


\maketitle
 



\section{Introduction}

Maxwell's two-dimensional fish eye is an optical lens with remarkable imaging properties. 
Light emitted from {\it any} point inside the lens refocuses at the antipodal point on the opposite side of the lens. Since J. C. Maxwell's original work that studied ray optics inside the lens \cite{Maxwell1854}, the properties of the fish eye have been analyzed in a variety contexts, including electromagnetic waves \cite{Tai1958,Rosu1994}, scalar waves \cite{Greenwood1999}, quantum mechanics \cite{Makowski2009} and supersymmetry \cite{Rosu1996}. 

More recently, it was proposed that Maxwell's fish eye lens may have the ability to perfectly refocus electromagnetic waves emerging from a point source \cite{Leonhardt2009,Leonhardt2010b,Ma2011}, thereby overcoming the diffraction limit \cite{Born1999}. The idea of perfect imaging with Maxwell's fish eye has generated vigorous debate \cite{Leonhardt2010,Leonhardt2010a,Leonhardt2011b,Tyc2011,Blaikie2010,Merlin2010,Blaikie2011,Kinsler2010,Gonzalez2011,Quevedo-Teruel2012,Tyc2014,Gonzalez2012,Xu2012,Ma2013,Sun2010,Sun2010b,Kinsler2011,Leonhardt2011,Pazynin2012,Liu2013,Tyc2014,Alonso2015,Horsley2015,He2015,Rosenblatt2017,Minano2011,Minano2014,Leonhardt2015,Leonhardt2015a}
. It has focused on how the presence of a point-like detector, placed at the focus point, changes the image formed and whether perfect imaging is an artifact of the detector. On the one hand, it has been argued that the presence of the detector, which can absorb the incoming radiation, is necessary to form a perfect image \cite{Leonhardt2009,Leonhardt2010b,Ma2011,Leonhardt2010,Leonhardt2010a,Leonhardt2011b,Tyc2011}. On the other hand, concerns have been raised that the detector itself would contribute electromagnetic waves to the image formed, giving rise to the apparent subwavelength focus point \cite{Blaikie2010,Tyc2011,Merlin2010,Blaikie2011}. Subsequently, the discussion about perfect imaging has shifted to finding a simple and realistic model for such detectors \cite{Kinsler2010,Gonzalez2011,Quevedo-Teruel2012,Tyc2014,Gonzalez2012,Xu2012,Ma2013}. More recently, it was suggested that perfect imaging may be possible when operating very close to the resonances of the fish eye lens \cite{Minano2011,Minano2014,Leonhardt2015,Leonhardt2015a}.

\begin{figure}[h!]
\begin{center}
\includegraphics[width=8.5cm]{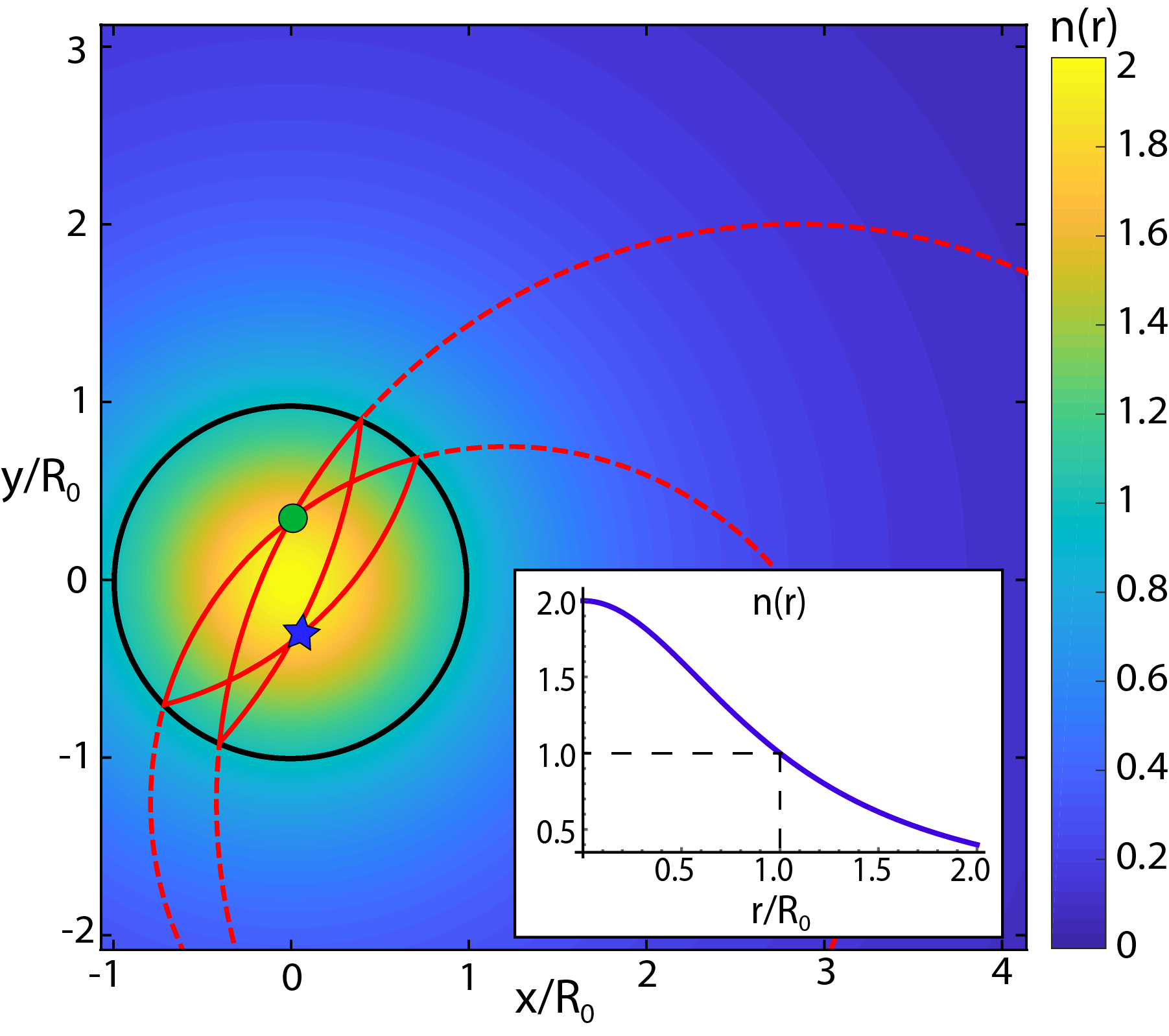}
\end{center}
\caption{(color online) Light rays propagating within the infinite 2D fish eye lens trace out perfect circles (dashed red lines). If a mirror of radius $R_0$ is introduced (black circle), the trajectories remain closed (solid red lines). All light rays emerging from an arbitrary point within the lens (green dot) refocus at the antipodal point (blue star). The color code and the inset show the spatial variation of the refractive index as a function of the radius, where we assume that $n_0=1$ in \refeq{FEn}. For $r>R_0$ the refractive index of the fish eye dips below 1.}
\label{geometry}
\end{figure}
   
In this paper, we study the imaging properties of Maxwell's two-dimensional (2D) fish eye lens at the single-photon level using single atoms. In particular, we assume that both the source and the detector of the photon are individual atoms and thus no ambiguity arises regarding their fundamental properties. One atom, initially in its excited state, emits the photon and the second atom, initially in its ground state, absorbs the photon, storing it in a metastable state for fluorescent readout. This is conceptually the simplest model for a source and a detector \cite{Leonhardt2015}.

We model the 2D lens as an effective photonic cavity filled with an inhomogeneous dielectric material and solve for the atom-photon dynamics inside the lens. Since the rate of photon exchange between the atoms is set by the local electric field strength, the atomic dynamics is a sensitive indicator of the electric field distribution of the photon during absorption. In particular, we find the the photon exchange rate between the two atoms, which are detuned from the cavity resonances, is well described by a simple model, which assumes that the photon is focused to a diffraction-limited area during absorption.


\begin{figure*}[!]
\includegraphics[width=16.5cm]{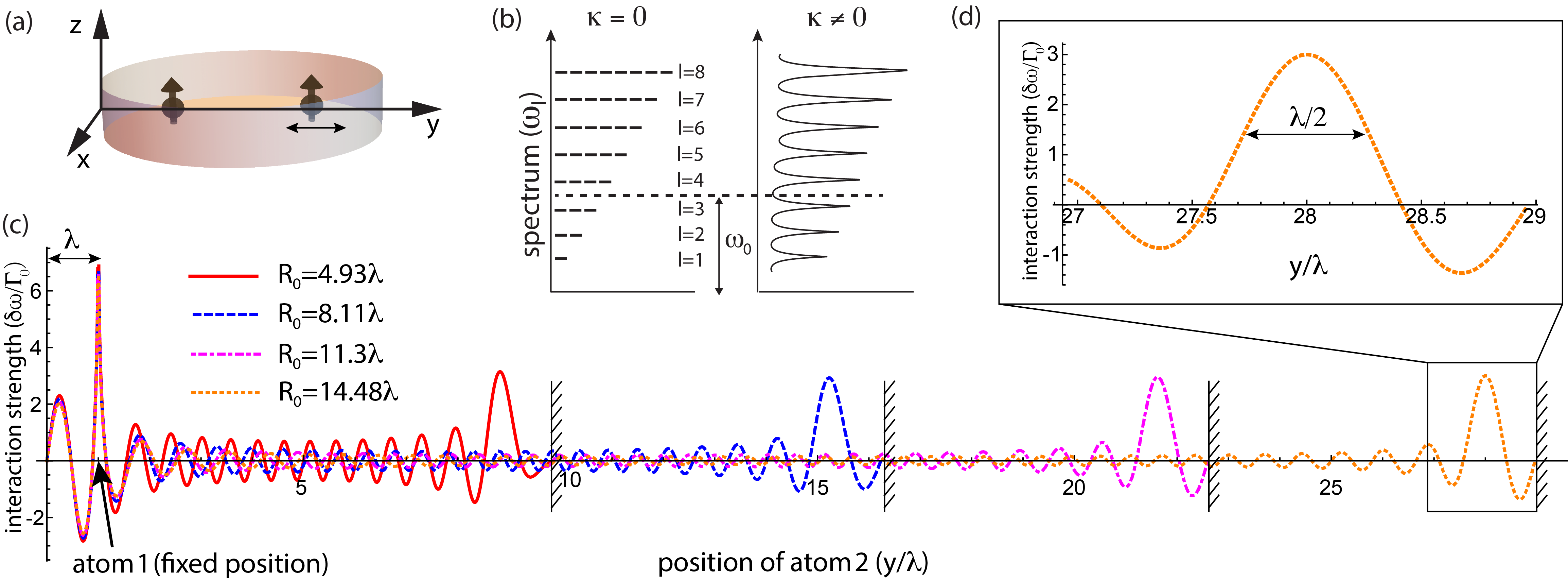}
\caption{(color online) (a) Schematic depiction of the two dipoles embedded in the fish eye cavity, which is surrounded by mirrors on all sides. (b) Spectrum of the cavity $(\omega_l=\sqrt{l(l+1)}c/(R_0n_0)$, $l=1,2,3\dots)$ in the absence ($\kappa = 0$) and the presence ($\kappa \neq 0$) of losses. The atomic resonant frequency $\omega_0$ is tuned between two resonances of the cavity. (c) Strength of the dipole-dipole interaction $\delta\omega(\br_1,\br_2)/\Gamma_0$ between two atoms for four different lens radii: (i) $R_0=4.93\lambda$, (ii) $R_0=8.11\lambda$, (iii) $R_0=11.3\lambda$ and (iv) $R_0=14.48\lambda$, assuming lens thickness $b=\lambda/10$ and $\Gamma_0=d_z^2\omega_0^3/(3\pi \vare_0\hbar c^3)$. The lens radii are chosen such that the transition frequency of the atoms $\omega_0=2\pi c/\lambda$ lies halfway between the resonances of the lens ($l =1,2,3\dots$). In particular, we chose the order parameters (i) $\nu=30.5$, (ii) $\nu=50.5$, (iii) $\nu=70.5$ and (iv) $\nu=90.5$, where $\nu=\fr{1}{2}(\sqrt{16\pi^2(R_0n_0/\lambda)^2+1}+1)$ and $n_0=1$. The atom on the left is positioned exactly $\lambda$ away from the mirror, whereas the position of the second atom is sweeped. The strength of the interaction peaks $\lambda$ away from the opposite mirror surface with a height that is independent of the radius of the lens and the interatomic distance. (d) Enlarged view of the dipole-dipole interaction near the antipodal point, showing that the width of the peak is approximately $\lambda/2$.}
\label{range}
\end{figure*}

We also analyze the capabilities of the fish eye to enhance the interaction between distant atoms. In particular, we show that the dipole-dipole interaction mediated by the fish eye lens is effectively infinite in range. This infinite-range interaction is a consequence of the unique focusing properties of the fish eye lens and is analogous to the infinite-range interactions mediated by quasi-1D waveguides, which have been the subject of extensive research in recent years in the context of hollow \cite{Gomez-Medina2001,Shahmoon2013}, plasmonic \cite{Chang2006,Gonzalez-Tudela2011,Dzsotjan2010}, microwave \cite{Lalumiere2013,vanLoo2013,Inomata2014} and dielectric \cite{Domokos2002,Horak2003,Klimov2004,LeKien2006,Chang2012,Douglas2015} waveguides. Within this model, we quantitatively evaluate entangling operations and discuss a realistic experimental realization.

This paper is organized as follows. In Section \ref{formalism} we discuss the general formalism behind our work and derive the dipole-dipole interaction mediated by the lens between atoms. In Section \ref{results} we discuss the entanglement of atoms within the lens. In Section \ref{realization} we discuss a possible physical realization of the 2D fish eye using transformational plasmon optics.  Key insights of our work are summarized in Section \ref{conclusion}.

\section{General Formalism}\label{formalism}

In this section we describe the general formalism behind our calculations for exploring the quantum optical properties of the system and calculate the dipole-dipole interaction between atoms placed inside the lens.

\subsection{Maxwell's Fish Eye Lens}


The two-dimensional fish eye lens is a dielectric medium of infinite size with refractive index \cite{Leonhardt2010b}
\be\label{FEn}
n(\br)=\fr{2n_0}{1+(r/R_0)^2},
\ee
where $r=\sqrt{x^2+y^2}$, $R_0$ is the natural length scale of the problem and $n_0\geq 1$ can be chosen arbitrarily. We assume $n_0=1$ for all numerical calculations in this paper. In the limit of geometric optics, light rays propagate in perfect circles (Fig.~\ref{geometry}, dashed circles). All rays emitted from a single point inside the lens ultimately meet at the antipodal point. For $|\br|>R_0$ the refractive index varies between ${n_0}$ and 0, which is difficult to achieve in practice. Thus the lens is modified by placing a mirror around the circle of radius $|\mathbf{r}|=R_0$ (black circle in Fig.~\ref{geometry}). In the presence of the mirror the trajectories still remain closed (solid red lines in Fig.~\ref{geometry}) \cite{Leonhardt2009}.

The 2D fish eye can be realized for {\it electromagnetic waves} in a thin disk of radius $R_0$ with a dielectric material of radially varying refractive index given in Eq.~\eqref{FEn}, which is constant along the $\hat{z}$ direction. If all surfaces of the disc are surrounded with mirrors, the TE modes with the lowest frequency will realize the ideal dynamics of the 2D fisheye \cite{Ma2011,Liu2013}. To ensure that a source radiating at frequency $\omega_0$ excites only the lowest TE modes of the lens, the thickness of the disk, $b$, is chosen such that the relevant optical frequency of the source is much smaller than $\pi c/b$, where c is the speed of light in vacuum \cite{Justice2006,Liu2013}. Later, we consider a realistic realization of the two-dimensional fish eye with surface plasmons, where the transverse confinement arises naturally from the confinement of the plasmons to the metal-dielectric interface \cite{Liu2010,Zentgraf2011}.

\subsection{Hamiltonian}

We model the atoms as two-level systems with ground and excited states denoted by $\ket{g}$ and $\ket{e}$, respectively. The Hamitonian describing the evolution of the system composed of the two atoms and the fish eye modes is given by
\be\label{Hamiltonian}
H=H_\te{atom}+H_\te{field}+V,
\ee
where the atoms evolve according to ${H_\te{atom}=\hbar\omega_0\sum_{i=1,2}\Ket{e_i}\Bra{e_i}}$ and the evolution of the electromagnetic field is described by $H_\te{field}=\sum_{l,m}\hbar\omega_l a\+_{l,m}a_{l,m}$, where $a_{l,m}$ is the annihilation operator of an eigenmode of the lens labelled by $(l,m)$. The interaction of the two atoms with the electromagnetic field is given by $V=-\sum_{i=1,2}\mbf{d}_i\cdot\mbf{E}(\br_i)$, where $\mbf{d}_i=d_z(\sigma^\dagger_i+\sigma_i)\hat{z}$ with $\sigma_i=\Ket{g_i}\Bra{e_i}$ and 
$d_z$ is the $z$-component of the dipole moment of the $e\to g$ transition of the atom, $\mbf{E}(\br_i)$ is the electric field operator at position $\br_i$ within the lens, and we neglect variations of the field over the size of the atoms. The two atoms are positioned at $\br_1$ and $\br_2$ (see Fig.~\ref{range}(a)). Note that \refeq{Hamiltonian} describes a closed lossless system composed of the lens and the two atoms with no coupling to free-space modes. Later we will consider how photon loss from the fish eye modes affects our results. 

\subsection{Quantization in the Fish Eye Lens}

We follow the quantization scheme of Glauber and Lewenstein \cite{Glauber1991} to write down the expression for the quantized electromagnetic field $\mbf{E}(\br_i)$ of the lens
\be\label{Efield}
\mbf{E}(\br_i)=i\sum_{l,m}\bigg(\fr{\hbar\omega_{l}}{2\vare_0}\bigg)^{1/2}[a_{l,m}\mbf{f}_{l,m}(\br_i)-a_{l,m}^{\dagger}\mbf{f}^*_{l,m}(\br_i)],
\ee
where $\mbf{f}_{l,m}$ are the classical eigenmodes of the cavity that are solutions of the wave equation
\be\label{waveEquation}
n(\br)^2\,\fr{\omega_{l,m}^2}{c^2}\mbf f_{l,m}(\br) - \nabla\times\left[\nabla\times \mbf f_{m,l}(\br)\right]=0,
\ee
subject to the transversality condition
\be\label{transversal}
\nabla\cdot \left[n(\br)^2\,\mbf f_{l,m}(\br)\right]=0,
\ee
together with the boundary condition that ${\mbf f_{l,m}\cdot \hat z=\mbf f_{l,m}\cdot \hat \phi=0}$ at ${|\br|=R_0}$ due to the presence of the mirror. The position-dependent refractive index $n(\br)$ is given by \refeq{FEn}. The solutions of \refeq{waveEquation} and \refeq{transversal} can be chosen to form an orthonormal set satisfying 
\be
\int_\mathcal{V}d^3r\; n(\br)^2\,  \mbf{f}_{l,m}(\br)\cdot \mbf{f}_{l',m'}^*(\br) = \delta_{ll'}\delta_{mm'},
\ee
where the integral is performed over the quantization volume $\mathcal{V}$. 

Solving these equations, the lowest TE modes of the fish eye take the following form
\be\label{modes}
\mbf{f}_{l,m}(r,\phi)=\sqrt{\fr{2}{bR_0^2n_0^2}}Y_l^m\bigg(\text{arccos}\bigg(\fr{|\br|^2-R_0^2}{|\br|^2+R_0^2}\bigg),\phi\bigg)\hat{z},
\ee
where $Y^m_l(\theta,\phi)$ are the spherical harmonic functions, ${\phi=\arccos(x/|\br|)}$ is the azimuthal angle associated with position $\br$ and the eigenfrequencies are ${\omega_l=c\sqrt{l(l+1)}/(R_0n_0)}$. The modes $\mbf{f}_{l,m}$ are labelled with the rescaled wavenumber $l=1,2,3\ldots$ and the angular momentum index $m$, where ${m(l)=-(l-1),-(l-3),...\,,(l-1)}$ is enforced by the boundary condition $\mbf{f}_{l,m}(R_0,\phi) = 0$. The discrete spectrum of the fish eye is schematically shown in Fig.~\ref{range}(b). The number of degenerate states increases linearly with $l$, since $\sum
_{m(l)}1=l$.

\subsection{Photon transfer between two atoms via dipole-dipole interaction}

In this section, we investigate the resonant transfer of a photon between two atoms via the dipole-dipole interaction, the strength of which we denote by $\delta\omega$.

In quantum optics, the most fundamental model for photon emission and detection assumes that one atom is initially in its excited state $\ket{e_1}$, while the second atom is in its ground state $\ket{g_2}$. When the system evolves coherently in time, the excited atom (virtually) emits the photon and after time $t_{\text{int}}\sim \pi/(2\delta\omega)$ the second atom fully absorbs the photon as its atomic population is transferred to the excited state $\ket{e_2}$ \cite{Milonni1974,Dzsotjan2010}. 

\begin{figure}[h!]
\begin{center}
\includegraphics[width=0.45 \textwidth]{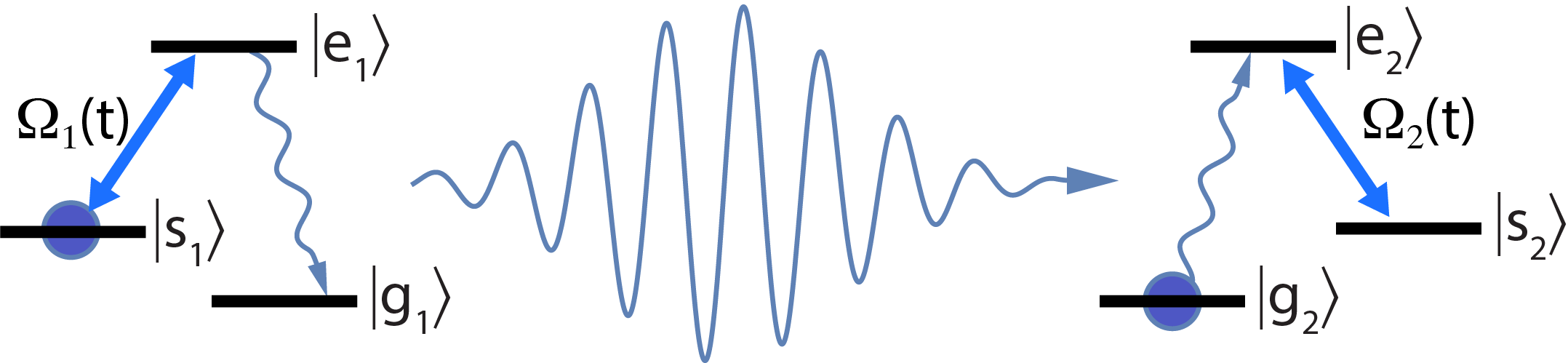}
\end{center}
\caption{(color online) Schematic depiction of a realistic scheme for the photon transfer between the two atoms. The first atom emits the photon, while the second atom fully absorbs it. By applying classical time-dependent control pulses $\Omega_1(t)$ and $\Omega_2(t)$, the transfer can be initiated and the photon can be captured in the metastable state of the second atom, from which the photon can be read out using fluorescence techniques.}
\label{levelstructure}
\end{figure}

Furthermore, by making use of additional metastable states $\ket{s_i}$ with $i=1,2$ (see \reffig{levelstructure}) that only couple to $\ket{e_i}$ via the time-dependent classical control pulse $\Omega_i(t)$ (such that $\Omega_i \gg \delta\omega$), the photon transfer can be performed in a controlled, realistic scheme \cite{Pellizzari1995,Cirac1997,Chang2007}. 
In particular, by adjusting $\Omega_1(t)$ and $\Omega_2(t)$, the photon transfer can be initiated via the excitation of $\ket{e_1}$ and, as the photon is reabsorbed, the atomic population of the second atom can be transferred to the metastable state $\ket{s_2}$. Then, by switching off $\Omega_2(t)$, reemission into the cavity can be prevented. From the metastable state the photon can be read out using standard fluorescence techniques \cite{Fuhrmanek2011,Gibbons2011}. This completes the detection of the photon.

In a standard quantum optical setting, the dipole-dipole interaction between two atoms with level spacing $\omega_0$ between ground $\ket{g_i}$ and excited states $\ket{e_i}$ in any environment can be expressed in terms of the classical Green's function components $G_{\alpha\beta}(\br_1,\br_2,\omega_0)$ (with $\alpha,\beta = x,y,z$) through the following expression \cite{Dung1998,Dung2002,Scheel2008,Buhmann2012}
\be\label{dd}
\delta\omega(\br_1,\br_2)=\fr{d_z^2\omega_0^2}{\hbar\vare_0c^2}\te{Re}\{G_{zz}(\br_1,\br_2,\omega_0)\},
\ee
where we assume that the two atoms are located at $\br_1$ and $\br_2$ and their dipole moments $d_z$ are oriented along the $z$-axis. Note that the real (imaginary) part of the Green's function $G_{zz}(\br_1,\br_2,\omega_0)$ has the simple interpretation of being the $z$-component of the in-phase (out-of-phase) component of the electric field generated at position $\br_2$ within the lens due to the presence of a $z$-oriented point-like dipole at position $\br_1$ radiating at frequency $\omega_0$.\\

We note that, when the classical Green's function of a problem is analytically known, it is typically a simple matter to evaluate \refeq{dd} and find the dipole-dipole interaction between atoms. However, for the fish eye there is debate about what Green's function correctly describes the imaging process. The subtlety of the issue arises from the fact that the fish eye, which models the closed surface of the sphere, is inherently a closed system from which radiation cannot escape in the absence of losses and detectors \cite{Leonhardt2009,Leonhardt2010b,Ma2011}. As mentioned previously, the accurate mathematical modeling of detectors has been a key focus of the discussion regarding perfect imaging \cite{Blaikie2010,Leonhardt2010,Kinsler2010,Leonhardt2010a,Sun2010,Sun2010b,Merlin2010,Blaikie2011,Kinsler2011,Tyc2011,Gonzalez2011,Leonhardt2011,Quevedo-Teruel2012,Pazynin2012,Gonzalez2012,Xu2012,Ma2013,Liu2013,Tyc2014,Alonso2015,Horsley2015,Leonhardt2015,Leonhardt2015a,He2015,Rosenblatt2017}.

Here, since we model both the `source' of the radiation and the `detector' as atoms, the exact expression for the dipole-dipole interaction can be obtained from the standard quantum optical master equation \cite{Gross1982}, where no ambiguity arises in the derivation of the results. Furthermore, as we show below, the expression obtained for the dipole-dipole interaction from the master equation exactly matches one of the two Green's functions discussed extensively in the fish eye literature, allowing us to directly use \refeq{dd}, which substantially simplifies numerical calculations. 

The quantum optical master equation in the Born-Markov approximation, which governs the evolution of the atoms inside the lens, takes the following form in the interaction picture \cite{Gross1982}
\bal\label{ME}
\fr{d{\tilde \rho}}{dt}=-\fr{1}{\hbar^2}\int_0^\infty \!\!\!d\tau\, \te{Tr}
\left[\tilde V(t),\left[\tilde V(t-\tau),\tilde \rho (t)\otimes \ket{0}\bra{0}\right]\right]\!\!,\quad
\eal
where $\ket{0}\bra{0}$ is a projector onto the vacuum state of the lens (i.e. no photons in the lens) and the trace is implied over all photonic Fock states of the lens, $\sum_n\bra{n}...\ket{n}$, and
\be\label{MEfirst}
\tilde \rho(t)=e^{iH_\te{atom}t/\hbar}\rho(t)e^{-iH_\te{atom}t/\hbar},
\ee
and
\be
\tilde V(t)=e^{i[H_\te{atom}+H_\te{field}]t/\hbar}V(t)e^{-i[H_\te{atom}+H_\te{field}]t/\hbar}.
\ee
In \refeq{ME}, the Born approximation was performed by writing the density matrix for the system in the form $\tilde \rho (t-\tau)\otimes \ket{0}\bra{0}$, which amounts to neglecting correlations between the atoms and the electromagnetic modes of the lens  \cite{Gross1982}. The Markov approximation was made by replacing $\tilde \rho(t-\tau)$ by $\tilde \rho (t)$, which is based on the assumption that the atom-field correlation time is negligibly short compared to the time scale on which the system evolves \cite{Gross1982}. The Markov approximation allowed us to self-consistently extend to infinity the upper limit of the integration with respect to $d\tau$. We confirm the validity of the Born-Markov approximation in Section \ref{BornMarkov}.

After performing the trace over the modes of the fish eye lens, we need to evaluate the following standard integral
\be\label{integralDP}
\int_0^\infty d\tau e^{-i(\omega_l\mp \omega_0)\tau}=\pi\delta(\omega_0\mp\omega_l)\pm iP\fr{1}{\omega_0\mp\omega_l},
\ee
where $\delta(x)$ stands for the Dirac delta and $Pf(x)$ denotes the principal value component of the function $f(x)$. Since the spectrum of the fish eye modes (which act as the reservoir for the atoms) is discreet, the Dirac delta and the principal value do not contribute away from resonances and we may simply replace the right-hand side of \refeq{integralDP} with $\pm i/(\omega_0\mp\omega_l)$. More specifically, in the absence of any mechanism for photon loss that would broaden the energy levels, the atoms experience no spontaneous decay or cooperative emission when their transition frequency does not coincide with the resonant frequencies of the lens. The master equation then describes the fully coherent, lossless evolution of the atoms and takes the form
\be
\fr{d\rho}{dt}=\fr{1}{i\hbar}[H_\te{at},\rho]-i\!\!\!\sum\limits_{\substack{i,j=1,2\\i\neq j}} \delta\omega(\br_i,\br_j)\,[\sigma_i^\dagger\sigma_j,\rho],
\ee
where the dipole-dipole interaction between the atoms is given by
\be\label{sumdd}
\delta\omega(\mbf{r}_i,\br_j)  =  \fr{d_z^2}{\hbar\vare_0}\sum_{l,m}\fr{\omega_l^2}{\omega_l^2-\omega_0^2} \,f_{l,m}^*(\mbf{r}_i)f_{l,m}(\mbf{r}_j),
\ee
where the fish eye modes $f_{l,m}(\br)$ are given by \refeq{modes} and the summation runs over all eigenmodes of the fish eye. Given the summation over an infinite number of modes, it is difficult to work directly with the expression given in \refeq{sumdd} and it is desirable to replace it with a simple, closed-form expression. 

As shown in Appendix \ref{transformation}, the right-hand side of \refeq{sumdd} can indeed be replaced by an expression of the same form as \refeq{dd} using a Green's function, where the Green's function is given by the following expression
\bal\label{analytic}
G_{zz}(\br_1, \br_2, \omega) =-\fr{P_\nu(\xi(\alpha_1, \alpha_2)) -P_\nu(\xi(\alpha_1, 1/\alpha_2\s))}{4b\sin(\pi \nu)},\quad\;\;\;
\eal
where $P_\nu$ is the Legendre function of (non-integer) order ${\nu  = \fr{1}{2}(\sqrt{16\pi^2\,( R_0n_0/\lambda)^2  + 1}- 1)}$. Note that the order parameter $\nu$ depends on the atom frequency $\omega_0$ through the free-space wavelength $\lambda=2\pi c/\omega_0$ and the order parameters with integer values ($\nu = 1,2,3\dots$) correspond to the resonances of the lens. We have also defined ${\xi(\alpha_1, \alpha_2) = (|\zeta(\alpha_1, \alpha_2)|^2-1)/(|\zeta(\alpha_1, \alpha_2)|^2 + 1)}$ and ${\zeta(\alpha_1, \alpha_2) = (\alpha_1 -
\alpha_2)/(\alpha_1\alpha_2\s + 1)}$, with ${\alpha_j =\frac{r_j}{R_0} e^{i\phi_j}}$, where $(r_j,\phi_j)$ are the cylindrical coordinates of the positions of the two atoms ($j=1,2$) within the lens. In Eq.~\eqref{analytic} the second term on the right hand side accounts for the presence of the mirror at $|\br|=R_0$, ensuring that the electric field goes to zero \cite{Leonhardt2009}. This Green's function was first derived in Ref.~\cite{Leonhardt2011b}, and is obtained from the canonical equation of the dyadic Green's function in the presence of a single source term \cite{Leonhardt2009,Leonhardt2010a,Ma2011}. This Green's function has been used previously to describe the static electric field distribution inside the lens for the case when a diffraction-limited image forms at the antipodal point in the presence of a classical source and in the absence of a `drain' \cite{Blaikie2010,Leonhardt2010a}. 

Using \refeq{dd} and \refeq{analytic}, the dipole-dipole interaction can be calculated in a straightforward manner within the lens. In Fig.~\ref{range}(c) we plot the strength of the dipole-dipole interaction between two atoms. The position of the first atom is fixed exactly one wavelength away from the mirror and the position of the second atom is varied across the lens. We plot the interaction strength for four different radii of the fish eye. As Fig.~\ref{range}(c) shows, the strength of the dipole-dipole interaction peaks at the antipodal point, exactly one wavelength away from the mirror. 

As noted at the start of this section, in quantum optics the strength of the dipole-dipole interaction sets the rate at which a photon can be resonantly transferred from one atom to the other. Physically, this exchange rate depends on the strength of the photon field at the location of the second atom that absorbs the photon. In general, the smaller the volume the photon is focused to, the larger the field strength gets. Thus, the dipole-dipole exchange rate depends sensitively on the area the photon is focused to. Fig.~\ref{range}(d) provides an enlarged view that shows the dipole-dipole interaction rate -- and thus the electric field strength -- experienced by the second atom near the antipodal point \footnote{Note that since the Green's function in \refeq{analytic} is purely real, the dipole-dipole interaction is simply proportional to the electric field strength.}. The width of the peak is approximately $\lambda/2$, suggesting that the photon is focused to a diffraction-limited area at the location of the second atom. 
These results for the rate of photon transfer are numerically confirmed in Section \ref{BornMarkov}. 

Fig.~\ref{range}(c) also shows that the height of the peak remains constant as the radius of the fish eye and, therefore, the distance between the two atoms is increased. The photon emitted by an atom anywhere within the cavity gets refocused at the antipodal point regardless of the size of the lens. Such infinite range dipole-dipole interaction is a well-known feature of quasi-1D waveguides \cite{Domokos2002,Horak2003,Klimov2004,LeKien2006,Chang2012,Douglas2015,Chang2006,Gonzalez-Tudela2011,Dzsotjan2010,Gomez-Medina2001,Shahmoon2013,Lalumiere2013,vanLoo2013,Inomata2014}. Intuitively, the 2D fish eye lens acts as quasi-1D system due to the fact that the lens mimics the propagation of light on the surface of a sphere \cite{Leonhardt2010b}. Just as in 1D light is confined to propagate along a single axis without dispersion, the same way light emitted from a point on the 2D surface of a sphere is constrained to propagate along the geodesics of the sphere and refocuses at the antipodal point without any dispersion.  

The functional form of the dipole-dipole interaction can also be understood analytically by considering the asymptotic behavior of the Green's function near the source and image points. In particular, note that the source and image points in the lens correspond to $\xi(\alpha_1, \alpha_2)=-1$ and $\xi(\alpha_1, 1/\alpha_2\s)=+1$ respectively \cite{Leonhardt2009}. As $\xi \to -1$ we obtain the asymptotic expansion  \cite{Leonhardt2009,Erdelyi1953}
\bal\label{expansionAtSource}
P_\nu(\xi)  \to \frac{\sin(\nu \pi)}{\pi} \Bigg[&&\log\left(\fr{1+\xi}{2}\right)+F(\nu)\Bigg], 
\eal
where we have defined the function
\bal
F(\nu)=\gamma+2\psi(\nu+1)+\pi \cot(\nu \pi).
\eal
Here $\gamma$ is Euler's constant and $\psi$ is the digamma function. In addition, when $\xi \to 1$  we obtain the asymptotics $P_\nu(\xi)\to 1$ \cite{Leonhardt2009,Erdelyi1953}. Thus, near the source point the first term dominates in \refeq{analytic} and a logarithmic divergence is formed. In contrast, near the image point, the second term dominates and 
we can analytically approximate the Green's function as
\be\label{GreenApproxNew}
G_{zz}\approx -\fr{1}{4b\sin(\pi\nu)}.
\ee
This shows that the absolute value of the Green's function is maximized when the frequency falls half-way between two resonances such that $\nu  = m+0.5$, where $m\in \mathbb N$. Furthermore, this expression also shows that the height of the peak at a given frequency only depends on the transverse confinement of the modes $b$ and is independent of the lens radius $R_0$. Finally, we note that \refeq{GreenApproxNew} also shows that the dipole-dipole interaction is independent of where we place the atoms within the lens as long as they are situated at antipodal points.

\subsection{Spontaneous and cooperative decay of atoms}\label{spontAndCoop}

In all calculations so far, we assumed that the fish eye lens is completely isolated from its surrounding environment and the photon cannot leak out of the cavity. Here, we next consider the situation when the lifetime of the eigenmodes of the fish eye are finite e.g. due to the imperfection of the mirrors and dissipation in the dielectrics. We account for the gradual loss of photons from the fish eye modes by modifying the Hamiltonian in \refeq{Hamiltonian} with a non-Hermitian term of the following form
\be\label{modeLosses}
H_\te{field}=\sum_{l,m}\hbar(\omega_l-i\kappa) a\+_{l,m}a_{l,m},
\ee
where $2\kappa$ sets the rate of decay from the modes, which is assumed to be frequency-independent in the range of interest. The decay of the cavity modes broadens the discrete energy levels of the fish eye, creating a continuous spectrum, as shown schematically in Fig.~\ref{range}(b).   

With this modification, we can re-derive the master equation from \refeq{MEfirst}. We evaluate the following integral
\be
\int_0^\infty d\tau e^{-i(\omega_l\mp \omega_0)\tau}e^{-\kappa\tau}=\fr{1}{i(\omega_l\mp\omega_0)+\kappa},
\ee
and after neglecting the off-resonant decay terms \cite{Gross1982} we obtain the master equation in the following form
\bal
&&\fr{d\rho}{dt}=\fr{1}{i\hbar}[H_\te{at},\rho]-i\!\!\!\sum\limits_{\substack{i,j=1,2\\i\neq j}} \delta\omega(\br_i,\br_j)\,[\sigma_i^\dagger\sigma_j,\rho]\nonumber\\
&&\qquad\quad-\sum\limits_{\substack{i,j=1,2\\i\neq j}}\Gamma (\br_i,\br_j)\left(\sigma_i\rho\sigma_j^\dagger-\fr{1}{2}\left\{\sigma_i^\dagger\sigma_j,\rho\right\}\right)\!,\quad
\eal
where the rate of decay is given by
\bal\label{Gcoop}
\Gamma(\mbf{r}_i,\br_j)=\fr{d_z^2}{\hbar\vare_0}\sum_{l,m} \kappa\,f_{l,m}^*(\mbf{r}_i)f_{l,m}(\mbf{r}_j) (L_l^++L_l^-),\quad\;\;\;
\eal
and the modified dipole-dipole interaction is given by
\bal\label{lossdd}
\delta\omega(\mbf{r}_1,\br_2) = \fr{d_z^2}{2\hbar\vare_0}\sum_{l,m}\omega_l \,f_{l,m}^*(\mbf{r}_1)f_{l,m}(\mbf{r}_2)\big(D_l^++D_l^-\big),\qquad
\eal
where we have defined
\bal
L_l^\pm=\fr{\mp\omega_l}{\kappa^2+(\omega_l\pm\omega_0)^2}\;\; \te{and}\;\; D_l^\pm=\fr{\omega_l\pm\omega_0}{\kappa^2+(\omega_l\pm\omega_0)^2}.\quad\;\;
\eal
Since we are now including losses in the system, the excited states of the two atoms can irreversibly decay into the eigenmodes of the lens and leave the cavity, leading to non-zero single atom decay $\gamma(\br_i)=\Gamma(\br_i,\br_i)$ (with ${i=1,2}$) and cooperative decay $\gamma_\text{coop}(\br_1,\br_2)=\Gamma(\br_1,\br_2)$. The single atom decay $\gamma$ describes how quickly an excitation decays from state $\ket{e}$ of an individual atom to the fish eye modes, whereas the cooperative decay $\gamma_\text{coop}$ governs the coherent joint emission of the two atoms into the modes leading to super ($\gamma+\gamma_\te{coop}$) and subradiant decay ($\gamma-\gamma_\te{coop}$) of the symmetric and anti-symmetric superpositions of the two atoms, respectively \cite{Gross1982}. 

As for the lossless case, it is desirable to find closed-form expressions to replace the expressions that involve infinite summations on the right-hand side of \refeq{Gcoop} and \refeq{lossdd}. As shown in Appendix \ref{transformation}, the decay rates and the dipole-dipole interaction can be expressed using the Green's function of \refeq{analytic} in the following form
\bal\label{acoopMain}
\Gamma(\mbf{r}_i,\br_j)\!=\!\fr{2d_z^2}{\hbar\vare_0 c^2}\te{Im}\{(\omega_0+i\kappa)^2G_{zz}(\br_i,\br_j,\omega_0+i\kappa) \},\quad\;\;
\eal
and
\bal\label{addMain}
\delta\omega(\mbf{r}_i,\br_j)\!=\!\fr{d_z^2}{\hbar\vare_0c^2}\te{Re}\{(\omega_0+i\kappa)^2G_{zz}(\br_i,\br_j,\omega_0+i\kappa)\}.\;\quad\;
\eal
These simple, analytic expressions provide a convenient way to calculate the quantum optical properties of atoms inside the lossy fish eye lens and to study the atomic dynamics. 

We also note that when $\kappa\ll \omega_0$, Eq.~\eqref{acoopMain} and Eq.~\eqref{addMain} can be approximated as
\bal\label{acoopMainApprox}
\Gamma(\mbf{r}_i,\br_j)\approx\fr{2d_z^2\omega_0^2}{\hbar\vare_0 c^2}\te{Im}\{G_{zz}(\br_i,\br_j,\omega_0+i\kappa) \},\quad
\eal
and
\bal\label{addMainApprox}
\delta\omega(\mbf{r}_i,\br_j)\approx\fr{d_z^2\omega_0^2}{\hbar\vare_0c^2}\te{Re}\{G_{zz}(\br_i,\br_j,\omega_0+i\kappa)\}.\quad
\eal
Eq.~\eqref{acoopMainApprox} and Eq.~\eqref{addMainApprox} suggest an alternative way of accounting for the loss of photons from the modes of the fish eye. In particular, it can be shown (see Appendix \ref{transformation}) that ${G_{zz}(\br_i,\br_j,\omega_0+i\kappa)}$ is the Green's function of the fish eye lens with the following complex refractive index
\be
\tilde n(\br)=n(\br)(1+i\alpha),
\ee
where 
\be
\alpha=\kappa/\omega_0,
\ee
and $n(\br)$ is given by \refeq{FEn}. Therefore, the loss of photons from the modes of the fish eye can also be thought to arise from material absorption in the dielectric \cite{Leonhardt2009}. This is a key observation, which allows us to associate a $\kappa$ value with material absorption and, therefore, treat all losses that contribute to photon decay from the fish eye modes in a unified manner. In particular, even if different loss processes are present, e.g. material absorption and leakage through the mirror, we can still associate a $\kappa$ value with each of these processes and calculate the total decay rate via
\be
\kappa_\text{total} = \kappa_\text{abs}+\kappa_\text{mirror},
\ee
which can be substituted into \refeq{acoopMainApprox} and \refeq{addMainApprox} to calculate the relevant atomic properties in the lossy lens. This will be particularly useful when we consider a possible physical realizations of the fish eye lens with plasmons (see Section \ref{realization}).    

Furthermore, we can also find how $\nu$, $\Gamma$ and $\delta\omega$ scale with $\alpha$ for system parameters of interest. First, we note that ${16\pi^2(R_0/\lambda)^2\gg 1}$, whenever ${\lambda\lesssim R_0}$. Assuming $\alpha \ll 1$, to first order in $\alpha$ we find that
\be\label{nuApprox}
\nu \approx \fr{2\pi R_0}{\lambda} (1+i\alpha).
\ee
Assuming that $\text{Re}[\nu]=m+0.5$ with $m\in \mathbb N$ (which corresponds to tuning the atomic frequency between two resonances), from \refeq{GreenApproxNew} we obtain that, to lowest order in $\alpha$, the following approximation holds at the image point ($\br_1=-\br_2$) 
\bal
G_{zz}(\br,-\br,\omega_0+i\kappa)&\approx& -\fr{1}{4b\sin(\pi\nu)}\nonumber\\
 &\approx& \mp \fr{1}{4b(1+(2\pi^2R_0\alpha/\lambda)^2)},
\eal
where the choice of sign $\mp$ depends on whether $m$ is even or odd. This is a purely real quantity and, therefore, from \refeq{acoopMainApprox} and \refeq{addMainApprox} we find that the cooperative decay is given by
\bal\label{coopscale}
\gamma_\text{coop}=\Gamma(\mbf{r},-\br)\approx 0,
\eal
and the dipole-dipole interaction takes the form
\bal\label{ddscale}
\delta\omega(\mbf{r},-\br)\approx \mp\fr{d_z^2\omega_0^2}{\hbar\vare_0c^2} \fr{1}{4b(1+(2\pi^2R_0\alpha/\lambda)^2)}.\quad
\eal

Finally, we can find the single atom decay rate $\gamma$ by substituting $\br_i=\br_j$ into \refeq{acoopMainApprox} and substituting  \refeq{expansionAtSource} and \refeq{nuApprox} into \refeq{analytic}. We find that to leading order in $\alpha$ the following approximation holds
\be\label{decayscale}
\gamma=\Gamma(\mbf{r},\br)\approx \fr{d_z^2\omega_0^2}{\hbar\vare_0c^2}\fr{\pi^2R_0\alpha}{ b \lambda}.
\ee

\section{Entanglement of atoms}\label{results}

Structures that mediate long-range dipole-dipole interactions are of significant interest in quantum information processing, as such interactions make it possible to entangle \cite{Shahmoon2013} and perform deterministic phase gates between distant atoms \cite{Dzsotjan2010}. In this section, we characterize the potential of the fish eye to entangle distant atomic quits. We focus on the simple case of a single excitation being exchanged between two atoms due to the dipole-dipole interaction. In what follows, for simplicity we assume that the two atoms are located at antipodal points (i.e. $|\br_1|=|\br_2|$ and $\phi_1=\phi_2+\pi$) and, therefore, $\gamma=\Gamma(\br_1,\br_1)=\Gamma(\br_2,\br_2)$.

In the absence of a driving field, the no-jump evolution of the system can be described by a non-Hermitian effective Hamiltonian of the form \cite{Brennen2000}
\bal\label{nonHermitian}
H_0&=&(\hbar\omega_{0}-i\gamma)\ket{e_1,e_2}\bra{e_1,e_2}\nonumber\\
&+&\left(\delta\omega-i(\gamma+\gamma_\te{coop})/2\right)\ket{+}\bra{+}\nonumber\\
&+&\left(-\delta\omega-i(\gamma-\gamma_\te{coop})/2\right)\ket{-}\bra{-},
\eal
where we have defined ${\ket{\pm}=(\ket{e_1,g_2}\pm\ket{g_1,e_2})/\sqrt{2}}$, and recall from the previous section that $\gamma_\text{coop}=\Gamma(\br_1,\br_2)$ and $\delta\omega(\br_1,\br_2)$ stand for the cooperative decay and dipole-dipole interaction of the atoms, respectively. 
Note that the overall decrease of population in \refeq{nonHermitian} due to the non-Hermitian terms reflects the gradual loss of the photonic excitation from the cavity.

\begin{figure}[h!]
\begin{center}
\includegraphics[width=8.5cm]{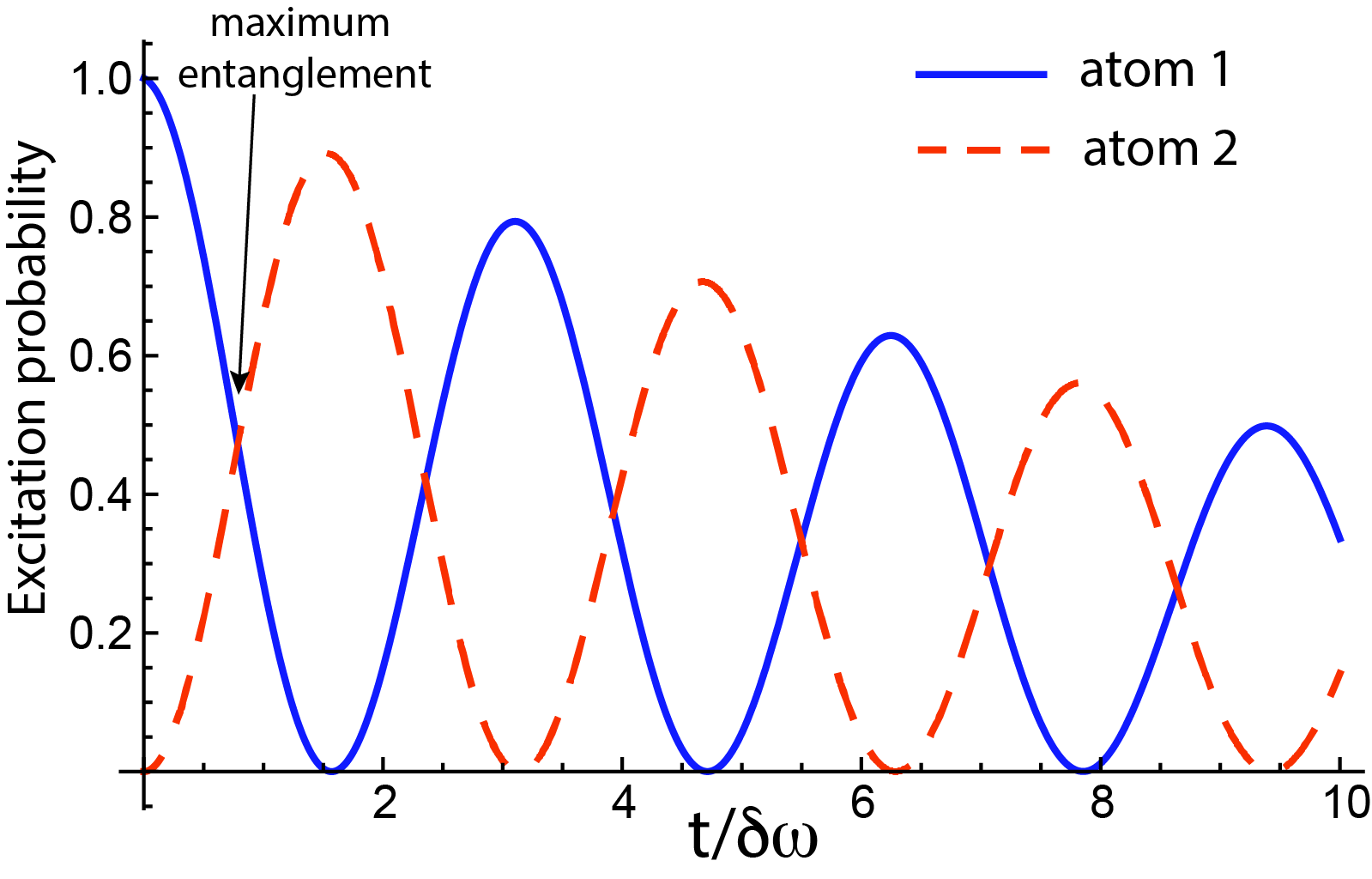}
\end{center}
\caption{(color online) Excitation probability of two atoms within the cavity as a function of time. Initially, atom 1 is excited and atom 2 is in its ground state. As the system evolves, the two atoms repeatedly exchange a photon via the dipole-dipole interaction. The photon gradually decays from the cavity modes, leaving the atoms in their ground states. A fully entangled state with maximal fidelity is formed at  $ t=\pi/(4\delta\omega)$ (see arrow). The plot was obtained for $R_0=3.34\lambda$ with a cavity loss rate of $\alpha = \kappa/\omega_0=5\times 10^{-4}$, assuming that the two atoms are located at two antipodal points within the lens such that $|\br_1|=|\br_2|=0.27R_0$ and $\phi_1=\phi_2+\pi$.   }
\label{trajectory}
\end{figure}

Assuming that at $t=0$ the two atoms are in the state ${\ket{\psi(0)}=\big|e_1,g_2\big>=(\ket{+}+\ket{-})/\sqrt{2}}$, the time evolution of the atomic wavefunction is governed by
\bal
\ket{\psi(t)}&=&\fr{1}{\sqrt{2}}\Big(e^{-i[\delta\omega-\fr{i}{2}(\gamma+\gamma_\text{coop})]t}\ket{+}\nonumber\\
&&\qquad\qquad\quad+e^{-i[-\delta\omega-\fr{i}{2}(\gamma-\gamma_\text{coop})]t}\ket{-}\Big),
\eal
which, upon substitution, yields
\be
\ket{\psi(t)}=C_+(t)\ket{e_1,g_2}+C_-(t)\ket{g_1,e_2},
\ee
where
\be
\left|C_\pm(t)\right|^2=\fr{e^{-\gamma t}}{2}\left[\te{cosh}(\gamma_\text{coop} t)\pm\te{cos}(2\delta\omega t)\right].
\ee
The expressions $\left|C_+\right|^2$ and $\left|C_-\right|^2$ give the excitation probability of atom 1 and atom 2, respectively, as a function of time. In \reffig{trajectory} we plot the excitation probability of the two atoms as a function of time. As the plots shows, the photon is coherently exchanged a number of times between the two atoms before it gradually decays from the cavity modes.  

\begin{figure}[h!]
\begin{center}
\includegraphics[width=8.5cm]{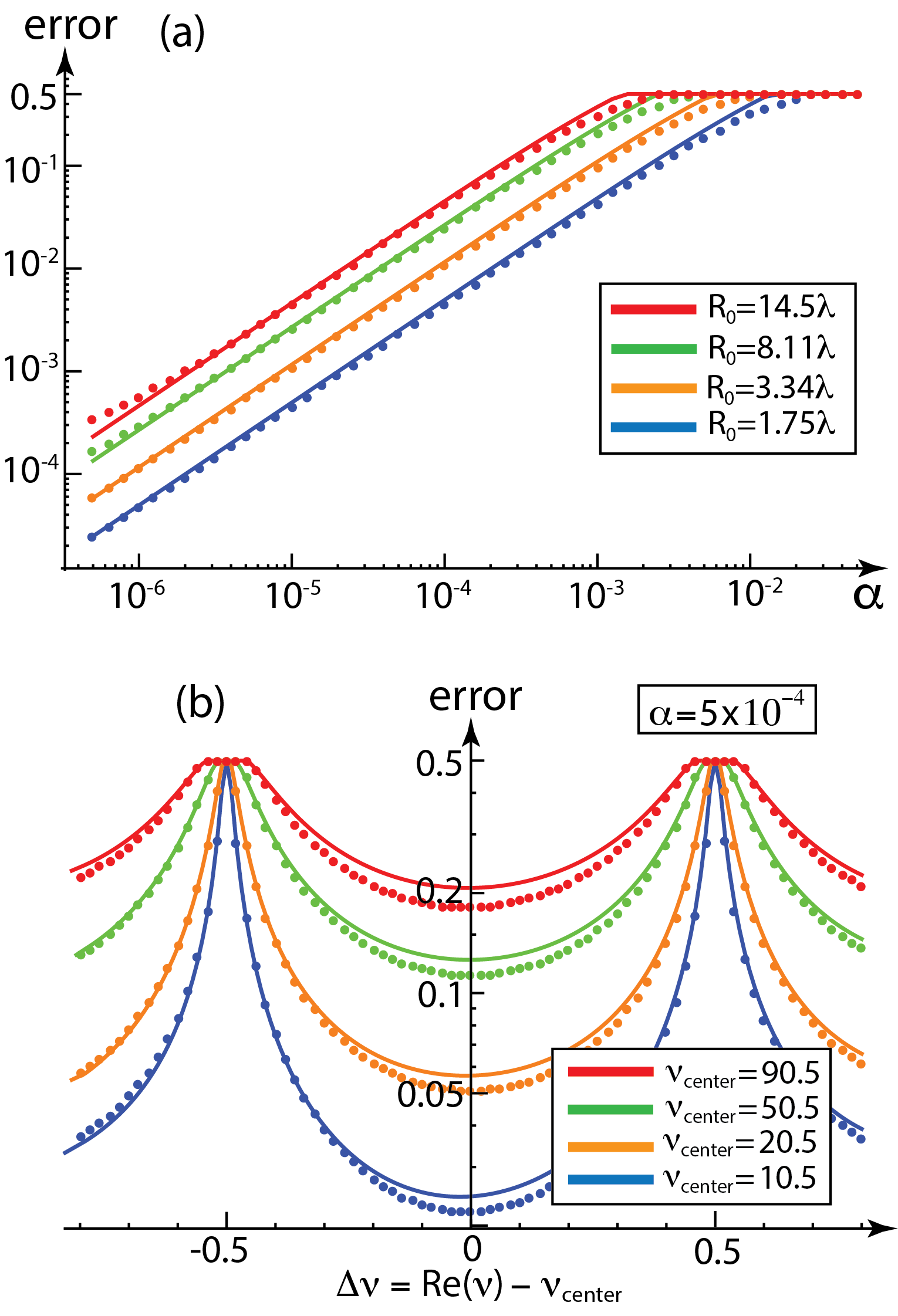}
\end{center}
\caption{(color online) Error $(1-F)$ of the entangling operation between two qubits located at two antipodal points within the lens ($|\br_1|=|\br_2|=0.27R_0$ and $\phi_1=\phi_2+\pi$). (a) Error of the entangling operation as a function of the cavity loss rate $\alpha=\kappa/\omega_0$ for four different lens sizes $R_0\in\{1.75,3.34,8.11,14.5\}\lambda$, where the $R_0/\lambda$ ratio was chosen such that ${\text{Re}(\nu)=\fr{1}{2}(\sqrt{16\pi^2(R_0/\lambda)^2+1}+1)\in \{10.5,20.5,50.5,90.5\}}$. The error increases as the losses and lens radii increase. (b) Error of the entangling operation for a fixed loss rate $\alpha=5\times 10^{-4}$ as a function of the detuning $\Delta\nu=\text{Re}(\nu)-\nu_\te{center}$, where $\nu_\te{center}\in\{10.5,20.5,50.5,90.5\}$. Error is plotted for the same four lens radii as in (a). The error increases with radius and as the frequency approaches one of the resonances. Analytic (numerical) results are shown with solid (dotted) lines. Good agreement is obtained between analytic and numerical data, confirming the validity of the Born-Markov analysis.}
\label{compare}
\end{figure}

During time evolution, the state $\ket{\psi(t)}$ will have maximal overlap with the maximally entangled state ${\ket{\xi}=\left(\ket{e_1,g_1}-i\ket{g_1,e_1}\right)/\sqrt{2}}$ when $\left|C_+(t)\right|=\left|C_-(t)\right|$, which happens when $2\delta\omega t\approx\fr{\pi}{2}+m\pi$, where $m\in \mathbb{Z}$. Since in the presence of losses the fidelity decreases over time, we choose $m=0$.  Thus, the time needed to reach the maximal overlap with the entangled state is $t_0=\pi/(4\delta\omega)$  (see arrow in \reffig{trajectory}) and the maximum fidelity of the entanglement operation will be
\bal\label{fidelityE}
\mbf{{\it F}}=\big|\langle\xi\ket{\psi(t_0)}\big|^2
=\te{exp}\left(-\fr{\pi}{4}\left|\fr{\gamma}{\delta\omega}\right|\right)\te{cosh}\left(\fr{\pi}{4}\left|\fr{\gamma_\te{coop}}{\delta\omega}\right|\right)\!.\qquad
\eal
Eq.~\eqref{fidelityE} gives a simple, analytic expression for the fidelity of the entangling operation in terms of $\gamma=\Gamma(\br_i,\br_i)$, $\gamma_\te{coop}=\Gamma_\te{coop}(\br_i,\br_j)$ and $\delta\omega(\br_i,\br_j)$, which can be evaluated analytically through \refeq{acoopMain} and \refeq{addMain}. Here, the key figure of merit is the ratio $\beta=\delta\omega/(\gamma+\gamma_\te{coop})$. If the frequency of the atoms is chosen to lie half-way between two resonances of the fish eye (see Fig.~\ref{range}(b)), the single atom decay $\gamma$ and the cooperative decay $\gamma_\te{coop}$ are small and the dipole-dipole interaction dominates \cite{Brennen2000}. Intuitively, in the absence of losses ($\gamma = \gamma_\te{coop}=0$), the fidelity of the entangling operation is 1.

In Fig.~\ref{compare}(a) we plot the error in the entangling operation $(1-F)$ for four different lens radii as a function of $\alpha$, where $\alpha=\kappa/\omega_0=1/Q$ is the inverse of the cavity Q-factor, characterizing the ratio of the lifetime of the eigenmodes of the lens to the frequency of the excitation. For all lens sizes, the position of the two atoms is fixed at two antipodal points such that $|\br_1|=|\br_2|=0.27R_0$ and $\phi_1=\phi_2+\pi$. 
The ratio of the lens radius to the transition wavelength ($R_0/\lambda$) was chosen such that the real part of the order parameter $\nu=\fr{1}{2}(\sqrt{16\pi^2(R_0/\lambda)^2+1}+1)$ associated with the atomic frequency falls half-way between two resonances of the fish eye for all four lens radii (i.e. $\text{Re}(\nu)=q+0.5$ with $q\in\{10,20,50,90\}$, where note that for $\text{Re}(\nu)=1,2,3\dots$ the transition frequency $\omega_0$ is resonant with one of the eigenenergies $\omega_l$ of the lens). Clearly, the error increases with increasing $\alpha$ and increasing $R_0$ (i.e. increasing interatomic distance). The maximal value of the error is $0.5$, which is reached when $\beta$ becomes so small that the initial state has the highest fidelity ($F=|\langle\xi\ket{\psi(0)}|^2=0.5$).

Fig.~\ref{compare}(b) shows the error for a fixed value of ${\alpha=5\times 10^{-4}}$ for the same four lens radii as in (a) and the same antipodal atomic positions. The error is now plotted as a function of the detuning $\Delta\nu=(\text{Re}(\nu)-\nu_\te{center})$, where $\nu_\te{center}=q+0.5$ with $q\in\{10,20,50,90\}$. Clearly, the error is minimal half-way between the resonances and increases as the frequency approaches the resonances.

\begin{figure}[h!]
\begin{center}
\includegraphics[width=8.5cm]{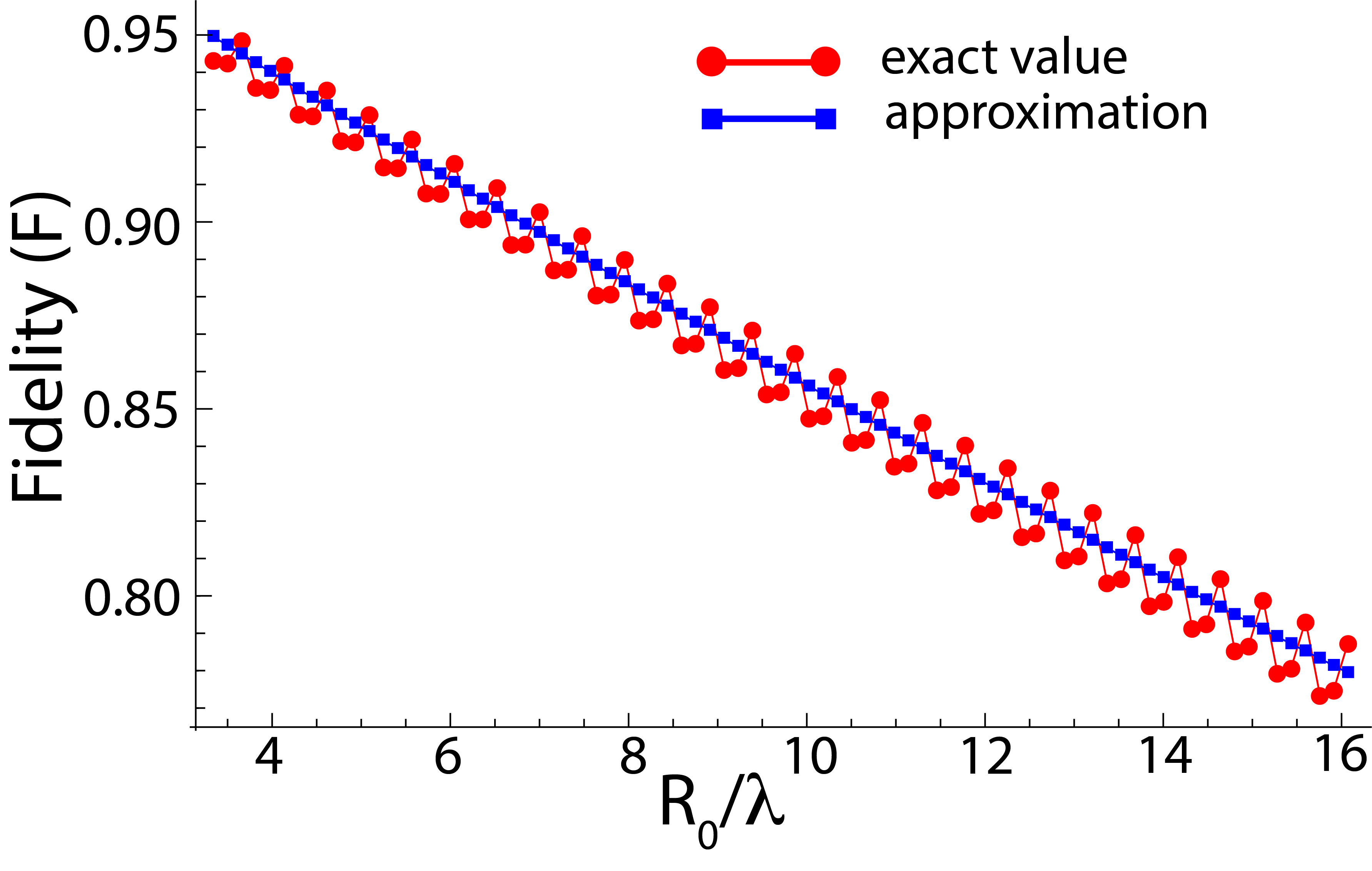}
\end{center}
\caption{(color online) Maximum fidelity of the entanglement operation as a function of the lens radius. The fidelity was evaluated at discrete values of $R_0/\lambda$ that correspond to tuning the atomic frequency half-way between two resonances, i.e. ${\text{Re}[\nu]=m+0.5}$, where $m$ is an integer. The red line marked with circles was obtained from the exact analytical expression in \refeq{fidelityE}, whereas the blue line marked with squares was obtained from the approximate expression in \refeq{fidelityApprox}. Good agreement is obtained between the two curves. The loss rate was assumed to be $\alpha = \kappa/\omega_0=5\times 10^{-4}$.   }
\label{fidelity_scaling}
\end{figure}

To gain further insight, we assume that the atomic frequencies lie between two resonances of the lens and obtain the scaling of the fidelity with system parameters by substituting \refeq{coopscale}, \refeq{ddscale} and \refeq{decayscale} into \refeq{fidelityE}. We obtain the following simple expression
\be\label{fidelityApprox}
F = e^{-\pi^3R_0\alpha/\lambda}.
\ee
In Fig.~\ref{fidelity_scaling} we plot the fidelity of the entangling operation as a function of the lens radius using both the exact expression in \refeq{fidelityE} and the analytic approximation in \refeq{fidelityApprox}. Very good agreement is observed between the two curves.

Finally, we note that the fish eye lens could be used for entangling many pairs of atoms simultaneously. As the radius of the fish eye is increased, the dipole-dipole interaction at all points further than $\lambda/2$ away from the antipodal point monotonically decreases. For lens radii with $R_0>5\lambda$, the dipole-dipole interaction at the antipodal point is an order of magnitude larger than anywhere else in the cavity (see Fig.~\ref{range}(c)). Thus, by placing numerous pairs of atoms into the cavity simultaneously, they can be entangled pairwise, without substantial interaction between the different pairs.

\section{Validity of the Born-Markov approximation}\label{BornMarkov}

In our derivation of \refeq{sumdd}, \refeq{Gcoop} and \refeq{lossdd} we made use of the Born-Markov approximation, which presupposes that the environment is large and the correlation time of the environment is very short compared to the evolution of the atomic states \cite{Gross1982}. Since in our formalism the role of the `environment' is played by the modes of the finite cavity, the validity of these assumptions needs to be evaluated carefully.

In order to verify the validity of the above results, we numerically solve the Sch\"odinger equation, where the Hamiltonian is given by Eq.~\eqref{Hamiltonian} together with the non-Hermitian term introduced in \refeq{modeLosses}. The form of $V$ is considerably simplified when the two atoms are placed at two antipodal points within the lens such that $|\br_1|=|\br_2|$ and $\phi_1=\phi_2+\pi$. In this case the in-phase combination of the atomic dipole moments ($d_z(\sigma_1+\sigma_2)/\sqrt{2}$ + h.c.) only couples to the odd modes ($l=1,3,5\ldots$) and the out-of-phase combination of the dipole moments ($d_z(\sigma_1-\sigma_2)/\sqrt{2}$ + h.c.) only couples to the even modes $(l=2,4,6\ldots)$ of the fish eye (see Appendix \ref{numericalSchroedinger}). This reduces the size of the Hilbert space, making it possible to efficiently simulate the system while including a large number of the eigenmodes of the lens with frequencies close to $\omega_0$. We further restrict the Hilbert space to have at most a single excitation in the system. 

We numerically determine the time-evolution, starting from the state $\ket{\psi(0)}=\ket{e_1,g_2}$ via the operator $U(t) = \exp[-iH t/\hbar]$. To obtain the maximum fidelity of the entangling operation, the overlap of the time-evolved atomic state is calculated with the maximally entangled state $(\ket{e,g}-i\ket{g,e})/\sqrt{2}$. In Figs.~\ref{compare}(a) and \ref{compare}(b) we plot the numerically obtained values for the error $(1-F)$ (dotted lines) for different lens radii as a function of losses and atom frequencies, respectively. Even though the analytical results were derived using the Born-Markov approximation and neglecting retardation \cite{Milonni1974}, good agreement is obtained between the analytic results and numerical data. This confirms the validity of the analytical formalism described in previous sections. 

\begin{figure}[h!]
\begin{center}
\includegraphics[width=8cm]{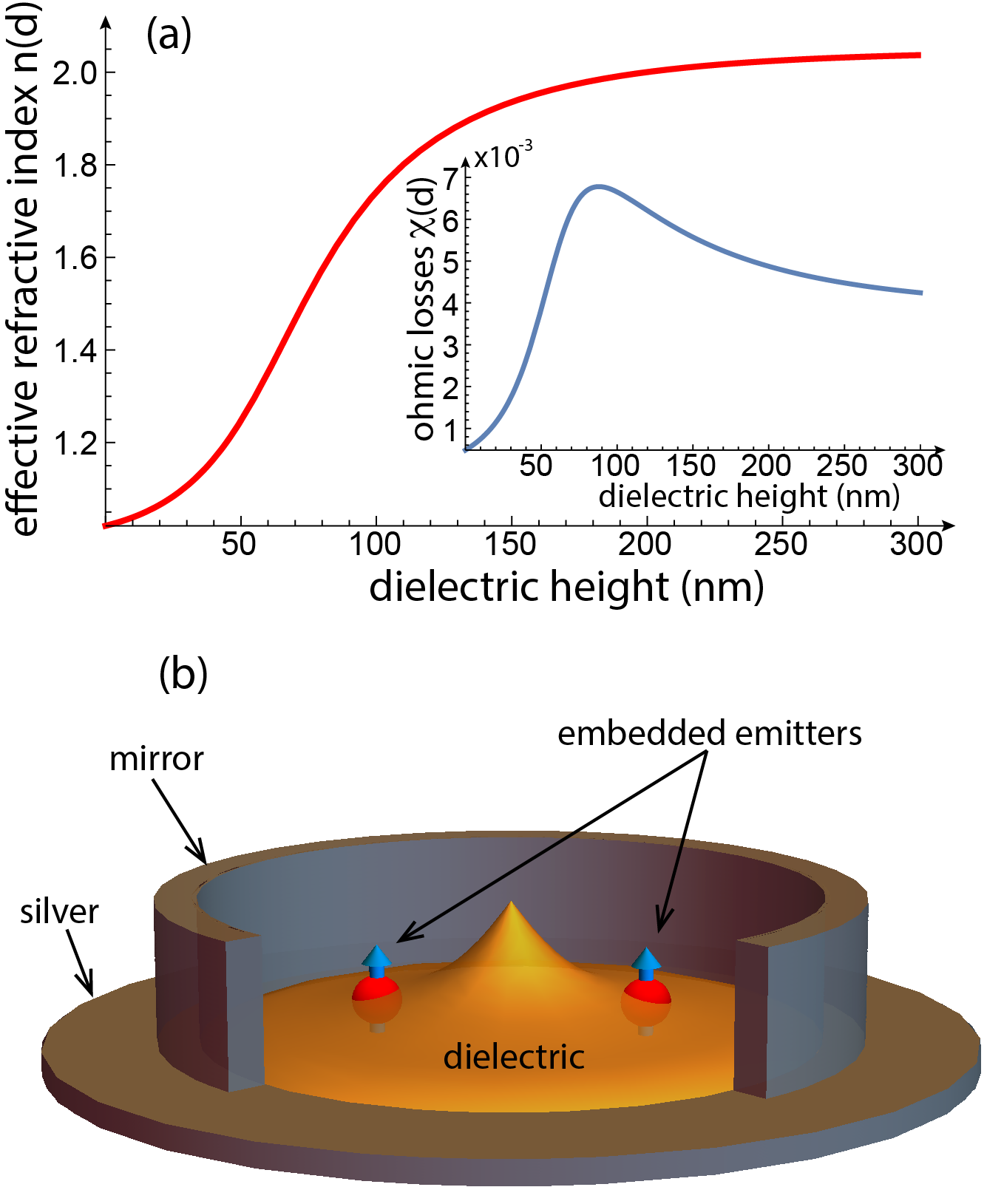}
\end{center}
\caption{(color online) Physical realization of the fish eye lens using transformation plasmon optics. (a) Effective refractive index ${n(d)=
\text{Re}\{\tilde n(d)\}}$ created as a function of the height of the dielectric $d$ deposited on the silver surface. The inset shows the material losses ${\chi(d)=\text{Im}\{\tilde n(d)\}}$ as a function of the dielectric height $d$. (b) Schematic depiction of the plasmonic fish eye lens. The two emitters are embedded in the dielectric. The height of the dielectric varies across the lens, which creates the effective refractive index distribution of the fish eye lens. The lens is surrounded by mirrors from all sides (the front part of the mirror has been removed to show the interior).   }
\label{realizationFig}
\end{figure}

\section{Possible experimental realization}\label{realization}

A promising way to realize the fish eye lens is via transformational plasmon optics \cite{Liu2010,Zentgraf2011}. The idea behind this approach is to engineer an effective refractive index distribution for surface plasmon polaritons by depositing a layer of high-index dielectric on top a 2D silver surface (see Fig.~\ref{realizationFig}). By varying the height of the dielectric layer on the surface, the effective refractive index seen by the plasmons can be changed. In particular, when there is no dielectric on top of the silver, the effective refractive index seen by the plasmons is close to 1, whereas in the presence of a thick dielectric layer, the effective plasmonic refractive index will be close to the refractive index of the dielectric itself. Through this experimental technique, complex spatially-varying refractive index profiles can be engineered \cite{Liu2010}. Crucially, the behavior of plasmons in a plasmonic lens with a particular refractive index profile closely mimics the predicted behavior of classical light rays in the corresponding 2D lens. This correspondence between 2D classical lenses and quasi-2D plasmonic lenses was theoretically established in Ref.~\cite{Liu2010} and experimentally confirmed for the nanoscale Luneburg and Eaton lenses \cite{Zentgraf2011}. 

We expect that the plasmonic version of the nanoscale fish eye lens could be experimentally realized analogously to the Luneburg and Eaton lenses. A dielectric layer of varying height could be deposited on a flat silver surface while the lens is surrounded by a circular mirror (see Fig.~\ref{realizationFig}(b)). To explore the quantum optical properties of the fish eye, atom-like color defects in diamond could be used as quantum emitters. Subwavelength positioning and coherent manipulation of such color defects has been experimentally demonstrated previously \cite{Dolde2013,Sipahigil2014,Kolkowitz2015,Iwasaki2015}. Recently, the entanglement of two silicon-vacancy (SiV) color defects inside a nanoscale cavity was also demonstrated \cite{Sipahigil2016}.

For illustration, we provide here an estimate of the entanglement fidelity of two atoms inside a particular example of a plasmonic fish eye lens. We assume that the lens operates at $406.706$THz, which is the zero-phonon resonance of SiVs corresponding to a vacuum wavelength of $\lambda_\text{SiV}=737$nm. Furthermore, we assume that the lens has a radius of $R_0=1.749\,\lambda_\text{SiV}$, which ensures that the SiV resonance falls between two resonant modes of cavity ($\text{Re}(\nu)=10.5$). We also assume that the flat silver substrate is made of single-crystal silver \cite{High2015}, which at the SiV resonance frequency has a permittivity of $\epsilon_m=-25.23+0.589i$ and gives rise to plasmonic propagation distances on the order of $\sim 160\, \lambda_\text{Siv}$. It is also assumed that there is a thin ($\sim 10-15$nm) diamond layer  on top of the metal that has two SiVs implanted at two antipodal points such that $|\br_1|=|\br_2|=0.27R_0$ and $\phi_1=\phi_2+\pi$, as schematically shown in Fig.~\ref{realizationFig}(b). Due to their proximity to the silver surface, the two $\hat z$-polarized emitters will couple strongly to the surface plasmons, which are tightly confined to the metal-dielectric interface. 

The spatially varying refractive index $n(\br)$ of the fish (\refeq{FEn} with $n_0=1$) could be experimentally realized by depositing a dielectric of permittivity $\epsilon_d=3.6$ on top of thin diamond layer. By varying the height of the dielectric between 0 and 200nm, the effective refractive index seen by the plasmons can be varied between 1 and 2. 
The refractive index of the dielectric ($n_d=\sqrt{\epsilon_d}=1.9$) was chosen such that the effective index can reach 2, but a dielectric with even higher index (such as diamond with $\epsilon_\text{diamond}=5.76$) was avoided to ensure that the plasmons are not confined unnecessarily tightly to the silver surface, which would give rise to significantly higher ohmic losses.  

The direct relationship between the height of the dielectric layer $d$ and the resulting (complex) refractive index ${\tilde n}(d)=n(d)+i\chi(d)$ can be obtained from the following implicit equation \cite{Liu2010}
\be
\te{tanh}(k_d\epsilon_dd)=-\fr{k_\te{air}k_d+k_dk_m}{k_d^2+k_\te{air}k_m},
\ee
where  
\be
k_\te{air}=\sqrt{({\tilde n}k_0)^2-k_0^2},
\ee
\be
k_{d}=\fr{\sqrt{({\tilde n}k_0)^2-\vare_{d}k_0^2}}{\vare_{d}},
\ee
and
\be
k_{m}=\fr{\sqrt{({\tilde n}k_0)^2-\vare_{m}k_0^2}}{\vare_{m}},
\ee
where $k_0=2\pi/\lambda_\text{SiV}$ and in our calculation we ignored, for simplicity, the presence of the diamond layer, as it does not significantly modify the effective index seen by the plasmons as long as the diamond layer is much thiner than the transverse confinement of the plasmons, which is on the order of a wavelength. 

Fig.~\ref{realizationFig}(a) shows the real part $n(d)$ and imaginary part $\chi(d)$ (inset) of the complex refractive index ${\tilde n}(d)$ seen by the plasmons as the thickness of the dielectric $d$ is varied. The effective refractive index increases monotonically with the thickness of the dielectric layer. Since the refractive index of the fish eye increases radially inward (see \refeq{FEn}), the dielectric layer in the fish eye lens takes a conical shape as shown in Fig.~\ref{realizationFig}(b). 

From the imaginary part of the effective refractive index $\chi(d)$, we can estimate the average photon loss rate due to ohmic losses via the relation ${\kappa_\text{abs}(\br)/\omega_0=\chi(\br)/n(\br)}$ (see Section \ref{spontAndCoop}). Since this loss rate varies significantly across the lens, we numerically average $\chi(\br)/n(\br)$ over the radius of the lens and obtain the averaged quantity $\overline{\kappa_\text{abs}(\br)}/\omega_0\approx 3\times 10^{-3}$. This is the leading order contribution to the photon loss.

Photons can also be lost from the lens by leaking out through the mirror. Assuming that the reflectivity of the mirror is $r^2$, we can estimate the loss rate $\kappa_\text{mirror}/\omega_0$. In the absence of other loss mechanisms, the photon would bounce off the mirror $\sim1/t^2$ times before being lost, where $t^2=1-r^2$. The time interval between two bounces is approximately $(2R_0)/(c/\bar{n})$, where $R_0$ is the radius of the lens, $c$ is the speed of light in vacuum and $\bar{n}$ is the average index of refraction in the lens. Thus the lifetime of the photon due to the finite mirror reflectivity is
\be
\tau_\text{mirror}\sim\fr{1}{\kappa_\text{mirror}}\sim\fr{2R_0}{c/\bar{n}}\fr{1}{t^2}.
\ee
Making the conservative estimate that $r^2=0.95$, we obtain the following loss rate
\be
\fr{\kappa_\text{mirror}}{\omega_0}\sim \fr{1}{4\pi}\fr{1}{\bar{n}}\fr{\lambda_0 t^2}{R_0}\sim 4\times 10^{-4},
\ee
where we have used $\overline{n(\br)}= 1.57$, which is obtained by numerically averaging the refractive index over the radius of the lens. Note that this shows that the losses due to the finite reflectivity of the mirror are an order of magnitude smaller than the ohmic losses. 

Next, we consider emission into free space $\gamma_0$. In the close proximity of a metal surface, the rate of decay of the emitter into plasmonic modes $\gamma$ can significantly exceed the rate of emission into free-space modes ${\gamma_0=d_z^2\omega_0^3/(3\pi\vare_0\hbar c^3)}$ \cite{Chang2006,Francs2016}. Here, we take the Purcell factor to be $\eta=\gamma/\gamma_0\approx 3$, which is the approximate value for a z-oriented dipole 10-15 nm away from a flat silver surface emitting radiation at 737nm \footnote{We calculate the Purcell factor exactly near a flat metal surface by evaluating ${\gamma_\text{pl}=2d_z^2\omega_0^2/(\hbar\vare_0c^2)\text{Im}\{G_{zz}\}}$, where $G_{zz}$ is the exact Green's function of a z-oriented dipole near the surface \cite{Sipe1987}}. Furthermore, we also make the conservative estimate that the emission to free space is reduced by a factor of two due to the presence of the silver surface \cite{Arcari2014}. In order to account for the presence of this additional decay channel, we need to make the replacement $\gamma\to \gamma+\gamma_0/2$ in \refeq{nonHermitian}, and thus \refeq{fidelityApprox} becomes
\be\label{fidelityApproxNew}
F = e^{-\pi^3\left(1+\fr{1}{2\eta}\right)\fr{R_0}{\lambda}\alpha}\;.
\ee
Note that this equation holds only if the atomic frequencies fall half-way between two resonances and the atoms are placed at two antipodal points anywhere in the lens.
Substituting $R_0/\lambda_\text{SiV}=1.749$, $\alpha={(\overline{\kappa_\text{abs}}+\kappa_\text{mirror})/\omega_0}=3.4\times 10^{-3}$  and $\eta=3$ into \refeq{fidelityApproxNew}, we obtain that the fidelity of the entangling operation would be approximately $F=80\%$. We note that this fidelity could be further improved by utilizing the adiabatic passage of a dark state in a Raman scheme \cite{Pellizzari1995}.

\section{Conclusion}\label{conclusion}

In conclusion, we have investigated the single-photon dynamics of atoms inside the fish eye lens. We demonstrated that the lens mediates long-range interactions between distant emitters. The dipole-dipole interaction has an infinite range, limited only by the decay rate of the cavity modes. Furthermore, our results show that the fish eye focuses a single photon to a diffraction-limited area during the exchange of a photon between two antipodal atoms, whose frequency is tuned between two resonances of the cavity. We derived closed-form expressions for the decay rates and dipole-dipole interaction of atoms in the presence of losses and studied the fidelity of entangling operations. We confirmed the validity of our analysis, which relied on the Born-Markov approximation, by numerically solving the Schr\"odinger equation. Finally, we proposed a possible realization for the fish eye lens using tranformational plasmon optics and silicon-vacancy centers that could open up the fish eye for practical applications.

\section{Acknowledgements}

We thank U. Leonhardt, E. Shahmoon, M. Mezei, S. Horsley, K. Lalumi\`ere, W. M. R. Simpson, A. Levy, S. Choi, D. Wild, and M. Kan\'{a}sz-Nagy for stimulating discussions. We acknowledge funding from the MIT-Harvard CUA, NSF and the Hungary Initiatives Foundation.

\onecolumngrid

\section*{Appendix}

\appendix

\section{Derivation of a closed-form expression for the dipole-dipole interaction}\label{transformation}

In this section we show that the single-source Green function derived by Leonhardt \cite{Leonhardt2009,Leonhardt2010a} for the 2D fish eye lens can be written as a sum over the eigenmodes of the lens (\refeq{modes}). This result enables us to connect the Green's function to the expressions obtained for the atomic properties from the master equation treatment.  

\subsection{Green's function of the 2D fish eye}

The single-source Green's function of Maxwell's 2D fish eye (of radius $R_0$, thickness $b$
and refractive index profile $n(r) = \frac{2n_0}{1+(r/R_0)^2}$) is a solution of the following equation
\bal\label{greensEquation}
\left(\pa_\alpha\pa_\nu-\delta_{\alpha\nu}\pa_\eta\pa_\eta\right)G_{\nu\beta}(\br,\br',\omega_0)-\vare(\br)\,\fr{\omega_0^2}{c^2} G_{\alpha\beta}(\br,\br',\omega_0)=\delta_{\alpha\beta}\delta(\br-\br'),
\eal
where $\alpha,\beta,\mu,\nu=x,y,z$ and summation is implied over repeated indices and $\vare(\br)=n(\br)^2=\left[2n_0/(1+(r/R_0)^2)\right]^2$ is the position-dependent electric permittivity. When $b$ is chosen such that $\omega_0\ll \pi c/b$, only the lowest TE polarized mode of the fish eye can be excited and the electric field is invariant along the $z$-axis ($\pa_z\mbf E(\br)=0$). The explicit expression for the $zz$-components of the Green's function (\refeq{analytic}) is then given by 
\be
	G_{zz}(\br_1, \br_2, \omega)= F(\alpha_1, \alpha_2) - F(\alpha_1, 1/\alpha_2\s),
	\qquad \text{where} \quad F(\alpha_1, \alpha_2) =
	-\frac{P_\nu(\xi(\alpha_1, \alpha_2))}{4b\sin(\pi \nu)},
\ee
where $P_\nu$ is the Legendre function of (non-integer) oder $\nu$,
\ba
	\nu  =  \frac{1}{2}\left[\sqrt{4\frac{\omega^2}{c^2} R_0^2n_0^2  + 1}
	- 1\right] \notin \mathds{Z} \qquad
	\text{and} \qquad &&
	\xi(\alpha_1, \alpha_2) = \frac{|\zeta(\alpha_1, \alpha_2)|^2
	-1}{|\zeta(\alpha_1, \alpha_2)|^2 + 1},\\
	\text{where}\qquad && \zeta(\alpha_1, \alpha_2) = \frac{\alpha_1 -
	\alpha_2}{\alpha_1\alpha_2\s + 1},\qquad\text{and}\quad	\alpha_j =
	\underbrace{\frac{r_j}{R_0}}_{\rho_j} e^{i\phi_j} \quad (j=1,2).
\ea

\subsection{Virtual coordiantes}
The stereographic transformation \cite{Leonhardt2010b}
\be
	r_j \quad \mapsto \quad \frac{r_j^2 - R_0^2}{r_j^2 + R_0^2} = \frac{\rho_j^2
	- 1}{\rho_j^2 + 1} \quad =:\quad \cos\theta_j,
\ee
can be used to map any point $(r,\phi)$ on the real plane to a point
$(\theta,\phi)$ on the surface of a virtual sphere (where $\phi$ is the same
value in both coordinate systems). Using this transformation, we can simplify
the definition of the Green's function
\ba
	\zeta(\alpha_1, \alpha_2) &=& \frac{(\rho_1 \cos\phi_1 - \rho_2 \cos\phi_2) +
	i(\rho_1\sin\phi_1 - \rho_2\sin\phi_2)}{\rho_1\rho_2 \cos(\phi_1 - \phi_2) + 1
	+ i\rho_1\rho_2 \sin(\phi_1 - \phi_2)}
	\\
	|\zeta(\alpha_1, \alpha_2)|^2 &=& \frac{\rho_1^2 + \rho_2^2 - 2\rho_1\rho_2\cos(\phi_1 -
	\phi_2)}{(\rho_1\rho_2)^1 + 1 + 2\rho_1\rho_2\cos(\phi_1 - \phi_2)}
	\\
	\xi(\alpha_1, \alpha_2) &=& \frac{\rho_1^2 + \rho_2^2 - (\rho_1\rho_2)^2 -1 -
	4\rho_1 \rho_2 \cos(\phi_1 - \phi_2)}{\rho_1^2 + \rho_2^2 + (\rho_1\rho_2)^2 +1} = 
	\\
	&=& - \left[\left(\frac{\rho_1^2 - 1}{\rho_1^2 + 1}\right)
	\left(\frac{\rho_2^2 -1}{\rho_2^2 + 1}\right) + \left(\frac{2\rho_1}{\rho_1^2 +
	1}\right) \left(\frac{2\rho_2}{\rho_2^2 + 1}\right)\cos(\phi_1 - \phi_2)\right]
	=
	\\
	&=& - \big[\cos\theta_1 \cos\theta_2 + \sin\theta_1 \sin\theta_2
	\cos(\phi_1 - \phi_2)\big] = -\cos\theta_{12},
\ea
where $\theta_{12}$ is the spherical distance beetween two
points, $(\theta_1, \phi_1)$ and $(\theta_2, \phi_2)$, on the surface of a unit
sphere, since
\be
	\cos\theta_{12} = \mathbf{x}_1 \mathbf{x}_2 = \threevector{\sin\theta_1
	\cos\phi_1}{\sin\theta_1\sin\phi_1}{\cos\theta_1}\cdot
	\threevector{\sin\theta_2 \cos\phi_2}{\sin\theta_2\sin\phi_2}{\cos\theta_2} = \cos\theta_1 \cos\theta_2 + \sin\theta_1 \sin\theta_2
	\cos(\phi_1 - \phi_2).
\ee

Similarly,
\ba
	\zeta(\alpha_1, 1/\alpha_2\s) &=& \frac{(\rho_1 \cos\phi_1 - \frac{1}{\rho_2}
	\cos\phi_2) + i(\rho_1\sin\phi_1 -
	\frac{1}{\rho_2}\sin\phi_2)}{\frac{\rho_1}{\rho_2} \cos(\phi_1 - \phi_2) + 1 +
	i\frac{\rho_1}{\rho_2} \sin(\phi_1 - \phi_2)},
	\\
	|\zeta(\alpha_1, 1/\alpha_2\s)|^2 &=& \frac{\rho_1^2 + \frac{1}{\rho_2^2} -
	2\frac{\rho_1}{\rho_2}\cos(\phi_1 - \phi_2)}{(\frac{\rho_1}{\rho_2})^1 + 1 +
	2\frac{\rho_1}{\rho_2}\cos(\phi_1 - \phi_2)},
	\\
	\xi(\alpha_1, 1/\alpha_2\s) &=& \frac{\rho_1^2 +
	\left(\frac{1}{\rho_2}\right)^2 - (\frac{\rho_1}{\rho_2})^2 -1 - 4\frac{\rho_1}{\rho_2} \cos(\phi_1 - \phi_2)}{\rho_1^2 +
	\left(\frac{1}{\rho_2}\right)^2 + (\frac{\rho_1}{\rho_2})^2 +1} =
	\\
	&=& - \left[\left(\frac{\rho_1^2 - 1}{\rho_1^2 + 1}\right)
	\left(\frac{1-\rho_2^2}{\rho_2^2 + 1}\right) + \left(\frac{2\rho_1}{\rho_1^2
	+ 1}\right) \left(\frac{2\rho_2}{\rho_2^2 + 1}\right)\cos(\phi_1 - \phi_2)\right]
	=
	\\
	&=& - \big[\cos\theta_1 \cos(\pi-\theta_2) + \sin\theta_1 \sin(\pi - \theta_2)
	\cos(\phi_1 - \phi_2)\big] = -\cos\theta'_{12},
\ea
where, now, $\theta_{12}'$ is the spherical distance between the points
$(\theta_1, \phi_1)$ and $(\pi - \theta_2, \phi_2)$.

Now, we can write the Green's function as
\bel
	G_{zz}(\br_1, \br_2, \omega) = - \frac{P_\nu(-\cos\theta_{12}) -
	P_\nu(-\cos\theta_{12}')}{4b\sin(\pi\nu)}.
	\label{eq:G_1}
\eel

\subsection{Expansion in Spherical harmonics}
\subsubsection{Expansion with respect to $l$}
The full set of Legendre polynomials, $P_l$, form a complete, orthogonal basis
on the space of smooth $[-1,1]\rightarrow \mathds{R}$ functions. This allows us to
expand the Legendre function $P_{\nu}$ in terms of the Legendre polynomials
$P_l$
\be
	P_\nu(x) = \sum_{l=0}^\infty c_l P_l(x),\qquad \text{where}\qquad
	c_l = \frac{2l + 1}{2} \intop_{-1}^{+1} \d{x} P_l(x) P_\nu(x).
\ee	
According to Abramowitz \& Stegun, Section 8.14 \cite{Abramowitz1970}
\be
	\intop_{-1}^{+1}\d{x} P_\eta(x)P_\nu(x) = \frac{2}{\pi^2}
	\frac{2\sin(\pi\eta)\sin(\pi\nu)[\psi(\eta+1) - \psi(\nu+1)] +
	\pi\sin(\pi\nu - \pi\eta)}{(\nu-\eta)(\nu+\eta+1)},
\ee
where $\psi$ is the digamma function and which expression, in case of $\eta = l \in \mathds{N}$, simplifies to
\be
	\intop_{-1}^{+1}\d{x} P_l(x)P_\nu(x) =
	\frac{2}{\pi}\frac{(-1)^l\sin(\pi\nu)}{\nu(\nu+1) - l(l+1)},\qquad
	\text{if}\quad l \in \mathds{N}.
\ee
This means that
\be
	P_\nu(x) = \frac{\sin(\pi\nu)}{\pi}\sum_{l=0}^\infty 
	(-1)^l\frac{2l+1}{\nu(\nu+1) - l(l+1)} P_l(x), 
\ee
and we can write the Green's function as
\be
	G_{zz}(\br_1, \br_2, \omega) = -\frac{1}{4\pi b}\sum_{l=0}^\infty(-1)^l
	\frac{2l+1}{\nu(\nu+1) - l(l+1)} \big[P_l(-\cos\theta_{12}) -
	P_l(-\cos\theta_{12}')\big].
\ee

\subsubsection{Expansion with respect to $m$}
According to the addition
theorem of spherical harmonics,
\be
	P_l(\mathbf{x}_1\cdot \mathbf{x}_2) = P_l(\cos\theta_{12}) =
	\frac{4\pi}{2l+1}\sum_{m=-l}^{+l} Y_l^{m\ast}(\theta_1, \phi_1)Y_l^m(\theta_2,\phi_2),
\ee
where the spherical harmonics are defined by
\be
	Y_l^m(\theta,\phi) =
	\sqrt{\frac{2l+1}{4\pi}\frac{(l-m)!}{(l+m)!}}P_l^m(\cos\theta)e^{im\phi},
\ee
where $P_l^m$ are the associated Legendre polynomials.

By using this theorem, and the property that $P_l(-x) = (-1)^l P_l(x)$, we can
write
\ba
	P(-\cos\theta_{12}) &=& (-1)^l \frac{4\pi}{2l+1} \sum_{m=-l}^{+l}
	Y_l^{m\ast}(\theta_1, \phi_1) Y_l^m(\theta_2, \phi_2),
	\\
	P(-\cos\theta_{12}') &=& (-1)^l \frac{4\pi}{2l+1} \sum_{m=-l}^{+l} Y_l^{m\ast}(\theta_1,
	\phi_1) \underbrace{Y_l^m(\pi-\theta_2, \phi_2)}_{(-1)^{l-m} Y_l^m(\theta_2,
	\phi_2)},
	\\
	P(-\cos\theta_{12}) - P(-\cos\theta_{12}') &=& (-1)^l \frac{4\pi}{2l+1}
	\sum_{m=-l}^{+l} \Big[1 - (-1)^{l-m}\Big] Y_l^{m\ast}(\theta_1,
	\phi_1) Y_l^m(\theta_2, \phi_2).
\ea
The expression inside the square brackets is zero if $l$ and $m$ have the same
parity, and 2, if they have different parity. The set of $m$ values for which
the corresponding term is non-zero is $M_l = \{-(l-1), -(l-3), \dots, (l-3),
(l-1)\}$. Using this notation, we can write the Green's function as
\bel
	G_{zz}(\br_1, \br_2, \omega) = -\fr{2}{b}\sum_{l=0}^\infty \sum_{m\in
	M_l}\frac{Y_l^{m\ast}(\theta_1, \phi_1) Y_l^m(\theta_2, \phi_2)}{\nu(\nu+1) -
	l(l+1)}
	\label{eq:G_2}
\eel

\subsection{Expansion in cavity modes}
Recall from \refeq{modes} that the TE eigenmodes of Maxwell's fish eye with radius $R_0$ and width $b$ are
\be
	f_{l,m}(\br) = \sqrt{\frac{2}{bR_0^2n_0^2}}
	\sqrt{\frac{2l+1}{4\pi}\frac{(l-m)!}{(l+m)!}} P_l^m\left(\frac{r^2 -
	R_0^2}{r^2 + R_0^2} \right)e^{im\phi} = \sqrt{\frac{2}{bR_0^2n_0^2}}
	(-1)^m Y_l^m(\theta, \phi),
\ee
where $r$ and $\phi$ are polar coordinates of $\br$ and $\cos\theta = \frac{r^2
- R_0^2}{r^2 + R_0^2}$. They satisfy the orthonormality condition,
\be
	\delta_{l,l'}\delta_{m,m'} =
	\intop_0^{R_0}\d{r}r\intop_0^{2\pi}\d{\phi}\intop_0^b\d{z} 
	n^2(r)f_{l,m}^{\ast}(r,\phi) f_{l',m'}(r,\phi).
\ee
The corresponding (partially degenerate) eigenfrequencies are
\be
	\omega_{l,m} = \omega_l = \frac{c}{R_0 n_0} \sqrt{l(l+1)} =: ck_l. 
\ee

Now, we can write the Green's function in \refeq{eq:G_2} as
\bal
	G_{zz}(\br_1, \br_2, \omega/c) &=&  - R_0^2n_0^2 \sum_{l=0}^\infty \sum_{m\in
	M_l}\frac{f_{l,m}^{\ast}(\br_1) f_{l,m}(\br_2)}{\nu(\nu+1) -
	l(l+1)}
	\nonumber\\
	&=&
	- \sum_{l=1}^\infty \sum_{m\in
	M_l}\frac{f_{l,m}^{\ast}(\br_1) f_{l,m}(\br_2)}{
	(\omega/c)^2-k_l^2},
	\label{eq:G_3}
\eal
where we used the connection between $\omega_l$ and $l$, and $\omega$ and $\nu$. Using \refeq{dd}, we can then write the dipole-dipole interaction within the fish eye in the following form
\be
\delta\omega(\br_1,\br_2)=\fr{d_z^2\omega_0^2}{\hbar\vare_0c^2}\te{Re}\{G_{zz}(\br_1,\br_2,\omega_0)\}=\fr{d_z^2\omega_0^2}{\hbar\vare_0c^2} \sum_{l=1}^\infty \sum_{m\in
	M_l}\frac{f_{l,m}^{\ast}(\br_1) f_{l,m}(\br_2)}{
	k_l^2-(\omega_0/c)^2}.\label{ddPeter}
\ee
We note that this decomposition of the Green's function in terms of the eigenmodes of the fish eye is a particular example of Fredholm's theorem \cite{Fredholm1903}.

\subsection{Comparison with master equation results}\label{comparison}

Recall that the dipole-dipole interaction obtained from the master equation (see \refeq{sumdd}) has the form 
\be
\delta\omega(\mbf{r}_1,\br_2)  =  \fr{d_z^2}{\hbar\vare_0}\sum_{l,m}\fr{\omega_l^2}{\omega_l^2-\omega_0^2} \,f_{l,m}^*(\mbf{r}_1)f_{l,m}(\mbf{r}_2).
\ee

Making use of the transformation
\be\label{transEasy}
	\frac{\omega_l^2}{\omega_l^2 - \omega_0^2} = \left[1 +
	\frac{\omega_0^2}{\omega_l^2 - \omega_0^2}\right],
\ee
and the dipole-dipole interaction becomes
\be
	\delta\omega (\br_1,\br_2)= \frac{d_z^2}{\hbar \eps_0}
	\left[\sum_{l,m}f_{l,m}(\br_1)f_{l,m}(\br_2) + \omega_0^2
	\sum_{l,m}\frac{f_{l,m}^*(\br_1)f_{l,m}(\br_2)}{\omega_l^2 - \omega_0^2} \right].
\ee
Since the modes $f_{l,m}$ form a complete basis, i.e.
\be\label{orthonorm}
	\sum_{l,m}f_{l,m}^*(\br_1)f_{l,m}(\br_2) = \delta^{(3)}(\br_1 - \br_2),
\ee
the first term inside the square brackets does not contribute if $\br_1 \neq
\br_2$, thus
\be
	\delta\omega (\br_1,\br_2)= \frac{d^2}{\hbar \eps_0}
	 \fr{\omega_0^2}{c^2}
	\sum_{l,m}\frac{f_{l,m}^*(\br_1)f_{l,m}(\br_2)}{k_l^2 - (\omega_0/c)^2} ,\qquad \text{if}\quad \br_1 \neq \br_2,
\ee
which is identical to \refeq{ddPeter}. This shows that the right-hand side of \refeq{sumdd} can indeed be replaced by \refeq{dd} and
\refeq{analytic}.

More generally, using \refeq{transEasy} and \refeq{orthonorm} we can express \refeq{eq:G_3} in the form
\bal\label{Gold}
\fr{d^2}{\hbar\vare_0}\fr{\omega_0^2}{c^2}G_{zz}(\br_1,\br_2,\omega_0)= \fr{d^2}{2\hbar\vare_0}\sum_{l,m}\omega_l f_{l,m}^*(\br_1)f_{l,m}(\br_2)\bigg( \fr{1}{\omega_0+\omega_l}-\fr{1}{\omega_0-\omega_l} \bigg),
\eal
from which it is straighforward to show that 
\bal\label{full}
\fr{d^2}{\hbar\vare_0}\fr{(\omega_0+i\kappa)^2}{c^2}G_{zz}(\br_1,\br_2,\omega_0+i\kappa) 
&=& \fr{d^2}{2\hbar\vare_0}\sum_{l,m}\omega_l \,f_{l,m}^*(\mbf{r}_1)f_{l,m}(\mbf{r}_2)\bigg\{  
\fr{(\omega_l+\omega_0)}{\kappa^2+(\omega_l+\omega_0)^2}+\fr{(\omega_l-\omega_0)}{\kappa^2+(\omega_l-\omega_0)^2}
\\
&&-i\bigg(\fr{\kappa}{\kappa^2+(\omega_l+\omega_0)^2}-\fr{\kappa}{\kappa^2+(\omega_l-\omega_0)^2}\bigg)\bigg \},
\eal
which allows us to express the decay rates \refeq{Gcoop} and the dipole-dipole interaction (\refeq{lossdd}) in the presence of losses in the following closed form
\be\label{acoop}
\Gamma(\br_i,\br_j)=\fr{2d_z^2}{\hbar\vare_0 c^2}\te{Im}\{(\omega_0+i\kappa)^2G_{zz}(\br_i,\br_j,\omega_0+i\kappa) \}.
\ee
and
\be\label{add}
\delta\omega(\br_i,\br_j)=\fr{d_z^2}{\hbar\vare_0c^2}\te{Re}\{(\omega+i\kappa)^2G_{zz}(\br_i,\br_j,\omega_0+i\kappa)\}.
\ee
We note that $G_{zz}(\br_i,\br_j,\omega_0+i\kappa)$ is the solution of the following equation
\bal\label{greensEquationLossy}
\left(\pa_\alpha\pa_\nu-\delta_{\alpha\nu}\pa_\eta\pa_\eta\right)G_{\nu\beta}(\br,\br',\omega_0+i\kappa)-\fr{n(\br)^2\,(\omega_0+i\kappa)^2}{c^2} G_{\alpha\beta}(\br,\br',\omega_0+i\kappa)=\delta_{\alpha\beta}\delta(\br-\br'),
\eal
which can be thought of as the dyadic equation for the fish eye lens with the complex refractive index ${\tilde n(\br)=n(\br)(1+i\kappa/\omega_0)}$, since
\be
n(\br)^2(\omega_0+i\kappa)^2=n(\br)^2(1+i\kappa/\omega_0)^2\omega_0^2=\tilde n(\br)^2\omega_0^2.
\ee
Noting that for $\kappa\ll\omega_0$ the following approximations hold
\be\label{acoopApprox}
\Gamma(\br_i,\br_j)\approx\fr{2d_z^2\omega_0^2}{\hbar\vare_0 c^2}\te{Im}\{G_{zz}(\br_i,\br_j,\omega_0+i\kappa) \}
\quad\text{and}\quad
\delta\omega(\br_i,\br_j)\approx\fr{d_z^2\omega_0^2}{\hbar\vare_0c^2}\te{Re}\{G_{zz}(\br_i,\br_j,\omega_0+i\kappa)\},
\ee
we find that photon loss from the modes of the fish eye of the form of \refeq{modeLosses} can simply be modeled with the complex refractive index profile $\tilde n(\br)$.

\section{Numerical Solution of the Schr\"odinger Equation}\label{numericalSchroedinger}

In this Appendix we describe an efficient way to numerically solve the Schr\"odinger Equation while including the two atoms and the modes of the fish eye in the dynamics. 

\subsection{Hamiltonian}

\subsubsection{Electric dipole interaction of a single atom}
Recall that the electric dipole coupling of a single atom, placed at $\br_i$, to the electromagnetic field modes of the fish eye is given by 
\bel\label{VAppendix}
	V=-\mbf{d}_i\cdot\mbf{E}(\br_i)
	,\quad \text{where}\quad  \mbf d_i = d_z\hat z\big(\sigma_i\+
	 + \sigma_i \big),\qquad \text{where}\quad \sigma_i
	= \ket{g_i}\bra{e_i},
\eel
where $d_z$ is the magnitude of the transition dipole moment between the two
states of the atom $\ket{e_i}$ and $\ket{g_i}$, whose energy difference
is $\hbar \omega_0$. 
Substituting \refeq{Efield} into \refeq{VAppendix} and neglecting the counter-rotating terms in $V$, we arrive
at
\bel
	V_{RW\!A}=
	\sum_{l,m}id\sqrt{\frac{\hbar \omega_l}{bR_0^2 \eps_0}}
	\big[a_{l,m}\sigma\+ Y_{l,m}(\theta,\phi) - a_{l,m}\+
	\sigma Y_{l,m}(\theta,\phi)\big],
\eel
where $n_0=1$ was assumed.

\subsubsection{Two atoms}
Assuming that there are two identical atoms positioned at $\br_1$ and at
$\br_2$, the interaction term takes the form $V_{RW\!A}(\br_1)+V_{RW\!A}(\br_2)$.
The total Hamiltonian then becomes
\bel
	H =\hbar\omega_0\sigma_1^\dagger\sigma_1+\hbar\omega_0\sigma_2^\dagger\sigma_2+ \sum_{l,m} \hbar\omega_l a_{l,m}\+ a_{l,m} +
	\sum_{l}
	\hbar g_l\left[\sigma_1\+ \sum_m a_{l,m}Y_{l,m}(\theta_1, \phi_1)  + \sigma_2\+ \sum_m a_{l,m}
	Y_{l,m}(\theta_2, \phi_2)\right]  + \text{h.c.},
\eel
where $g_l = \frac{id_z}{\sqrt{\hbar bR_0^2\eps_0}}\sqrt{\omega_l}$.

The diagonal terms $\hbar\omega_0\left(\sigma_1^\dagger\sigma_1+\sigma_2^\dagger\sigma_2+\sum_{l,m} a_{l,m}\+ a_{l,m}\right)$, simply give a constant energy shift to all eigenvectors in the subspace of interest and can, therefore, be subtracted from the Hamiltonian. The modified Hamiltonian then takes the form
\bel
	H/\hbar = \sum_{l} \delta_l \sum_m a_{l,m}\+ a_{l,m} + \sum_{l}
	g_l\left[\sigma_1\+ \sum_m a_{l,m}Y_{l,m}(\theta_1, \phi_1)  + \sigma_2\+ \sum_m a_{l,m}
	Y_{l,m}(\theta_2, \phi_2)\right]  + \text{h.c.},
\eel
where $\delta_l = \omega_l - \omega_0$.

\subsubsection{Opposite positions}
If the two atoms are placed at opposite positions ($\theta = \theta_1 =
\theta_2$ and $\phi = \phi_1 = -\phi_2$), then we can write the interaction part
of $H$ as
\bel
	V/\hbar = \sum_l g_l \sum_m \left[ \sigma_1\+ a_{l,m} y_{l,m} +
	\sigma_2\+ a_{l,m}(-1)^m y_{l,m} + \text{h.c.}\right],
\eel
where $y_{l,m} = Y_{l,m}(\theta, \phi)$. Here we used that $Y_{l,m}(\theta,
-\phi) = (-1)^mY_{l,m}(\theta, \phi)$. Since the summation of $m$ goes over $m =
-l+1, -l+3, \ldots l-3, l-1$, $m$ and $l$ always have opposite parity, and we can pull
out $(-1)^m = (-1)^{l+1}$ from the summation, giving
\bel
	V/\hbar = \sum_l g_l \left[\sigma_1\+ + (-1)^{l+1}\sigma_2\+ \right] \sum_{m}
	a_{l,m} y_{l,m} + \text{h.c.}
\eel

We define an incomplete  set of new modes,
\bel
	A_l = \frac{\sum_m a_{l,m}y_{l,m}}{N_l},\qquad N_l^2 =
	\sum_{m}|y_{l,m}|^2,\quad [A_l,A_l\+] = 1.
\eel
The normalization factor can be calculated as follows.
\bel
	\sum_{m\in M} |Y_{l,m}(\theta,\phi)|^2 = \sum_{m = -l}^{+l}
	\frac{1-(-1)^{l-m}}{2} |Y_{l,m}(\theta,\phi)|^2,\qquad \text{where}\quad M =
	\{-l+1, -l+3, \ldots l-3, l-1\}.
\eel
Recall the sum rule:
\bel\label{sumRule}
	\sum_{m=-l}^{+l} Y\s_{l,m}(\theta_1,\phi_1) Y_{l,m}(\theta_2,\phi_2) =
	\frac{2l + 1}{4\pi} P_l(\cos\theta_{12}),
\eel
where $\theta_{12}$ is the angle between point 1 and 2 on the unit sphere, and
$P_l$ is the $l$th Legendre polynomial. We use \refeq{sumRule} to evaluate the two series:
\bal
	\sum_{m=-l}^{+l} |Y_{l,m}(\theta,\phi)|^2 
	&=& 
	\frac{2l+1}{4\pi} P_l(1) =
	\frac{2l+1}{4\pi},
	\\
	\sum_{m=-l}^{+l} (-1)^{l-m}|Y_{l,m}(\theta,\phi)|^2
	&=&
	\sum_{m=-l}^{+l} Y\s_{l,m}(\pi - \theta, \phi) Y_{l,m}(\theta,\phi)
	=
	\frac{2l + 1}{4\pi}P_l\big(\cos(\pi - 2\theta)\big),
\eal
where we used that $(-1)^{l-m}Y_{l,m}(\theta,\phi) = Y_{l,m}(\pi - \theta,
\phi)$ and that the angle between point $(\theta, \phi)$ and point $(\pi -
\theta, \phi)$ is $\theta_{12} = \pi - 2\theta$.

Using these new $A_l$ modes, the interaction can be written as
\bel
	V/\hbar = \sum_l g_l  N_l \left[\sigma_1\+ + (-1)^{l+1}\sigma_2\+\right]A_l +
	\text{h.c.}
\eel
Modes $A_l$ with different $l$ parity couple to different combination of the two
atoms. Let us define
\bel
	\sigma_o = \frac{\sigma_1 + \sigma_2}{\sqrt{2}},\qquad
	\sigma_e = \frac{\sigma_1 - \sigma_2}{\sqrt{2}}, 
\eel
and write
\bel
	V/\hbar = \sum_{l\in\text{odd}} G_l (A_l \sigma_o\+ + A_l\+ \sigma_o) +
	\sum_{l\in\text{even}} G_l (A_l \sigma_e\+ + A_l\+ \sigma_e ),
\eel
where
\bel
	G_l = \sqrt{2} g_l N_l = \sqrt{\frac{d_z^2 c}{\hbar bR_0^3 \eps_0}}
	\sqrt{\frac{(2l+1)\sqrt{l(l+1)}}{4\pi} \big[1- P_l( \cos(\pi - 2\theta) )
	\big]}.
\eel

\subsection{Numerical analysis}
\subsubsection{Hilbert space}
We are interested in the dynamics of a single excitation, i.e. we truncate the
Hilbert space to
\bel
	\mathcal{H} = \text{Span}\big\{ 
	\underbrace{\ket{e}\ket{g}\ket{\text{vac}}}_{\ket{a}},\;
	\underbrace{\ket{g}\ket{e}\ket{\text{vac}}}_{\ket{b}},\;
	 \{
	 \underbrace{\ket{g}\ket{g}A_l\+\ket{\text{vac}}}_{\ket{l}}
	 \;:\;
	l=1,2,\ldots l_\text{max}\} \big\},
\eel
and separate it into two subspaces
\bal
	\mathcal{H}_o 
	&=& 
	\text{Span}\big\{ 
	\underbrace{\sigma_o\+\ket{g}\ket{g}\ket{\text{vac}}}_{\ket{o}}, \; 
	\{
	\underbrace{A_l\+\ket{g}\ket{g}\ket{\text{vac}}}_{\ket{l}}\;:\; l =
	1,3,5,\ldots\} \big\}
	 \\
	\mathcal{H}_e 
	&=& 
	\text{Span}\big\{
	\underbrace{\sigma_e\+\ket{g}\ket{g}\ket{\text{vac}}}_{\ket{e}}, \; 
	\{
	\underbrace{A_l\+\ket{g}\ket{g}\ket{\text{vac}}}_{\ket{l}}\;:\; l =
	2,4,6,\ldots\} \big\},
\eal
each of which is governed by its own Hamiltonian block.

\subsubsection{Hamiltonian}
The following $H_o, H_e$ act as two independent blocks on $\mathcal{H}_o$ and
$\mathcal{H}_e$.
\bal
	H_o &=& \sum_{l\in\text{odd}} \left[\delta_l A_l\+ A_l  + G_l(A_l \sigma_o\+ +
	A_l\+ \sigma_o) \right] = \sum_{l\in\text{odd}} \big[\delta_l
	\ket{l}\bra{l} + G_l(\ket{o}\bra{l} + \ket{l}\bra{o})\big]
	\\
	H_e &=& \sum_{l\in\text{even}} \left[\delta_l A_l\+ A_l  + G_l(A_l \sigma_e\+ +
	A_l\+ \sigma_e) \right] = \sum_{l\in\text{even}} \big[\delta_l
	\ket{l}\bra{l} + G_l(\ket{e}\bra{l} + \ket{l}\bra{e})\big],
\eal
where 
\bal
	\delta_l &=& \frac{c}{R_0}\left[\sqrt{l(l+1)} - \sqrt{l_0(l_0 + 1)}\right],
\eal
where $l_0$ stands for the atomic frequency, i.e. $\omega_0 =
\frac{c}{R_0}\sqrt{l_0 (l_0 + 1)}$ and
\bal
 	G_l &=&  \sqrt{\frac{d_z^2 c}{\hbar bR_0^3 \eps_0}} \sqrt{\frac{(2l+1)\sqrt{l(l+1)}}{4\pi} \big[1- P_l(
 	\cos(\pi - 2\theta) ) \big]}.
\eal

\subsubsection{Results: Time series}
We start the system from
$
	\ket{\psi(0)} = \ket{e}\ket{g}\ket{\text{vac}} = \ket{a} = \frac{\ket{o} +
	\ket{e}}{\sqrt{2}},
$
evolve it with $U(t) = \exp[-iH t/\hbar]$, to get
\bal
	\ket{\psi(t)} &=& \frac{1}{\sqrt{2}}\big[ e^{-iH_ot/\hbar} \ket{o} + e^{-iH_et/\hbar} \ket{e}
	\big]
	= \frac{1}{\sqrt{2}}\big[ \sum_j \ev{\phi_{o,j}|o}e^{-i\Omega_{o,j} t}
	\ket{\phi_{o,j}} + \sum_k\ev{\phi_{e,k}|e}e^{-i\Omega_{e,k} t} \ket{\phi_{e,k}}
	\big],
\eal	
where $\Omega_{o,j}$, $\ket{\phi_{o,j}}$ and $\Omega_{e,k}$, $\ket{\phi_{e,k}}$
are eigenvalues and eigenstates of $H_{o}/\hbar$ and $H_e/\hbar$, respectively.

Finally, we note that the numerical results shown in \reffig{compare} are independent of the atomic parameters and the thickness of the lens as only ratios of the dipole-dipole interaction, spontaneous decay and cooperative decay are considered (each of which is proportional to the square of the prefactor $\sqrt{d_z^2c/\hbar bR_0^3\epsilon_0}$).

\twocolumngrid


\begin{thebibliography}{82}%
\makeatletter
\providecommand \@ifxundefined [1]{%
 \@ifx{#1\undefined}
}%
\providecommand \@ifnum [1]{%
 \ifnum #1\expandafter \@firstoftwo
 \else \expandafter \@secondoftwo
 \fi
}%
\providecommand \@ifx [1]{%
 \ifx #1\expandafter \@firstoftwo
 \else \expandafter \@secondoftwo
 \fi
}%
\providecommand \natexlab [1]{#1}%
\providecommand \enquote  [1]{``#1''}%
\providecommand \bibnamefont  [1]{#1}%
\providecommand \bibfnamefont [1]{#1}%
\providecommand \citenamefont [1]{#1}%
\providecommand \href@noop [0]{\@secondoftwo}%
\providecommand \href [0]{\begingroup \@sanitize@url \@href}%
\providecommand \@href[1]{\@@startlink{#1}\@@href}%
\providecommand \@@href[1]{\endgroup#1\@@endlink}%
\providecommand \@sanitize@url [0]{\catcode `\\12\catcode `\$12\catcode
  `\&12\catcode `\#12\catcode `\^12\catcode `\_12\catcode `\%12\relax}%
\providecommand \@@startlink[1]{}%
\providecommand \@@endlink[0]{}%
\providecommand \url  [0]{\begingroup\@sanitize@url \@url }%
\providecommand \@url [1]{\endgroup\@href {#1}{\urlprefix }}%
\providecommand \urlprefix  [0]{URL }%
\providecommand \Eprint [0]{\href }%
\providecommand \doibase [0]{http://dx.doi.org/}%
\providecommand \selectlanguage [0]{\@gobble}%
\providecommand \bibinfo  [0]{\@secondoftwo}%
\providecommand \bibfield  [0]{\@secondoftwo}%
\providecommand \translation [1]{[#1]}%
\providecommand \BibitemOpen [0]{}%
\providecommand \bibitemStop [0]{}%
\providecommand \bibitemNoStop [0]{.\EOS\space}%
\providecommand \EOS [0]{\spacefactor3000\relax}%
\providecommand \BibitemShut  [1]{\csname bibitem#1\endcsname}%
\let\auto@bib@innerbib\@empty
\bibitem [{\citenamefont {Maxwell}(1854)}]{Maxwell1854}%
  \BibitemOpen
  \bibfield  {author} {\bibinfo {author} {\bibfnamefont {J.~C.}\ \bibnamefont
  {Maxwell}},\ }\href@noop {} {\bibfield  {journal} {\bibinfo  {journal} {Camb.
  Dublin Math. J}\ }\textbf {\bibinfo {volume} {8}} (\bibinfo {year}
  {1854})}\BibitemShut {NoStop}%
\bibitem [{\citenamefont {Tai}(1958)}]{Tai1958}%
  \BibitemOpen
  \bibfield  {author} {\bibinfo {author} {\bibfnamefont {C.~T.}\ \bibnamefont
  {Tai}},\ }\href@noop {} {\bibfield  {journal} {\bibinfo  {journal} {Nature}\
  }\textbf {\bibinfo {volume} {182}} (\bibinfo {year} {1958})}\BibitemShut
  {NoStop}%
\bibitem [{\citenamefont {Rosu}\ and\ \citenamefont {Reyes}(1994)}]{Rosu1994}%
  \BibitemOpen
  \bibfield  {author} {\bibinfo {author} {\bibfnamefont {H.}~\bibnamefont
  {Rosu}}\ and\ \bibinfo {author} {\bibfnamefont {M.}~\bibnamefont {Reyes}},\
  }\href {http://arxiv.org/abs/atom-ph/9604006
  http://dx.doi.org/10.1007/BF02463739
  http://link.springer.com/10.1007/BF02463739} {\bibfield  {journal} {\bibinfo
  {journal} {Il Nuovo Cimento D}\ }\textbf {\bibinfo {volume} {16}},\ \bibinfo
  {pages} {517} (\bibinfo {year} {1994})}\BibitemShut {NoStop}%
\bibitem [{\citenamefont {Greenwood}\ and\ \citenamefont {{Jian-Ming
  Jin}}(1999)}]{Greenwood1999}%
  \BibitemOpen
  \bibfield  {author} {\bibinfo {author} {\bibfnamefont {A.}~\bibnamefont
  {Greenwood}}\ and\ \bibinfo {author} {\bibnamefont {{Jian-Ming Jin}}},\
  }\href {http://ieeexplore.ieee.org/document/801510/} {\bibfield  {journal}
  {\bibinfo  {journal} {IEEE Antennas and Propagation Magazine}\ }\textbf
  {\bibinfo {volume} {41}},\ \bibinfo {pages} {9} (\bibinfo {year}
  {1999})}\BibitemShut {NoStop}%
\bibitem [{\citenamefont {Makowski}\ and\ \citenamefont
  {G{\'{o}}rska}(2009)}]{Makowski2009}%
  \BibitemOpen
  \bibfield  {author} {\bibinfo {author} {\bibfnamefont {A.~J.}\ \bibnamefont
  {Makowski}}\ and\ \bibinfo {author} {\bibfnamefont {K.~J.}\ \bibnamefont
  {G{\'{o}}rska}},\ }\href
  {https://link.aps.org/doi/10.1103/PhysRevA.79.052116} {\bibfield  {journal}
  {\bibinfo  {journal} {Physical Review A}\ }\textbf {\bibinfo {volume} {79}},\
  \bibinfo {pages} {052116} (\bibinfo {year} {2009})}\BibitemShut {NoStop}%
\bibitem [{\citenamefont {Rosu}\ \emph {et~al.}(1996)\citenamefont {Rosu},
  \citenamefont {Reyes}, \citenamefont {Wolf},\ and\ \citenamefont
  {Obregon}}]{Rosu1996}%
  \BibitemOpen
  \bibfield  {author} {\bibinfo {author} {\bibfnamefont {H.~C.}\ \bibnamefont
  {Rosu}}, \bibinfo {author} {\bibfnamefont {M.}~\bibnamefont {Reyes}},
  \bibinfo {author} {\bibfnamefont {K.~B.}\ \bibnamefont {Wolf}}, \ and\
  \bibinfo {author} {\bibfnamefont {O.}~\bibnamefont {Obregon}},\ }in\ \href
  {http://arxiv.org/abs/quant-ph/9504018 http://dx.doi.org/10.1117/12.231113
  http://proceedings.spiedigitallibrary.org/proceeding.aspx?articleid=1018141}
  {\emph {\bibinfo {booktitle} {Proc. SPIE 2730, Second Iberoamerican Meeting
  on Optics, 436 (February 5, 1996)}}},\ \bibinfo {editor} {edited by\ \bibinfo
  {editor} {\bibfnamefont {D.}~\bibnamefont {Malacara-Hernandez}}, \bibinfo
  {editor} {\bibfnamefont {S.~E.}\ \bibnamefont {Acosta-Ortiz}}, \bibinfo
  {editor} {\bibfnamefont {R.}~\bibnamefont {Rodriguez-Vera}}, \bibinfo
  {editor} {\bibfnamefont {Z.}~\bibnamefont {Malacara}}, \ and\ \bibinfo
  {editor} {\bibfnamefont {A.~A.}\ \bibnamefont {Morales}}}\ (\bibinfo {year}
  {1996})\ pp.\ \bibinfo {pages} {436--439}\BibitemShut {NoStop}%
\bibitem [{\citenamefont {Leonhardt}(2009)}]{Leonhardt2009}%
  \BibitemOpen
  \bibfield  {author} {\bibinfo {author} {\bibfnamefont {U.}~\bibnamefont
  {Leonhardt}},\ }\href
  {http://stacks.iop.org/1367-2630/11/i=9/a=093040?key=crossref.c63a62fca2ec4e072986ce40b843057b}
  {\bibfield  {journal} {\bibinfo  {journal} {New Journal of Physics}\ }\textbf
  {\bibinfo {volume} {11}},\ \bibinfo {pages} {093040} (\bibinfo {year}
  {2009})}\BibitemShut {NoStop}%
\bibitem [{\citenamefont {Leonhardt}\ and\ \citenamefont
  {Philbin}(2010{\natexlab{a}})}]{Leonhardt2010b}%
  \BibitemOpen
  \bibfield  {author} {\bibinfo {author} {\bibfnamefont {U.}~\bibnamefont
  {Leonhardt}}\ and\ \bibinfo {author} {\bibfnamefont {T.}~\bibnamefont
  {Philbin}},\ }\href@noop {} {\emph {\bibinfo {title} {{Geometry and light:
  the science of invisibility}}}}\ (\bibinfo  {publisher} {Dover
  Publications},\ \bibinfo {year} {2010})\BibitemShut {NoStop}%
\bibitem [{\citenamefont {Ma}\ \emph {et~al.}(2011)\citenamefont {Ma},
  \citenamefont {Sahebdivan}, \citenamefont {Ong}, \citenamefont {Tyc},\ and\
  \citenamefont {Leonhardt}}]{Ma2011}%
  \BibitemOpen
  \bibfield  {author} {\bibinfo {author} {\bibfnamefont {Y.~G.}\ \bibnamefont
  {Ma}}, \bibinfo {author} {\bibfnamefont {S.}~\bibnamefont {Sahebdivan}},
  \bibinfo {author} {\bibfnamefont {C.~K.}\ \bibnamefont {Ong}}, \bibinfo
  {author} {\bibfnamefont {T.}~\bibnamefont {Tyc}}, \ and\ \bibinfo {author}
  {\bibfnamefont {U.}~\bibnamefont {Leonhardt}},\ }\href
  {http://stacks.iop.org/1367-2630/13/i=3/a=033016?key=crossref.f912cd2516e17dc8bed6f63941ba64f9}
  {\bibfield  {journal} {\bibinfo  {journal} {New Journal of Physics}\ }\textbf
  {\bibinfo {volume} {13}},\ \bibinfo {pages} {033016} (\bibinfo {year}
  {2011})}\BibitemShut {NoStop}%
\bibitem [{\citenamefont {Born}\ and\ \citenamefont {Wolf}(1999)}]{Born1999}%
  \BibitemOpen
  \bibfield  {author} {\bibinfo {author} {\bibfnamefont {M.}~\bibnamefont
  {Born}}\ and\ \bibinfo {author} {\bibfnamefont {E.}~\bibnamefont {Wolf}},\
  }\href@noop {} {\emph {\bibinfo {title} {{Principles of optics:
  electromagnetic theory of propagation, interference and diffraction of
  light}}}}\ (\bibinfo  {publisher} {Cambridge University Press},\ \bibinfo
  {year} {1999})\BibitemShut {NoStop}%
\bibitem [{\citenamefont {Leonhardt}\ and\ \citenamefont
  {Philbin}(2010{\natexlab{b}})}]{Leonhardt2010}%
  \BibitemOpen
  \bibfield  {author} {\bibinfo {author} {\bibfnamefont {U.}~\bibnamefont
  {Leonhardt}}\ and\ \bibinfo {author} {\bibfnamefont {T.~G.}\ \bibnamefont
  {Philbin}},\ }\href {https://link.aps.org/doi/10.1103/PhysRevA.82.057802}
  {\bibfield  {journal} {\bibinfo  {journal} {Physical Review A}\ }\textbf
  {\bibinfo {volume} {82}},\ \bibinfo {pages} {057802} (\bibinfo {year}
  {2010}{\natexlab{b}})}\BibitemShut {NoStop}%
\bibitem [{\citenamefont {Leonhardt}(2010)}]{Leonhardt2010a}%
  \BibitemOpen
  \bibfield  {author} {\bibinfo {author} {\bibfnamefont {U.}~\bibnamefont
  {Leonhardt}},\ }\href
  {http://stacks.iop.org/1367-2630/12/i=5/a=058002?key=crossref.260587c815c6e14c0e0a5472bfe5ecdc}
  {\bibfield  {journal} {\bibinfo  {journal} {New Journal of Physics}\ }\textbf
  {\bibinfo {volume} {12}},\ \bibinfo {pages} {058002} (\bibinfo {year}
  {2010})}\BibitemShut {NoStop}%
\bibitem [{\citenamefont {Leonhardt}(2011)}]{Leonhardt2011b}%
  \BibitemOpen
  \bibfield  {author} {\bibinfo {author} {\bibfnamefont {U.}~\bibnamefont
  {Leonhardt}},\ }\href {\doibase 10.1088/1367-2630/13/2/028002} {\bibfield
  {journal} {\bibinfo  {journal} {New Journal of Physics}\ }\textbf {\bibinfo
  {volume} {13}},\ \bibinfo {pages} {028002} (\bibinfo {year}
  {2011})}\BibitemShut {NoStop}%
\bibitem [{\citenamefont {Tyc}\ and\ \citenamefont {Zhang}(2011)}]{Tyc2011}%
  \BibitemOpen
  \bibfield  {author} {\bibinfo {author} {\bibfnamefont {T.}~\bibnamefont
  {Tyc}}\ and\ \bibinfo {author} {\bibfnamefont {X.}~\bibnamefont {Zhang}},\
  }\href {http://cat.inist.fr/?aModele=afficheN{\&}cpsidt=24798160
  http://www.nature.com/doifinder/10.1038/480042a} {\bibfield  {journal}
  {\bibinfo  {journal} {Nature}\ }\textbf {\bibinfo {volume} {480}},\ \bibinfo
  {pages} {42} (\bibinfo {year} {2011})}\BibitemShut {NoStop}%
\bibitem [{\citenamefont {Blaikie}(2010)}]{Blaikie2010}%
  \BibitemOpen
  \bibfield  {author} {\bibinfo {author} {\bibfnamefont {R.~J.}\ \bibnamefont
  {Blaikie}},\ }\href
  {http://stacks.iop.org/1367-2630/12/i=5/a=058001?key=crossref.cc827fa2cfe54a5742e15bb8b0891308}
  {\bibfield  {journal} {\bibinfo  {journal} {New Journal of Physics}\ }\textbf
  {\bibinfo {volume} {12}},\ \bibinfo {pages} {058001} (\bibinfo {year}
  {2010})}\BibitemShut {NoStop}%
\bibitem [{\citenamefont {Merlin}(2010)}]{Merlin2010}%
  \BibitemOpen
  \bibfield  {author} {\bibinfo {author} {\bibfnamefont {R.}~\bibnamefont
  {Merlin}},\ }\href {\doibase 10.1103/PhysRevA.82.057801} {\bibfield
  {journal} {\bibinfo  {journal} {Physical Review A}\ }\textbf {\bibinfo
  {volume} {82}},\ \bibinfo {pages} {057801} (\bibinfo {year}
  {2010})}\BibitemShut {NoStop}%
\bibitem [{\citenamefont {Blaikie}(2011)}]{Blaikie2011}%
  \BibitemOpen
  \bibfield  {author} {\bibinfo {author} {\bibfnamefont {R.~J.}\ \bibnamefont
  {Blaikie}},\ }\href {\doibase 10.1088/1367-2630/13/12/125006} {\bibfield
  {journal} {\bibinfo  {journal} {New Journal of Physics}\ }\textbf {\bibinfo
  {volume} {13}},\ \bibinfo {pages} {125006} (\bibinfo {year}
  {2011})}\BibitemShut {NoStop}%
\bibitem [{\citenamefont {Kinsler}(2010)}]{Kinsler2010}%
  \BibitemOpen
  \bibfield  {author} {\bibinfo {author} {\bibfnamefont {P.}~\bibnamefont
  {Kinsler}},\ }\href {https://link.aps.org/doi/10.1103/PhysRevA.82.055804}
  {\bibfield  {journal} {\bibinfo  {journal} {Physical Review A}\ }\textbf
  {\bibinfo {volume} {82}},\ \bibinfo {pages} {055804} (\bibinfo {year}
  {2010})}\BibitemShut {NoStop}%
\bibitem [{\citenamefont {Gonz{\'{a}}lez}\ \emph {et~al.}(2011)\citenamefont
  {Gonz{\'{a}}lez}, \citenamefont {Ben{\'{i}}tez},\ and\ \citenamefont
  {Mi{\~{n}}ano}}]{Gonzalez2011}%
  \BibitemOpen
  \bibfield  {author} {\bibinfo {author} {\bibfnamefont {J.~C.}\ \bibnamefont
  {Gonz{\'{a}}lez}}, \bibinfo {author} {\bibfnamefont {P.}~\bibnamefont
  {Ben{\'{i}}tez}}, \ and\ \bibinfo {author} {\bibfnamefont {J.~C.}\
  \bibnamefont {Mi{\~{n}}ano}},\ }\href {\doibase
  10.1088/1367-2630/13/2/023038} {\bibfield  {journal} {\bibinfo  {journal}
  {New Journal of Physics}\ }\textbf {\bibinfo {volume} {13}},\ \bibinfo
  {pages} {023038} (\bibinfo {year} {2011})}\BibitemShut {NoStop}%
\bibitem [{\citenamefont {Quevedo-Teruel}\ \emph {et~al.}(2012)\citenamefont
  {Quevedo-Teruel}, \citenamefont {Mitchell-Thomas},\ and\ \citenamefont
  {Hao}}]{Quevedo-Teruel2012}%
  \BibitemOpen
  \bibfield  {author} {\bibinfo {author} {\bibfnamefont {O.}~\bibnamefont
  {Quevedo-Teruel}}, \bibinfo {author} {\bibfnamefont {R.~C.}\ \bibnamefont
  {Mitchell-Thomas}}, \ and\ \bibinfo {author} {\bibfnamefont {Y.}~\bibnamefont
  {Hao}},\ }\href {https://link.aps.org/doi/10.1103/PhysRevA.86.053817}
  {\bibfield  {journal} {\bibinfo  {journal} {Physical Review A}\ }\textbf
  {\bibinfo {volume} {86}},\ \bibinfo {pages} {053817} (\bibinfo {year}
  {2012})}\BibitemShut {NoStop}%
\bibitem [{\citenamefont {Tyc}\ and\ \citenamefont {Danner}(2014)}]{Tyc2014}%
  \BibitemOpen
  \bibfield  {author} {\bibinfo {author} {\bibfnamefont {T.}~\bibnamefont
  {Tyc}}\ and\ \bibinfo {author} {\bibfnamefont {A.}~\bibnamefont {Danner}},\
  }\href {\doibase 10.1088/1367-2630/16/6/063001} {\bibfield  {journal}
  {\bibinfo  {journal} {New Journal of Physics}\ }\textbf {\bibinfo {volume}
  {16}},\ \bibinfo {pages} {063001} (\bibinfo {year} {2014})}\BibitemShut
  {NoStop}%
\bibitem [{\citenamefont {Gonz{\'{a}}lez}\ \emph {et~al.}(2012)\citenamefont
  {Gonz{\'{a}}lez}, \citenamefont {Grabovi{\v{c}}ki{\'{c}}}, \citenamefont
  {Ben{\'{i}}tez},\ and\ \citenamefont {Mi{\~{n}}ano}}]{Gonzalez2012}%
  \BibitemOpen
  \bibfield  {author} {\bibinfo {author} {\bibfnamefont {J.~C.}\ \bibnamefont
  {Gonz{\'{a}}lez}}, \bibinfo {author} {\bibfnamefont {D.}~\bibnamefont
  {Grabovi{\v{c}}ki{\'{c}}}}, \bibinfo {author} {\bibfnamefont
  {P.}~\bibnamefont {Ben{\'{i}}tez}}, \ and\ \bibinfo {author} {\bibfnamefont
  {J.~C.}\ \bibnamefont {Mi{\~{n}}ano}},\ }\href {\doibase
  10.1088/1367-2630/14/8/083033} {\bibfield  {journal} {\bibinfo  {journal}
  {New Journal of Physics}\ }\textbf {\bibinfo {volume} {14}},\ \bibinfo
  {pages} {083033} (\bibinfo {year} {2012})}\BibitemShut {NoStop}%
\bibitem [{\citenamefont {Xu}\ and\ \citenamefont {Chen}(2012)}]{Xu2012}%
  \BibitemOpen
  \bibfield  {author} {\bibinfo {author} {\bibfnamefont {L.}~\bibnamefont
  {Xu}}\ and\ \bibinfo {author} {\bibfnamefont {H.}~\bibnamefont {Chen}},\
  }\href {\doibase 10.1209/0295-5075/100/34001} {\bibfield  {journal} {\bibinfo
   {journal} {EPL (Europhysics Letters)}\ }\textbf {\bibinfo {volume} {100}},\
  \bibinfo {pages} {34001} (\bibinfo {year} {2012})}\BibitemShut {NoStop}%
\bibitem [{\citenamefont {Ma}\ \emph {et~al.}(2013)\citenamefont {Ma},
  \citenamefont {Wu},\ and\ \citenamefont {Ong}}]{Ma2013}%
  \BibitemOpen
  \bibfield  {author} {\bibinfo {author} {\bibfnamefont {Y.}~\bibnamefont
  {Ma}}, \bibinfo {author} {\bibfnamefont {T.}~\bibnamefont {Wu}}, \ and\
  \bibinfo {author} {\bibfnamefont {C.~K.}\ \bibnamefont {Ong}},\ }\href
  {\doibase 10.1088/2040-8978/15/12/125705} {\bibfield  {journal} {\bibinfo
  {journal} {Journal of Optics}\ }\textbf {\bibinfo {volume} {15}},\ \bibinfo
  {pages} {125705} (\bibinfo {year} {2013})}\BibitemShut {NoStop}%
\bibitem [{\citenamefont {Sun}\ and\ \citenamefont {He}(2010)}]{Sun2010}%
  \BibitemOpen
  \bibfield  {author} {\bibinfo {author} {\bibfnamefont {F.}~\bibnamefont
  {Sun}}\ and\ \bibinfo {author} {\bibfnamefont {S.}~\bibnamefont {He}},\
  }\href {\doibase 10.2528/PIER10091003} {\bibfield  {journal} {\bibinfo
  {journal} {Progress In Electromagnetics Research}\ }\textbf {\bibinfo
  {volume} {108}},\ \bibinfo {pages} {307} (\bibinfo {year}
  {2010})}\BibitemShut {NoStop}%
\bibitem [{\citenamefont {Sun}\ \emph {et~al.}(2010)\citenamefont {Sun},
  \citenamefont {Ge},\ and\ \citenamefont {He}}]{Sun2010b}%
  \BibitemOpen
  \bibfield  {author} {\bibinfo {author} {\bibfnamefont {F.}~\bibnamefont
  {Sun}}, \bibinfo {author} {\bibfnamefont {X.~C.}\ \bibnamefont {Ge}}, \ and\
  \bibinfo {author} {\bibfnamefont {S.}~\bibnamefont {He}},\ }\href {\doibase
  10.2528/PIER10110313} {\bibfield  {journal} {\bibinfo  {journal} {Progress In
  Electromagnetics Research}\ }\textbf {\bibinfo {volume} {110}},\ \bibinfo
  {pages} {313} (\bibinfo {year} {2010})}\BibitemShut {NoStop}%
\bibitem [{\citenamefont {Kinsler}\ and\ \citenamefont
  {Favaro}(2011)}]{Kinsler2011}%
  \BibitemOpen
  \bibfield  {author} {\bibinfo {author} {\bibfnamefont {P.}~\bibnamefont
  {Kinsler}}\ and\ \bibinfo {author} {\bibfnamefont {A.}~\bibnamefont
  {Favaro}},\ }\href
  {http://stacks.iop.org/1367-2630/13/i=2/a=028001?key=crossref.9eae278160a87dc9513d6a6c2ee38711}
  {\bibfield  {journal} {\bibinfo  {journal} {New Journal of Physics}\ }\textbf
  {\bibinfo {volume} {13}},\ \bibinfo {pages} {028001} (\bibinfo {year}
  {2011})}\BibitemShut {NoStop}%
\bibitem [{\citenamefont {Leonhardt}\ and\ \citenamefont
  {Sahebdivan}(2011)}]{Leonhardt2011}%
  \BibitemOpen
  \bibfield  {author} {\bibinfo {author} {\bibfnamefont {U.}~\bibnamefont
  {Leonhardt}}\ and\ \bibinfo {author} {\bibfnamefont {S.}~\bibnamefont
  {Sahebdivan}},\ }\href {\doibase 10.1088/2040-8978/13/2/024016} {\bibfield
  {journal} {\bibinfo  {journal} {Journal of Optics}\ }\textbf {\bibinfo
  {volume} {13}},\ \bibinfo {pages} {024016} (\bibinfo {year}
  {2011})}\BibitemShut {NoStop}%
\bibitem [{\citenamefont {Pazynin}\ and\ \citenamefont
  {Kryvchikova}(2012)}]{Pazynin2012}%
  \BibitemOpen
  \bibfield  {author} {\bibinfo {author} {\bibfnamefont {L.~A.}\ \bibnamefont
  {Pazynin}}\ and\ \bibinfo {author} {\bibfnamefont {G.~O.}\ \bibnamefont
  {Kryvchikova}},\ }\href {\doibase 10.2528/PIER12073009} {\bibfield  {journal}
  {\bibinfo  {journal} {Progress In Electromagnetics Research}\ }\textbf
  {\bibinfo {volume} {131}},\ \bibinfo {pages} {425} (\bibinfo {year}
  {2012})}\BibitemShut {NoStop}%
\bibitem [{\citenamefont {Liu}\ \emph {et~al.}(2013)\citenamefont {Liu},
  \citenamefont {Mendis},\ and\ \citenamefont {Mittleman}}]{Liu2013}%
  \BibitemOpen
  \bibfield  {author} {\bibinfo {author} {\bibfnamefont {J.}~\bibnamefont
  {Liu}}, \bibinfo {author} {\bibfnamefont {R.}~\bibnamefont {Mendis}}, \ and\
  \bibinfo {author} {\bibfnamefont {D.~M.}\ \bibnamefont {Mittleman}},\ }\href
  {http://aip.scitation.org/doi/10.1063/1.4813820} {\bibfield  {journal}
  {\bibinfo  {journal} {Applied Physics Letters}\ }\textbf {\bibinfo {volume}
  {103}},\ \bibinfo {pages} {031104} (\bibinfo {year} {2013})}\BibitemShut
  {NoStop}%
\bibitem [{\citenamefont {Alonso}(2015)}]{Alonso2015}%
  \BibitemOpen
  \bibfield  {author} {\bibinfo {author} {\bibfnamefont {M.~A.}\ \bibnamefont
  {Alonso}},\ }\href {\doibase 10.1088/1367-2630/17/7/073013} {\bibfield
  {journal} {\bibinfo  {journal} {New Journal of Physics}\ }\textbf {\bibinfo
  {volume} {17}},\ \bibinfo {pages} {073013} (\bibinfo {year}
  {2015})}\BibitemShut {NoStop}%
\bibitem [{\citenamefont {Horsley}\ \emph {et~al.}(2015)\citenamefont
  {Horsley}, \citenamefont {Foster}, \citenamefont {Tyc},\ and\ \citenamefont
  {Philbin}}]{Horsley2015}%
  \BibitemOpen
  \bibfield  {author} {\bibinfo {author} {\bibfnamefont {S.~A.~R.}\
  \bibnamefont {Horsley}}, \bibinfo {author} {\bibfnamefont {R.~N.}\
  \bibnamefont {Foster}}, \bibinfo {author} {\bibfnamefont {T.}~\bibnamefont
  {Tyc}}, \ and\ \bibinfo {author} {\bibfnamefont {T.~G.}\ \bibnamefont
  {Philbin}},\ }\href {\doibase 10.1088/1367-2630/17/5/053050} {\bibfield
  {journal} {\bibinfo  {journal} {New Journal of Physics}\ }\textbf {\bibinfo
  {volume} {17}},\ \bibinfo {pages} {053050} (\bibinfo {year}
  {2015})}\BibitemShut {NoStop}%
\bibitem [{\citenamefont {He}\ \emph {et~al.}(2015)\citenamefont {He},
  \citenamefont {Sun}, \citenamefont {Guo}, \citenamefont {Zhong},
  \citenamefont {Lan}, \citenamefont {Jiang}, \citenamefont {Ma},\ and\
  \citenamefont {Wu}}]{He2015}%
  \BibitemOpen
  \bibfield  {author} {\bibinfo {author} {\bibfnamefont {S.}~\bibnamefont
  {He}}, \bibinfo {author} {\bibfnamefont {F.}~\bibnamefont {Sun}}, \bibinfo
  {author} {\bibfnamefont {S.}~\bibnamefont {Guo}}, \bibinfo {author}
  {\bibfnamefont {S.}~\bibnamefont {Zhong}}, \bibinfo {author} {\bibfnamefont
  {L.}~\bibnamefont {Lan}}, \bibinfo {author} {\bibfnamefont {W.}~\bibnamefont
  {Jiang}}, \bibinfo {author} {\bibfnamefont {Y.}~\bibnamefont {Ma}}, \ and\
  \bibinfo {author} {\bibfnamefont {T.}~\bibnamefont {Wu}},\ }\href {\doibase
  10.2528/PIER15050101} {\bibfield  {journal} {\bibinfo  {journal} {Progress In
  Electromagnetics Research}\ }\textbf {\bibinfo {volume} {152}},\ \bibinfo
  {pages} {1} (\bibinfo {year} {2015})}\BibitemShut {NoStop}%
\bibitem [{\citenamefont {Rosenblatt}\ and\ \citenamefont
  {Orenstein}(2017)}]{Rosenblatt2017}%
  \BibitemOpen
  \bibfield  {author} {\bibinfo {author} {\bibfnamefont {G.}~\bibnamefont
  {Rosenblatt}}\ and\ \bibinfo {author} {\bibfnamefont {M.}~\bibnamefont
  {Orenstein}},\ }\href {\doibase 10.1103/PhysRevA.95.053857} {\bibfield
  {journal} {\bibinfo  {journal} {Physical Review A}\ }\textbf {\bibinfo
  {volume} {95}},\ \bibinfo {pages} {053857} (\bibinfo {year}
  {2017})}\BibitemShut {NoStop}%
\bibitem [{\citenamefont {Mi{\~{n}}ano}\ \emph {et~al.}(2011)\citenamefont
  {Mi{\~{n}}ano}, \citenamefont {Marqu{\'{e}}s}, \citenamefont
  {Gonz{\'{a}}lez}, \citenamefont {Ben{\'{i}}tez}, \citenamefont {Delgado},
  \citenamefont {Grabovickic},\ and\ \citenamefont {Freire}}]{Minano2011}%
  \BibitemOpen
  \bibfield  {author} {\bibinfo {author} {\bibfnamefont {J.~C.}\ \bibnamefont
  {Mi{\~{n}}ano}}, \bibinfo {author} {\bibfnamefont {R.}~\bibnamefont
  {Marqu{\'{e}}s}}, \bibinfo {author} {\bibfnamefont {J.~C.}\ \bibnamefont
  {Gonz{\'{a}}lez}}, \bibinfo {author} {\bibfnamefont {P.}~\bibnamefont
  {Ben{\'{i}}tez}}, \bibinfo {author} {\bibfnamefont {V.}~\bibnamefont
  {Delgado}}, \bibinfo {author} {\bibfnamefont {D.}~\bibnamefont
  {Grabovickic}}, \ and\ \bibinfo {author} {\bibfnamefont {M.}~\bibnamefont
  {Freire}},\ }\href {\doibase 10.1088/1367-2630/13/12/125009} {\bibfield
  {journal} {\bibinfo  {journal} {New Journal of Physics}\ }\textbf {\bibinfo
  {volume} {13}},\ \bibinfo {pages} {125009} (\bibinfo {year}
  {2011})}\BibitemShut {NoStop}%
\bibitem [{\citenamefont {Mi{\~{n}}ano}\ \emph {et~al.}(2014)\citenamefont
  {Mi{\~{n}}ano}, \citenamefont {S{\'{a}}nchez-Dehesa}, \citenamefont
  {Gonz{\'{a}}lez}, \citenamefont {Ben{\'{i}}tez}, \citenamefont
  {Grabovi{\v{c}}ki{\'{c}}}, \citenamefont {Carbonell},\ and\ \citenamefont
  {Ahmadpanahi}}]{Minano2014}%
  \BibitemOpen
  \bibfield  {author} {\bibinfo {author} {\bibfnamefont {J.~C.}\ \bibnamefont
  {Mi{\~{n}}ano}}, \bibinfo {author} {\bibfnamefont {J.}~\bibnamefont
  {S{\'{a}}nchez-Dehesa}}, \bibinfo {author} {\bibfnamefont {J.~C.}\
  \bibnamefont {Gonz{\'{a}}lez}}, \bibinfo {author} {\bibfnamefont
  {P.}~\bibnamefont {Ben{\'{i}}tez}}, \bibinfo {author} {\bibfnamefont
  {D.}~\bibnamefont {Grabovi{\v{c}}ki{\'{c}}}}, \bibinfo {author}
  {\bibfnamefont {J.}~\bibnamefont {Carbonell}}, \ and\ \bibinfo {author}
  {\bibfnamefont {H.}~\bibnamefont {Ahmadpanahi}},\ }\href {\doibase
  10.1088/1367-2630/16/3/033015} {\bibfield  {journal} {\bibinfo  {journal}
  {New Journal of Physics}\ }\textbf {\bibinfo {volume} {16}},\ \bibinfo
  {pages} {033015} (\bibinfo {year} {2014})}\BibitemShut {NoStop}%
\bibitem [{\citenamefont {Leonhardt}\ \emph {et~al.}(2015)\citenamefont
  {Leonhardt}, \citenamefont {Sahebdivan}, \citenamefont {Kogan},\ and\
  \citenamefont {Tyc}}]{Leonhardt2015}%
  \BibitemOpen
  \bibfield  {author} {\bibinfo {author} {\bibfnamefont {U.}~\bibnamefont
  {Leonhardt}}, \bibinfo {author} {\bibfnamefont {S.}~\bibnamefont
  {Sahebdivan}}, \bibinfo {author} {\bibfnamefont {A.}~\bibnamefont {Kogan}}, \
  and\ \bibinfo {author} {\bibfnamefont {T.}~\bibnamefont {Tyc}},\ }\href
  {\doibase 10.1088/1367-2630/17/5/053007} {\bibfield  {journal} {\bibinfo
  {journal} {New Journal of Physics}\ }\textbf {\bibinfo {volume} {17}},\
  \bibinfo {pages} {053007} (\bibinfo {year} {2015})}\BibitemShut {NoStop}%
\bibitem [{\citenamefont {Leonhardt}\ and\ \citenamefont
  {Sahebdivan}(2015)}]{Leonhardt2015a}%
  \BibitemOpen
  \bibfield  {author} {\bibinfo {author} {\bibfnamefont {U.}~\bibnamefont
  {Leonhardt}}\ and\ \bibinfo {author} {\bibfnamefont {S.}~\bibnamefont
  {Sahebdivan}},\ }\href {\doibase 10.1103/PhysRevA.92.053848} {\bibfield
  {journal} {\bibinfo  {journal} {Physical Review A}\ }\textbf {\bibinfo
  {volume} {92}},\ \bibinfo {pages} {053848} (\bibinfo {year}
  {2015})}\BibitemShut {NoStop}%
\bibitem [{\citenamefont {Gomez-Medina}\ \emph {et~al.}(2001)\citenamefont
  {Gomez-Medina}, \citenamefont {{San Jose}}, \citenamefont {Garcia-Martin},
  \citenamefont {Lester}, \citenamefont {Nieto-Vesperinas},\ and\ \citenamefont
  {Saenz}}]{Gomez-Medina2001}%
  \BibitemOpen
  \bibfield  {author} {\bibinfo {author} {\bibfnamefont {R.}~\bibnamefont
  {Gomez-Medina}}, \bibinfo {author} {\bibfnamefont {P.}~\bibnamefont {{San
  Jose}}}, \bibinfo {author} {\bibfnamefont {A.}~\bibnamefont {Garcia-Martin}},
  \bibinfo {author} {\bibfnamefont {M.}~\bibnamefont {Lester}}, \bibinfo
  {author} {\bibfnamefont {M.}~\bibnamefont {Nieto-Vesperinas}}, \ and\
  \bibinfo {author} {\bibfnamefont {J.~J.}\ \bibnamefont {Saenz}},\ }\href
  {https://link.aps.org/doi/10.1103/PhysRevLett.86.4275} {\bibfield  {journal}
  {\bibinfo  {journal} {Physical Review Letters}\ }\textbf {\bibinfo {volume}
  {86}},\ \bibinfo {pages} {4275} (\bibinfo {year} {2001})}\BibitemShut
  {NoStop}%
\bibitem [{\citenamefont {Shahmoon}\ and\ \citenamefont
  {Kurizki}(2013)}]{Shahmoon2013}%
  \BibitemOpen
  \bibfield  {author} {\bibinfo {author} {\bibfnamefont {E.}~\bibnamefont
  {Shahmoon}}\ and\ \bibinfo {author} {\bibfnamefont {G.}~\bibnamefont
  {Kurizki}},\ }\href {https://link.aps.org/doi/10.1103/PhysRevA.87.033831}
  {\bibfield  {journal} {\bibinfo  {journal} {Physical Review A}\ }\textbf
  {\bibinfo {volume} {87}},\ \bibinfo {pages} {033831} (\bibinfo {year}
  {2013})}\BibitemShut {NoStop}%
\bibitem [{\citenamefont {Chang}\ \emph {et~al.}(2006)\citenamefont {Chang},
  \citenamefont {S{\o}rensen}, \citenamefont {Hemmer},\ and\ \citenamefont
  {Lukin}}]{Chang2006}%
  \BibitemOpen
  \bibfield  {author} {\bibinfo {author} {\bibfnamefont {D.~E.}\ \bibnamefont
  {Chang}}, \bibinfo {author} {\bibfnamefont {A.~S.}\ \bibnamefont
  {S{\o}rensen}}, \bibinfo {author} {\bibfnamefont {P.~R.}\ \bibnamefont
  {Hemmer}}, \ and\ \bibinfo {author} {\bibfnamefont {M.~D.}\ \bibnamefont
  {Lukin}},\ }\href {http://link.aps.org/doi/10.1103/PhysRevLett.97.053002}
  {\bibfield  {journal} {\bibinfo  {journal} {Physical Review Letters}\
  }\textbf {\bibinfo {volume} {97}},\ \bibinfo {pages} {053002} (\bibinfo
  {year} {2006})}\BibitemShut {NoStop}%
\bibitem [{\citenamefont {Gonzalez-Tudela}\ \emph {et~al.}(2011)\citenamefont
  {Gonzalez-Tudela}, \citenamefont {Martin-Cano}, \citenamefont {Moreno},
  \citenamefont {Martin-Moreno}, \citenamefont {Tejedor},\ and\ \citenamefont
  {Garcia-Vidal}}]{Gonzalez-Tudela2011}%
  \BibitemOpen
  \bibfield  {author} {\bibinfo {author} {\bibfnamefont {A.}~\bibnamefont
  {Gonzalez-Tudela}}, \bibinfo {author} {\bibfnamefont {D.}~\bibnamefont
  {Martin-Cano}}, \bibinfo {author} {\bibfnamefont {E.}~\bibnamefont {Moreno}},
  \bibinfo {author} {\bibfnamefont {L.}~\bibnamefont {Martin-Moreno}}, \bibinfo
  {author} {\bibfnamefont {C.}~\bibnamefont {Tejedor}}, \ and\ \bibinfo
  {author} {\bibfnamefont {F.~J.}\ \bibnamefont {Garcia-Vidal}},\ }\href
  {https://link.aps.org/doi/10.1103/PhysRevLett.106.020501} {\bibfield
  {journal} {\bibinfo  {journal} {Physical Review Letters}\ }\textbf {\bibinfo
  {volume} {106}},\ \bibinfo {pages} {020501} (\bibinfo {year}
  {2011})}\BibitemShut {NoStop}%
\bibitem [{\citenamefont {Dzsotjan}\ \emph {et~al.}(2010)\citenamefont
  {Dzsotjan}, \citenamefont {S{\o}rensen},\ and\ \citenamefont
  {Fleischhauer}}]{Dzsotjan2010}%
  \BibitemOpen
  \bibfield  {author} {\bibinfo {author} {\bibfnamefont {D.}~\bibnamefont
  {Dzsotjan}}, \bibinfo {author} {\bibfnamefont {A.~S.}\ \bibnamefont
  {S{\o}rensen}}, \ and\ \bibinfo {author} {\bibfnamefont {M.}~\bibnamefont
  {Fleischhauer}},\ }\href
  {https://link.aps.org/doi/10.1103/PhysRevB.82.075427} {\bibfield  {journal}
  {\bibinfo  {journal} {Physical Review B}\ }\textbf {\bibinfo {volume} {82}},\
  \bibinfo {pages} {075427} (\bibinfo {year} {2010})}\BibitemShut {NoStop}%
\bibitem [{\citenamefont {Lalumiere}\ \emph {et~al.}(2013)\citenamefont
  {Lalumiere}, \citenamefont {Sanders}, \citenamefont {van Loo}, \citenamefont
  {Fedorov}, \citenamefont {Wallraff},\ and\ \citenamefont
  {Blais}}]{Lalumiere2013}%
  \BibitemOpen
  \bibfield  {author} {\bibinfo {author} {\bibfnamefont {K.}~\bibnamefont
  {Lalumiere}}, \bibinfo {author} {\bibfnamefont {B.~C.}\ \bibnamefont
  {Sanders}}, \bibinfo {author} {\bibfnamefont {A.~F.}\ \bibnamefont {van
  Loo}}, \bibinfo {author} {\bibfnamefont {A.}~\bibnamefont {Fedorov}},
  \bibinfo {author} {\bibfnamefont {A.}~\bibnamefont {Wallraff}}, \ and\
  \bibinfo {author} {\bibfnamefont {A.}~\bibnamefont {Blais}},\ }\href
  {https://link.aps.org/doi/10.1103/PhysRevA.88.043806} {\bibfield  {journal}
  {\bibinfo  {journal} {Physical Review A}\ }\textbf {\bibinfo {volume} {88}},\
  \bibinfo {pages} {043806} (\bibinfo {year} {2013})}\BibitemShut {NoStop}%
\bibitem [{\citenamefont {van Loo}\ \emph {et~al.}(2013)\citenamefont {van
  Loo}, \citenamefont {Fedorov}, \citenamefont {Lalumi{\`{e}}re}, \citenamefont
  {Sanders}, \citenamefont {Blais},\ and\ \citenamefont
  {Wallraff}}]{vanLoo2013}%
  \BibitemOpen
  \bibfield  {author} {\bibinfo {author} {\bibfnamefont {A.~F.}\ \bibnamefont
  {van Loo}}, \bibinfo {author} {\bibfnamefont {A.}~\bibnamefont {Fedorov}},
  \bibinfo {author} {\bibfnamefont {K.}~\bibnamefont {Lalumi{\`{e}}re}},
  \bibinfo {author} {\bibfnamefont {B.~C.}\ \bibnamefont {Sanders}}, \bibinfo
  {author} {\bibfnamefont {A.}~\bibnamefont {Blais}}, \ and\ \bibinfo {author}
  {\bibfnamefont {A.}~\bibnamefont {Wallraff}},\ }\href
  {http://science.sciencemag.org/content/342/6165/1494} {\bibfield  {journal}
  {\bibinfo  {journal} {Science}\ }\textbf {\bibinfo {volume} {342}} (\bibinfo
  {year} {2013})}\BibitemShut {NoStop}%
\bibitem [{\citenamefont {Inomata}\ \emph {et~al.}(2014)\citenamefont
  {Inomata}, \citenamefont {Koshino}, \citenamefont {Lin}, \citenamefont
  {Oliver}, \citenamefont {Tsai}, \citenamefont {Nakamura},\ and\ \citenamefont
  {Yamamoto}}]{Inomata2014}%
  \BibitemOpen
  \bibfield  {author} {\bibinfo {author} {\bibfnamefont {K.}~\bibnamefont
  {Inomata}}, \bibinfo {author} {\bibfnamefont {K.}~\bibnamefont {Koshino}},
  \bibinfo {author} {\bibfnamefont {Z.}~\bibnamefont {Lin}}, \bibinfo {author}
  {\bibfnamefont {W.}~\bibnamefont {Oliver}}, \bibinfo {author} {\bibfnamefont
  {J.}~\bibnamefont {Tsai}}, \bibinfo {author} {\bibfnamefont {Y.}~\bibnamefont
  {Nakamura}}, \ and\ \bibinfo {author} {\bibfnamefont {T.}~\bibnamefont
  {Yamamoto}},\ }\href
  {https://link.aps.org/doi/10.1103/PhysRevLett.113.063604} {\bibfield
  {journal} {\bibinfo  {journal} {Physical Review Letters}\ }\textbf {\bibinfo
  {volume} {113}},\ \bibinfo {pages} {063604} (\bibinfo {year}
  {2014})}\BibitemShut {NoStop}%
\bibitem [{\citenamefont {Domokos}\ \emph {et~al.}(2002)\citenamefont
  {Domokos}, \citenamefont {Horak},\ and\ \citenamefont
  {Ritsch}}]{Domokos2002}%
  \BibitemOpen
  \bibfield  {author} {\bibinfo {author} {\bibfnamefont {P.}~\bibnamefont
  {Domokos}}, \bibinfo {author} {\bibfnamefont {P.}~\bibnamefont {Horak}}, \
  and\ \bibinfo {author} {\bibfnamefont {H.}~\bibnamefont {Ritsch}},\ }\href
  {https://link.aps.org/doi/10.1103/PhysRevA.65.033832} {\bibfield  {journal}
  {\bibinfo  {journal} {Physical Review A}\ }\textbf {\bibinfo {volume} {65}},\
  \bibinfo {pages} {033832} (\bibinfo {year} {2002})}\BibitemShut {NoStop}%
\bibitem [{\citenamefont {Horak}\ \emph {et~al.}(2003)\citenamefont {Horak},
  \citenamefont {Domokos},\ and\ \citenamefont {Ritsch}}]{Horak2003}%
  \BibitemOpen
  \bibfield  {author} {\bibinfo {author} {\bibfnamefont {P.}~\bibnamefont
  {Horak}}, \bibinfo {author} {\bibfnamefont {P.}~\bibnamefont {Domokos}}, \
  and\ \bibinfo {author} {\bibfnamefont {H.}~\bibnamefont {Ritsch}},\ }\href
  {http://stacks.iop.org/0295-5075/61/i=4/a=459?key=crossref.9ce75b1397cad502b0b5208379539e81}
  {\bibfield  {journal} {\bibinfo  {journal} {Europhysics Letters (EPL)}\
  }\textbf {\bibinfo {volume} {61}},\ \bibinfo {pages} {459} (\bibinfo {year}
  {2003})}\BibitemShut {NoStop}%
\bibitem [{\citenamefont {Klimov}\ and\ \citenamefont
  {Ducloy}(2004)}]{Klimov2004}%
  \BibitemOpen
  \bibfield  {author} {\bibinfo {author} {\bibfnamefont {V.~V.}\ \bibnamefont
  {Klimov}}\ and\ \bibinfo {author} {\bibfnamefont {M.}~\bibnamefont
  {Ducloy}},\ }\href {https://link.aps.org/doi/10.1103/PhysRevA.69.013812}
  {\bibfield  {journal} {\bibinfo  {journal} {Physical Review A}\ }\textbf
  {\bibinfo {volume} {69}},\ \bibinfo {pages} {013812} (\bibinfo {year}
  {2004})}\BibitemShut {NoStop}%
\bibitem [{\citenamefont {{Le Kien}}\ \emph {et~al.}(2006)\citenamefont {{Le
  Kien}}, \citenamefont {Balykin},\ and\ \citenamefont {Hakuta}}]{LeKien2006}%
  \BibitemOpen
  \bibfield  {author} {\bibinfo {author} {\bibfnamefont {F.}~\bibnamefont {{Le
  Kien}}}, \bibinfo {author} {\bibfnamefont {V.~I.}\ \bibnamefont {Balykin}}, \
  and\ \bibinfo {author} {\bibfnamefont {K.}~\bibnamefont {Hakuta}},\ }\href
  {https://link.aps.org/doi/10.1103/PhysRevA.73.013819} {\bibfield  {journal}
  {\bibinfo  {journal} {Physical Review A}\ }\textbf {\bibinfo {volume} {73}},\
  \bibinfo {pages} {013819} (\bibinfo {year} {2006})}\BibitemShut {NoStop}%
\bibitem [{\citenamefont {Chang}\ \emph {et~al.}(2012)\citenamefont {Chang},
  \citenamefont {Jiang}, \citenamefont {Gorshkov},\ and\ \citenamefont
  {Kimble}}]{Chang2012}%
  \BibitemOpen
  \bibfield  {author} {\bibinfo {author} {\bibfnamefont {D.~E.}\ \bibnamefont
  {Chang}}, \bibinfo {author} {\bibfnamefont {L.}~\bibnamefont {Jiang}},
  \bibinfo {author} {\bibfnamefont {A.~V.}\ \bibnamefont {Gorshkov}}, \ and\
  \bibinfo {author} {\bibfnamefont {H.~J.}\ \bibnamefont {Kimble}},\ }\href
  {http://stacks.iop.org/1367-2630/14/i=6/a=063003?key=crossref.fbdc41306514445fd3c981860d8fd8ec}
  {\bibfield  {journal} {\bibinfo  {journal} {New Journal of Physics}\ }\textbf
  {\bibinfo {volume} {14}},\ \bibinfo {pages} {063003} (\bibinfo {year}
  {2012})}\BibitemShut {NoStop}%
\bibitem [{\citenamefont {Douglas}\ \emph {et~al.}(2015)\citenamefont
  {Douglas}, \citenamefont {Habibian}, \citenamefont {Hung}, \citenamefont
  {Gorshkov}, \citenamefont {Kimble},\ and\ \citenamefont
  {Chang}}]{Douglas2015}%
  \BibitemOpen
  \bibfield  {author} {\bibinfo {author} {\bibfnamefont {J.~S.}\ \bibnamefont
  {Douglas}}, \bibinfo {author} {\bibfnamefont {H.}~\bibnamefont {Habibian}},
  \bibinfo {author} {\bibfnamefont {C.-L.}\ \bibnamefont {Hung}}, \bibinfo
  {author} {\bibfnamefont {A.~V.}\ \bibnamefont {Gorshkov}}, \bibinfo {author}
  {\bibfnamefont {H.~J.}\ \bibnamefont {Kimble}}, \ and\ \bibinfo {author}
  {\bibfnamefont {D.~E.}\ \bibnamefont {Chang}},\ }\href
  {http://dx.doi.org/10.1038/nphoton.2015.57
  http://www.nature.com/doifinder/10.1038/nphoton.2015.57} {\bibfield
  {journal} {\bibinfo  {journal} {Nature Photonics}\ }\textbf {\bibinfo
  {volume} {9}},\ \bibinfo {pages} {326} (\bibinfo {year} {2015})}\BibitemShut
  {NoStop}%
\bibitem [{\citenamefont {Justice}\ \emph {et~al.}(2006)\citenamefont
  {Justice}, \citenamefont {Mock}, \citenamefont {Guo}, \citenamefont
  {Degiron}, \citenamefont {Schurig},\ and\ \citenamefont
  {Smith}}]{Justice2006}%
  \BibitemOpen
  \bibfield  {author} {\bibinfo {author} {\bibfnamefont {B.~J.}\ \bibnamefont
  {Justice}}, \bibinfo {author} {\bibfnamefont {J.~J.}\ \bibnamefont {Mock}},
  \bibinfo {author} {\bibfnamefont {L.}~\bibnamefont {Guo}}, \bibinfo {author}
  {\bibfnamefont {A.}~\bibnamefont {Degiron}}, \bibinfo {author} {\bibfnamefont
  {D.}~\bibnamefont {Schurig}}, \ and\ \bibinfo {author} {\bibfnamefont
  {D.~R.}\ \bibnamefont {Smith}},\ }\href
  {https://www.osapublishing.org/abstract.cfm?URI=oe-14-19-8694} {\bibfield
  {journal} {\bibinfo  {journal} {Optics Express}\ }\textbf {\bibinfo {volume}
  {14}},\ \bibinfo {pages} {8694} (\bibinfo {year} {2006})}\BibitemShut
  {NoStop}%
\bibitem [{\citenamefont {Liu}\ \emph {et~al.}(2010)\citenamefont {Liu},
  \citenamefont {Zentgraf}, \citenamefont {Bartal},\ and\ \citenamefont
  {Zhang}}]{Liu2010}%
  \BibitemOpen
  \bibfield  {author} {\bibinfo {author} {\bibfnamefont {Y.}~\bibnamefont
  {Liu}}, \bibinfo {author} {\bibfnamefont {T.}~\bibnamefont {Zentgraf}},
  \bibinfo {author} {\bibfnamefont {G.}~\bibnamefont {Bartal}}, \ and\ \bibinfo
  {author} {\bibfnamefont {X.}~\bibnamefont {Zhang}},\ }\href
  {http://pubs.acs.org/doi/abs/10.1021/nl1008019} {\bibfield  {journal}
  {\bibinfo  {journal} {Nano Letters}\ }\textbf {\bibinfo {volume} {10}},\
  \bibinfo {pages} {1991} (\bibinfo {year} {2010})}\BibitemShut {NoStop}%
\bibitem [{\citenamefont {Zentgraf}\ \emph {et~al.}(2011)\citenamefont
  {Zentgraf}, \citenamefont {Liu}, \citenamefont {Mikkelsen}, \citenamefont
  {Valentine},\ and\ \citenamefont {Zhang}}]{Zentgraf2011}%
  \BibitemOpen
  \bibfield  {author} {\bibinfo {author} {\bibfnamefont {T.}~\bibnamefont
  {Zentgraf}}, \bibinfo {author} {\bibfnamefont {Y.}~\bibnamefont {Liu}},
  \bibinfo {author} {\bibfnamefont {M.~H.}\ \bibnamefont {Mikkelsen}}, \bibinfo
  {author} {\bibfnamefont {J.}~\bibnamefont {Valentine}}, \ and\ \bibinfo
  {author} {\bibfnamefont {X.}~\bibnamefont {Zhang}},\ }\href
  {http://www.nature.com/doifinder/10.1038/nnano.2010.282} {\bibfield
  {journal} {\bibinfo  {journal} {Nature Nanotechnology}\ }\textbf {\bibinfo
  {volume} {6}},\ \bibinfo {pages} {151} (\bibinfo {year} {2011})}\BibitemShut
  {NoStop}%
\bibitem [{\citenamefont {Glauber}\ and\ \citenamefont
  {Lewenstein}(1991)}]{Glauber1991}%
  \BibitemOpen
  \bibfield  {author} {\bibinfo {author} {\bibfnamefont {R.~J.}\ \bibnamefont
  {Glauber}}\ and\ \bibinfo {author} {\bibfnamefont {M.}~\bibnamefont
  {Lewenstein}},\ }\href {https://link.aps.org/doi/10.1103/PhysRevA.43.467}
  {\bibfield  {journal} {\bibinfo  {journal} {Physical Review A}\ }\textbf
  {\bibinfo {volume} {43}},\ \bibinfo {pages} {467} (\bibinfo {year}
  {1991})}\BibitemShut {NoStop}%
\bibitem [{\citenamefont {Milonni}\ and\ \citenamefont
  {Knight}(1974)}]{Milonni1974}%
  \BibitemOpen
  \bibfield  {author} {\bibinfo {author} {\bibfnamefont {P.~W.}\ \bibnamefont
  {Milonni}}\ and\ \bibinfo {author} {\bibfnamefont {P.~L.}\ \bibnamefont
  {Knight}},\ }\href {https://link.aps.org/doi/10.1103/PhysRevA.10.1096}
  {\bibfield  {journal} {\bibinfo  {journal} {Physical Review A}\ }\textbf
  {\bibinfo {volume} {10}},\ \bibinfo {pages} {1096} (\bibinfo {year}
  {1974})}\BibitemShut {NoStop}%
\bibitem [{\citenamefont {Pellizzari}\ \emph {et~al.}(1995)\citenamefont
  {Pellizzari}, \citenamefont {Gardiner}, \citenamefont {Cirac},\ and\
  \citenamefont {Zoller}}]{Pellizzari1995}%
  \BibitemOpen
  \bibfield  {author} {\bibinfo {author} {\bibfnamefont {T.}~\bibnamefont
  {Pellizzari}}, \bibinfo {author} {\bibfnamefont {S.~A.}\ \bibnamefont
  {Gardiner}}, \bibinfo {author} {\bibfnamefont {J.~I.}\ \bibnamefont {Cirac}},
  \ and\ \bibinfo {author} {\bibfnamefont {P.}~\bibnamefont {Zoller}},\ }\href
  {\doibase 10.1103/PhysRevLett.75.3788} {\bibfield  {journal} {\bibinfo
  {journal} {Physical Review Letters}\ }\textbf {\bibinfo {volume} {75}},\
  \bibinfo {pages} {3788} (\bibinfo {year} {1995})}\BibitemShut {NoStop}%
\bibitem [{\citenamefont {Cirac}\ \emph {et~al.}(1997)\citenamefont {Cirac},
  \citenamefont {Zoller}, \citenamefont {Kimble},\ and\ \citenamefont
  {Mabuchi}}]{Cirac1997}%
  \BibitemOpen
  \bibfield  {author} {\bibinfo {author} {\bibfnamefont {J.~I.}\ \bibnamefont
  {Cirac}}, \bibinfo {author} {\bibfnamefont {P.}~\bibnamefont {Zoller}},
  \bibinfo {author} {\bibfnamefont {H.~J.}\ \bibnamefont {Kimble}}, \ and\
  \bibinfo {author} {\bibfnamefont {H.}~\bibnamefont {Mabuchi}},\ }\href
  {\doibase 10.1103/PhysRevLett.78.3221} {\bibfield  {journal} {\bibinfo
  {journal} {Physical Review Letters}\ }\textbf {\bibinfo {volume} {78}},\
  \bibinfo {pages} {3221} (\bibinfo {year} {1997})}\BibitemShut {NoStop}%
\bibitem [{\citenamefont {Chang}\ \emph {et~al.}(2007)\citenamefont {Chang},
  \citenamefont {S{\o}rensen}, \citenamefont {Demler},\ and\ \citenamefont
  {Lukin}}]{Chang2007}%
  \BibitemOpen
  \bibfield  {author} {\bibinfo {author} {\bibfnamefont {D.~E.}\ \bibnamefont
  {Chang}}, \bibinfo {author} {\bibfnamefont {A.~S.}\ \bibnamefont
  {S{\o}rensen}}, \bibinfo {author} {\bibfnamefont {E.~A.}\ \bibnamefont
  {Demler}}, \ and\ \bibinfo {author} {\bibfnamefont {M.~D.}\ \bibnamefont
  {Lukin}},\ }\href {\doibase 10.1038/nphys708} {\bibfield  {journal} {\bibinfo
   {journal} {Nature Physics}\ }\textbf {\bibinfo {volume} {3}},\ \bibinfo
  {pages} {807} (\bibinfo {year} {2007})}\BibitemShut {NoStop}%
\bibitem [{\citenamefont {Fuhrmanek}\ \emph {et~al.}(2011)\citenamefont
  {Fuhrmanek}, \citenamefont {Bourgain}, \citenamefont {Sortais},\ and\
  \citenamefont {Browaeys}}]{Fuhrmanek2011}%
  \BibitemOpen
  \bibfield  {author} {\bibinfo {author} {\bibfnamefont {A.}~\bibnamefont
  {Fuhrmanek}}, \bibinfo {author} {\bibfnamefont {R.}~\bibnamefont {Bourgain}},
  \bibinfo {author} {\bibfnamefont {Y.~R.~P.}\ \bibnamefont {Sortais}}, \ and\
  \bibinfo {author} {\bibfnamefont {A.}~\bibnamefont {Browaeys}},\ }\href
  {\doibase 10.1103/PhysRevLett.106.133003} {\bibfield  {journal} {\bibinfo
  {journal} {Physical Review Letters}\ }\textbf {\bibinfo {volume} {106}},\
  \bibinfo {pages} {133003} (\bibinfo {year} {2011})}\BibitemShut {NoStop}%
\bibitem [{\citenamefont {Gibbons}\ \emph {et~al.}(2011)\citenamefont
  {Gibbons}, \citenamefont {Hamley}, \citenamefont {Shih},\ and\ \citenamefont
  {Chapman}}]{Gibbons2011}%
  \BibitemOpen
  \bibfield  {author} {\bibinfo {author} {\bibfnamefont {M.~J.}\ \bibnamefont
  {Gibbons}}, \bibinfo {author} {\bibfnamefont {C.~D.}\ \bibnamefont {Hamley}},
  \bibinfo {author} {\bibfnamefont {C.-Y.}\ \bibnamefont {Shih}}, \ and\
  \bibinfo {author} {\bibfnamefont {M.~S.}\ \bibnamefont {Chapman}},\ }\href
  {\doibase 10.1103/PhysRevLett.106.133002} {\bibfield  {journal} {\bibinfo
  {journal} {Physical Review Letters}\ }\textbf {\bibinfo {volume} {106}},\
  \bibinfo {pages} {133002} (\bibinfo {year} {2011})}\BibitemShut {NoStop}%
\bibitem [{\citenamefont {Dung}\ \emph {et~al.}(1998)\citenamefont {Dung},
  \citenamefont {Kn\"oll},\ and\ \citenamefont {Welsch}}]{Dung1998}%
  \BibitemOpen
  \bibfield  {author} {\bibinfo {author} {\bibfnamefont {H.~T.}\ \bibnamefont
  {Dung}}, \bibinfo {author} {\bibfnamefont {L.}~\bibnamefont {Kn\"oll}}, \
  and\ \bibinfo {author} {\bibfnamefont {D.-G.}\ \bibnamefont {Welsch}},\
  }\href {https://link.aps.org/doi/10.1103/PhysRevA.57.3931} {\bibfield
  {journal} {\bibinfo  {journal} {Physical Review A}\ }\textbf {\bibinfo
  {volume} {57}},\ \bibinfo {pages} {3931} (\bibinfo {year}
  {1998})}\BibitemShut {NoStop}%
\bibitem [{\citenamefont {Dung}\ \emph {et~al.}(2002)\citenamefont {Dung},
  \citenamefont {Kn\"oll},\ and\ \citenamefont {Welsch}}]{Dung2002}%
  \BibitemOpen
  \bibfield  {author} {\bibinfo {author} {\bibfnamefont {H.~T.}\ \bibnamefont
  {Dung}}, \bibinfo {author} {\bibfnamefont {L.}~\bibnamefont {Kn\"oll}}, \
  and\ \bibinfo {author} {\bibfnamefont {D.-G.}\ \bibnamefont {Welsch}},\
  }\href {https://link.aps.org/doi/10.1103/PhysRevA.66.063810} {\bibfield
  {journal} {\bibinfo  {journal} {Physical Review A}\ }\textbf {\bibinfo
  {volume} {66}},\ \bibinfo {pages} {063810} (\bibinfo {year}
  {2002})}\BibitemShut {NoStop}%
\bibitem [{\citenamefont {Scheel}\ and\ \citenamefont
  {Buhmann}(2008)}]{Scheel2008}%
  \BibitemOpen
  \bibfield  {author} {\bibinfo {author} {\bibfnamefont {S.}~\bibnamefont
  {Scheel}}\ and\ \bibinfo {author} {\bibfnamefont {S.~Y.}\ \bibnamefont
  {Buhmann}},\ }\href
  {http://scholar.google.com/scholar?hl=en{\&}btnG=Search{\&}q=intitle:acta+physica+slovaca{\#}1{\%}5Cnhttp://scholar.google.com/scholar?hl=en{\&}btnG=Search{\&}q=intitle:Acta+Physica+Slovaca{\#}1}
  {\bibfield  {journal} {\bibinfo  {journal} {Acta Physica Slovaca}\ }\textbf
  {\bibinfo {volume} {58}},\ \bibinfo {pages} {675} (\bibinfo {year}
  {2008})}\BibitemShut {NoStop}%
\bibitem [{\citenamefont {Buhmann}(2012)}]{Buhmann2012}%
  \BibitemOpen
  \bibfield  {author} {\bibinfo {author} {\bibfnamefont {S.~Y.}\ \bibnamefont
  {Buhmann}},\ }\href {http://link.springer.com/10.1007/978-3-642-32484-0{\_}2}
  {\emph {\bibinfo {title} {Dispersion Forces I. Springer Tracts in Modern
  Physics, vol 247}}}\ (\bibinfo  {publisher} {Springer, Berlin, Heidelberg},\
  \bibinfo {year} {2012})\BibitemShut {NoStop}%
\bibitem [{\citenamefont {Gross}\ and\ \citenamefont
  {Haroche}(1982)}]{Gross1982}%
  \BibitemOpen
  \bibfield  {author} {\bibinfo {author} {\bibfnamefont {M.}~\bibnamefont
  {Gross}}\ and\ \bibinfo {author} {\bibfnamefont {S.}~\bibnamefont
  {Haroche}},\ }\href
  {http://linkinghub.elsevier.com/retrieve/pii/0370157382901028} {\bibfield
  {journal} {\bibinfo  {journal} {Physics Reports}\ }\textbf {\bibinfo {volume}
  {93}},\ \bibinfo {pages} {301} (\bibinfo {year} {1982})}\BibitemShut
  {NoStop}%
\bibitem [{Note1()}]{Note1}%
  \BibitemOpen
  \bibinfo {note} {Note that since the Green's function in Eq.~(\ref
  {analytic}) is purely real, the dipole-dipole interaction is simply
  proportional to the electric field strength.}\BibitemShut {Stop}%
\bibitem [{\citenamefont {Erd{\'e}lyi}(1953)}]{Erdelyi1953}%
  \BibitemOpen
  \bibinfo {editor} {\bibfnamefont {A.}~\bibnamefont {Erd{\'e}lyi}},\ ed.,\
  \href@noop {} {\emph {\bibinfo {title} {Higher Transcendental Functions,}}},\
  Vol.~\bibinfo {volume} {1}\ (\bibinfo  {publisher} {McGraw-Hill, New
  York-London},\ \bibinfo {year} {1953})\BibitemShut {NoStop}%
\bibitem [{\citenamefont {Brennen}\ \emph {et~al.}(2000)\citenamefont
  {Brennen}, \citenamefont {Deutsch},\ and\ \citenamefont
  {Jessen}}]{Brennen2000}%
  \BibitemOpen
  \bibfield  {author} {\bibinfo {author} {\bibfnamefont {G.~K.}\ \bibnamefont
  {Brennen}}, \bibinfo {author} {\bibfnamefont {I.~H.}\ \bibnamefont
  {Deutsch}}, \ and\ \bibinfo {author} {\bibfnamefont {P.~S.}\ \bibnamefont
  {Jessen}},\ }\href {https://link.aps.org/doi/10.1103/PhysRevA.61.062309}
  {\bibfield  {journal} {\bibinfo  {journal} {Physical Review A}\ }\textbf
  {\bibinfo {volume} {61}},\ \bibinfo {pages} {062309} (\bibinfo {year}
  {2000})}\BibitemShut {NoStop}%
\bibitem [{\citenamefont {Dolde}\ \emph {et~al.}(2013)\citenamefont {Dolde},
  \citenamefont {Jakobi}, \citenamefont {Naydenov}, \citenamefont {Zhao},
  \citenamefont {Pezzagna}, \citenamefont {Trautmann}, \citenamefont {Meijer},
  \citenamefont {Neumann}, \citenamefont {Jelezko},\ and\ \citenamefont
  {Wrachtrup}}]{Dolde2013}%
  \BibitemOpen
  \bibfield  {author} {\bibinfo {author} {\bibfnamefont {F.}~\bibnamefont
  {Dolde}}, \bibinfo {author} {\bibfnamefont {I.}~\bibnamefont {Jakobi}},
  \bibinfo {author} {\bibfnamefont {B.}~\bibnamefont {Naydenov}}, \bibinfo
  {author} {\bibfnamefont {N.}~\bibnamefont {Zhao}}, \bibinfo {author}
  {\bibfnamefont {S.}~\bibnamefont {Pezzagna}}, \bibinfo {author}
  {\bibfnamefont {C.}~\bibnamefont {Trautmann}}, \bibinfo {author}
  {\bibfnamefont {J.}~\bibnamefont {Meijer}}, \bibinfo {author} {\bibfnamefont
  {P.}~\bibnamefont {Neumann}}, \bibinfo {author} {\bibfnamefont
  {F.}~\bibnamefont {Jelezko}}, \ and\ \bibinfo {author} {\bibfnamefont
  {J.}~\bibnamefont {Wrachtrup}},\ }\href
  {http://www.nature.com/doifinder/10.1038/nphys2545} {\bibfield  {journal}
  {\bibinfo  {journal} {Nature Physics}\ }\textbf {\bibinfo {volume} {9}},\
  \bibinfo {pages} {139} (\bibinfo {year} {2013})}\BibitemShut {NoStop}%
\bibitem [{\citenamefont {Sipahigil}\ \emph {et~al.}(2014)\citenamefont
  {Sipahigil}, \citenamefont {Jahnke}, \citenamefont {Rogers}, \citenamefont
  {Teraji}, \citenamefont {Isoya}, \citenamefont {Zibrov}, \citenamefont
  {Jelezko},\ and\ \citenamefont {Lukin}}]{Sipahigil2014}%
  \BibitemOpen
  \bibfield  {author} {\bibinfo {author} {\bibfnamefont {A.}~\bibnamefont
  {Sipahigil}}, \bibinfo {author} {\bibfnamefont {K.}~\bibnamefont {Jahnke}},
  \bibinfo {author} {\bibfnamefont {L.}~\bibnamefont {Rogers}}, \bibinfo
  {author} {\bibfnamefont {T.}~\bibnamefont {Teraji}}, \bibinfo {author}
  {\bibfnamefont {J.}~\bibnamefont {Isoya}}, \bibinfo {author} {\bibfnamefont
  {A.}~\bibnamefont {Zibrov}}, \bibinfo {author} {\bibfnamefont
  {F.}~\bibnamefont {Jelezko}}, \ and\ \bibinfo {author} {\bibfnamefont
  {M.}~\bibnamefont {Lukin}},\ }\href
  {https://link.aps.org/doi/10.1103/PhysRevLett.113.113602} {\bibfield
  {journal} {\bibinfo  {journal} {Physical Review Letters}\ }\textbf {\bibinfo
  {volume} {113}},\ \bibinfo {pages} {113602} (\bibinfo {year}
  {2014})}\BibitemShut {NoStop}%
\bibitem [{\citenamefont {Kolkowitz}\ \emph {et~al.}(2015)\citenamefont
  {Kolkowitz}, \citenamefont {Safira}, \citenamefont {High}, \citenamefont
  {Devlin}, \citenamefont {Choi}, \citenamefont {Unterreithmeier},
  \citenamefont {Patterson}, \citenamefont {Zibrov}, \citenamefont
  {Manucharyan}, \citenamefont {Park},\ and\ \citenamefont
  {Lukin}}]{Kolkowitz2015}%
  \BibitemOpen
  \bibfield  {author} {\bibinfo {author} {\bibfnamefont {S.}~\bibnamefont
  {Kolkowitz}}, \bibinfo {author} {\bibfnamefont {A.}~\bibnamefont {Safira}},
  \bibinfo {author} {\bibfnamefont {A.~A.}\ \bibnamefont {High}}, \bibinfo
  {author} {\bibfnamefont {R.~C.}\ \bibnamefont {Devlin}}, \bibinfo {author}
  {\bibfnamefont {S.}~\bibnamefont {Choi}}, \bibinfo {author} {\bibfnamefont
  {Q.~P.}\ \bibnamefont {Unterreithmeier}}, \bibinfo {author} {\bibfnamefont
  {D.}~\bibnamefont {Patterson}}, \bibinfo {author} {\bibfnamefont {A.~S.}\
  \bibnamefont {Zibrov}}, \bibinfo {author} {\bibfnamefont {V.~E.}\
  \bibnamefont {Manucharyan}}, \bibinfo {author} {\bibfnamefont
  {H.}~\bibnamefont {Park}}, \ and\ \bibinfo {author} {\bibfnamefont {M.~D.}\
  \bibnamefont {Lukin}},\ }\href@noop {} {\bibfield  {journal} {\bibinfo
  {journal} {Science}\ }\textbf {\bibinfo {volume} {347}} (\bibinfo {year}
  {2015})}\BibitemShut {NoStop}%
\bibitem [{\citenamefont {Iwasaki}\ \emph {et~al.}(2015)\citenamefont
  {Iwasaki}, \citenamefont {Ishibashi}, \citenamefont {Miyamoto}, \citenamefont
  {Doi}, \citenamefont {Kobayashi}, \citenamefont {Miyazaki}, \citenamefont
  {Tahara}, \citenamefont {Jahnke}, \citenamefont {D.~Rogers}, \citenamefont
  {J.~Naydenov}, \citenamefont {Jelezko}, \citenamefont {Yamasaki},
  \citenamefont {Nagamachi}, \citenamefont {Inubushi}, \citenamefont
  {Mizuochi},\ and\ \citenamefont {Hatano}}]{Iwasaki2015}%
  \BibitemOpen
  \bibfield  {author} {\bibinfo {author} {\bibfnamefont {T.}~\bibnamefont
  {Iwasaki}}, \bibinfo {author} {\bibfnamefont {F.}~\bibnamefont {Ishibashi}},
  \bibinfo {author} {\bibfnamefont {Y.}~\bibnamefont {Miyamoto}}, \bibinfo
  {author} {\bibfnamefont {Y.}~\bibnamefont {Doi}}, \bibinfo {author}
  {\bibfnamefont {S.}~\bibnamefont {Kobayashi}}, \bibinfo {author}
  {\bibfnamefont {T.}~\bibnamefont {Miyazaki}}, \bibinfo {author}
  {\bibfnamefont {K.}~\bibnamefont {Tahara}}, \bibinfo {author} {\bibfnamefont
  {K.}~\bibnamefont {Jahnke}}, \bibinfo {author} {\bibfnamefont
  {L.}~\bibnamefont {D.~Rogers}}, \bibinfo {author} {\bibfnamefont
  {B.}~\bibnamefont {J.~Naydenov}}, \bibinfo {author} {\bibfnamefont
  {F.}~\bibnamefont {Jelezko}}, \bibinfo {author} {\bibfnamefont
  {S.}~\bibnamefont {Yamasaki}}, \bibinfo {author} {\bibfnamefont
  {S.}~\bibnamefont {Nagamachi}}, \bibinfo {author} {\bibfnamefont
  {T.}~\bibnamefont {Inubushi}}, \bibinfo {author} {\bibfnamefont
  {N.}~\bibnamefont {Mizuochi}}, \ and\ \bibinfo {author} {\bibfnamefont
  {M.}~\bibnamefont {Hatano}},\ }\href
  {http://www.nature.com/articles/srep12882} {\bibfield  {journal} {\bibinfo
  {journal} {Scientific Reports}\ }\textbf {\bibinfo {volume} {5}},\ \bibinfo
  {pages} {12882} (\bibinfo {year} {2015})}\BibitemShut {NoStop}%
\bibitem [{\citenamefont {Sipahigil}\ \emph {et~al.}(2016)\citenamefont
  {Sipahigil}, \citenamefont {Evans}, \citenamefont {Sukachev}, \citenamefont
  {Burek}, \citenamefont {Borregaard}, \citenamefont {Bhaskar}, \citenamefont
  {Nguyen}, \citenamefont {Pacheco}, \citenamefont {Atikian}, \citenamefont
  {Meuwly}, \citenamefont {Camacho}, \citenamefont {Jelezko}, \citenamefont
  {Bielejec}, \citenamefont {Park}, \citenamefont {Lon{\v{c}}ar},\ and\
  \citenamefont {Lukin}}]{Sipahigil2016}%
  \BibitemOpen
  \bibfield  {author} {\bibinfo {author} {\bibfnamefont {A.}~\bibnamefont
  {Sipahigil}}, \bibinfo {author} {\bibfnamefont {R.~E.}\ \bibnamefont
  {Evans}}, \bibinfo {author} {\bibfnamefont {D.~D.}\ \bibnamefont {Sukachev}},
  \bibinfo {author} {\bibfnamefont {M.~J.}\ \bibnamefont {Burek}}, \bibinfo
  {author} {\bibfnamefont {J.}~\bibnamefont {Borregaard}}, \bibinfo {author}
  {\bibfnamefont {M.~K.}\ \bibnamefont {Bhaskar}}, \bibinfo {author}
  {\bibfnamefont {C.~T.}\ \bibnamefont {Nguyen}}, \bibinfo {author}
  {\bibfnamefont {J.~L.}\ \bibnamefont {Pacheco}}, \bibinfo {author}
  {\bibfnamefont {H.~A.}\ \bibnamefont {Atikian}}, \bibinfo {author}
  {\bibfnamefont {C.}~\bibnamefont {Meuwly}}, \bibinfo {author} {\bibfnamefont
  {R.~M.}\ \bibnamefont {Camacho}}, \bibinfo {author} {\bibfnamefont
  {F.}~\bibnamefont {Jelezko}}, \bibinfo {author} {\bibfnamefont
  {E.}~\bibnamefont {Bielejec}}, \bibinfo {author} {\bibfnamefont
  {H.}~\bibnamefont {Park}}, \bibinfo {author} {\bibfnamefont {M.}~\bibnamefont
  {Lon{\v{c}}ar}}, \ and\ \bibinfo {author} {\bibfnamefont {M.~D.}\
  \bibnamefont {Lukin}},\ }\href {http://arxiv.org/abs/1608.05147
  http://science.sciencemag.org/content/early/2016/10/12/science.aah6875
  http://www.sciencemag.org/lookup/doi/10.1126/science.aah6875} {\bibfield
  {journal} {\bibinfo  {journal} {Science}\ }\textbf {\bibinfo {volume}
  {354}},\ \bibinfo {pages} {847} (\bibinfo {year} {2016})}\BibitemShut
  {NoStop}%
\bibitem [{\citenamefont {High}\ \emph {et~al.}(2015)\citenamefont {High},
  \citenamefont {Devlin}, \citenamefont {Dibos}, \citenamefont {Polking},
  \citenamefont {Wild}, \citenamefont {Perczel}, \citenamefont {de~Leon},
  \citenamefont {Lukin},\ and\ \citenamefont {Park}}]{High2015}%
  \BibitemOpen
  \bibfield  {author} {\bibinfo {author} {\bibfnamefont {A.~A.}\ \bibnamefont
  {High}}, \bibinfo {author} {\bibfnamefont {R.~C.}\ \bibnamefont {Devlin}},
  \bibinfo {author} {\bibfnamefont {A.}~\bibnamefont {Dibos}}, \bibinfo
  {author} {\bibfnamefont {M.}~\bibnamefont {Polking}}, \bibinfo {author}
  {\bibfnamefont {D.~S.}\ \bibnamefont {Wild}}, \bibinfo {author}
  {\bibfnamefont {J.}~\bibnamefont {Perczel}}, \bibinfo {author} {\bibfnamefont
  {N.~P.}\ \bibnamefont {de~Leon}}, \bibinfo {author} {\bibfnamefont {M.~D.}\
  \bibnamefont {Lukin}}, \ and\ \bibinfo {author} {\bibfnamefont
  {H.}~\bibnamefont {Park}},\ }\href
  {http://www.nature.com/doifinder/10.1038/nature14477} {\bibfield  {journal}
  {\bibinfo  {journal} {Nature}\ }\textbf {\bibinfo {volume} {522}},\ \bibinfo
  {pages} {192} (\bibinfo {year} {2015})}\BibitemShut {NoStop}%
\bibitem [{\citenamefont {des Francs}\ \emph {et~al.}(2016)\citenamefont {des
  Francs}, \citenamefont {Barthes}, \citenamefont {Bouhelier}, \citenamefont
  {Weeber}, \citenamefont {Dereux}, \citenamefont {Cuche},\ and\ \citenamefont
  {Girard}}]{Francs2016}%
  \BibitemOpen
  \bibfield  {author} {\bibinfo {author} {\bibfnamefont {G.~C.}\ \bibnamefont
  {des Francs}}, \bibinfo {author} {\bibfnamefont {J.}~\bibnamefont {Barthes}},
  \bibinfo {author} {\bibfnamefont {A.}~\bibnamefont {Bouhelier}}, \bibinfo
  {author} {\bibfnamefont {J.~C.}\ \bibnamefont {Weeber}}, \bibinfo {author}
  {\bibfnamefont {A.}~\bibnamefont {Dereux}}, \bibinfo {author} {\bibfnamefont
  {A.}~\bibnamefont {Cuche}}, \ and\ \bibinfo {author} {\bibfnamefont
  {C.}~\bibnamefont {Girard}},\ }\href
  {http://stacks.iop.org/2040-8986/18/i=9/a=094005?key=crossref.64eda5ab8e6dcface0acde747ccdd224}
  {\bibfield  {journal} {\bibinfo  {journal} {Journal of Optics}\ }\textbf
  {\bibinfo {volume} {18}},\ \bibinfo {pages} {094005} (\bibinfo {year}
  {2016})}\BibitemShut {NoStop}%
\bibitem [{Note2()}]{Note2}%
  \BibitemOpen
  \bibinfo {note} {We calculate the Purcell factor exactly near a flat metal
  surface by evaluating ${\gamma _\protect \text {pl}=2d_z^2\omega _0^2/(\hbar
  \varepsilon _0c^2)\protect \text {Im}\protect \{G_{zz}\protect \}}$, where
  $G_{zz}$ is the exact Green's function of a z-oriented dipole near the
  surface \cite {Sipe1987}}\BibitemShut {NoStop}%
\bibitem [{\citenamefont {Arcari}\ \emph {et~al.}(2014)\citenamefont {Arcari},
  \citenamefont {Sollner}, \citenamefont {Javadi}, \citenamefont {{Lindskov
  Hansen}}, \citenamefont {Mahmoodian}, \citenamefont {Liu}, \citenamefont
  {Thyrrestrup}, \citenamefont {Lee}, \citenamefont {Song}, \citenamefont
  {Stobbe},\ and\ \citenamefont {Lodahl}}]{Arcari2014}%
  \BibitemOpen
  \bibfield  {author} {\bibinfo {author} {\bibfnamefont {M.}~\bibnamefont
  {Arcari}}, \bibinfo {author} {\bibfnamefont {I.}~\bibnamefont {Sollner}},
  \bibinfo {author} {\bibfnamefont {A.}~\bibnamefont {Javadi}}, \bibinfo
  {author} {\bibfnamefont {S.}~\bibnamefont {{Lindskov Hansen}}}, \bibinfo
  {author} {\bibfnamefont {S.}~\bibnamefont {Mahmoodian}}, \bibinfo {author}
  {\bibfnamefont {J.}~\bibnamefont {Liu}}, \bibinfo {author} {\bibfnamefont
  {H.}~\bibnamefont {Thyrrestrup}}, \bibinfo {author} {\bibfnamefont
  {E.}~\bibnamefont {Lee}}, \bibinfo {author} {\bibfnamefont {J.}~\bibnamefont
  {Song}}, \bibinfo {author} {\bibfnamefont {S.}~\bibnamefont {Stobbe}}, \ and\
  \bibinfo {author} {\bibfnamefont {P.}~\bibnamefont {Lodahl}},\ }\href
  {https://link.aps.org/doi/10.1103/PhysRevLett.113.093603} {\bibfield
  {journal} {\bibinfo  {journal} {Physical Review Letters}\ }\textbf {\bibinfo
  {volume} {113}},\ \bibinfo {pages} {093603} (\bibinfo {year}
  {2014})}\BibitemShut {NoStop}%
\bibitem [{\citenamefont {Abramowitz}\ and\ \citenamefont
  {Stegun}(1970)}]{Abramowitz1970}%
  \BibitemOpen
  \bibfield  {author} {\bibinfo {author} {\bibfnamefont {M.}~\bibnamefont
  {Abramowitz}}\ and\ \bibinfo {author} {\bibfnamefont {I.~A.}\ \bibnamefont
  {Stegun}},\ }\href@noop {} {\emph {\bibinfo {title} {{Handbook of
  mathematical functions: with formulas, graphs, and mathematical tables}}}}\
  (\bibinfo  {publisher} {Dover Publications},\ \bibinfo {year}
  {1970})\BibitemShut {NoStop}%
\bibitem [{\citenamefont {Fredholm}(1903)}]{Fredholm1903}%
  \BibitemOpen
  \bibfield  {author} {\bibinfo {author} {\bibfnamefont {I.}~\bibnamefont
  {Fredholm}},\ }\href {http://projecteuclid.org/euclid.acta/1485882170}
  {\bibfield  {journal} {\bibinfo  {journal} {Acta Mathematica}\ }\textbf
  {\bibinfo {volume} {27}},\ \bibinfo {pages} {365} (\bibinfo {year}
  {1903})}\BibitemShut {NoStop}%
\bibitem [{\citenamefont {Sipe}(1987)}]{Sipe1987}%
  \BibitemOpen
  \bibfield  {author} {\bibinfo {author} {\bibfnamefont {J.~E.}\ \bibnamefont
  {Sipe}},\ }\href@noop {} {\bibfield  {journal} {\bibinfo  {journal} {Journal
  of the Optical Society of America B}\ }\textbf {\bibinfo {volume} {4}},\
  \bibinfo {pages} {481} (\bibinfo {year} {1987})}\BibitemShut {NoStop}%
\end{thebibliography}
\end{document}